 \documentclass[authoryear,preprint,review,12pt]{elsarticle}

 \usepackage{dblfloatfix}
 \usepackage{graphicx}

\usepackage{pifont,latexsym,ifthen,rotating,calc,textcase,booktabs,color}
\usepackage{amssymb}
\usepackage{amsfonts}
\usepackage{amsbsy}
\usepackage{amsmath}

\usepackage{float}
\usepackage{upgreek}
\usepackage{subfigure}
\usepackage{tikz}

\usepackage{textcomp}

\usepackage{placeins}

\usepackage{ifthen}
\usepackage{bigstrut}
\def\path{\string~/}
\usepackage{bm}

\journal{Journal of Loss Prevention in the Process Industries}

\begin{document}

\begin{frontmatter}

\title{Simulation of the Flow of an Explosive Atmosphere Exposed to a Hot Surface}

\author[label1,label2]{Subrahmanyeswara Velagala}
\author[label1,label2]{Priyank Raval}
\author[label1,label2]{Sai Charan Singh Chowhan}
\author[label1]{Ghazaleh Esmaeelzade}
\author[label1]{Michael Beyer}
\author[label1,label2]{Holger Grosshans\corref{cor1}}
\ead{holger.grosshans@ptb.de}
\cortext[cor1]{Corresponding author}

\address[label1]{Physikalisch-Technische Bundesanstalt (PTB), Braunschweig, Germany}
\address[label2]{Institute of Apparatus- and Environmental Technology, Otto von Guericke University of Magdeburg, Germany}

\begin{abstract}
The accidental ignition of combustible atmospheres by hot surfaces is of great concern for chemical and process plant safety.
In this paper, we present our research regarding the evolution of thermal plumes originating from hot hemispheres and discs.
In particular, we focus on the effect of the orientation of the surface on the ignition process.
The auto-ignition temperatures and ignition locations were studied experimentally.
To get further insight, we conducted detailed numerical simulations and validated \textcolor{black}{them} with measurements.
Three-dimensional simulations were performed on hot hemispheres and hot discs for different orientations ranging from 0$^{\circ}$ to 180$^{\circ}$.
The solver employs a transient, implicit scheme which is based on the coupled heat transfer and flow equations.
The mesh in the vicinity of the hot surfaces is refined to resolve the steep temperature gradients and to capture the boundary layer separation. 
The influence of the orientation on critical hot spots in the gas mixture is analyzed by examining the flow structures and the temperature evolution of the buoyancy-driven flow.
Using the obtained results, we discuss the change of the onset and location of the ignition.
\end{abstract}


\end{frontmatter}


\section{Introduction}

The accidental ignition of combustible atmospheres by hot surfaces which causes disastrous fires or explosions is a great concern in many industries~\citep{ECKHOFF1994281}.
For this reason, it is important to understand the underlying physics of the heat transfer from a hot surface towards the flammable mixture.
A measure to characterize the ignition hazard of an atmosphere in the absence of a spark or flame is the auto-ignition temperature (AIT).
The AIT is defined as the maximum acceptable surface temperature in a particular area to prevent fires and explosions~\citep{Chen,ECKHOFF1994281}.
Typically, the AIT is measured by placing the substance in a half-liter vessel and inside a temperature-controlled oven, where the rate of heating is not relevant.
At first, the fuel/air mixture in the vicinity of the hot surface heats up and the surrounding air remains at the ambient temperatures.
The resulting temperature gradient and, thus, the density gradient near the surface induces buoyant forces acting on the gas.
Then, the affected gas acquires velocity and convects in the opposite direction to the gravity vector away from the high-temperature region.
In order to ignite, a heated fuel/air packet needs to obtain a sufficiently high temperature for a certain amount of time which is called the residence time.
However, the concept of the AIT is defined assuming the fuel/air packet to remain in a specific condition for an indefinitely long period.
Therefore, ignition in real applications requires a higher temperature than the AIT in an environment where because of buoyancy the gas is constantly accelerated~\citep{Gro20a,Babrauskas2008} 

The earliest experimental study by \citet{Coward1927} concerned the effect of the material of the hot surface on ignition.
Through the years a range of researchers shed light on different aspects of the topic:
\citet{ashman1961ignition} conducted investigations using a cylindrical wire located in the combustible atmosphere and heated it until the mixture ignited.
The ignition of gases through a suddenly heated circular rod has been examined by \citet{adomeit1965ignition} for different equivalence ratios and gas pressures.
Then, the influence of the hot surface area and the diameter of the hot sphere were studied experimentally by \citet{Laurendeau1982}.
More recently, \citet{Babrauskas2008} concluded that the actual surface temperature required for ignition does highly depend on the degree of enclosedness.
More specifically, higher temperatures are needed to start the ignition process as the degree of enclosedness is increased.
Another recent experimental work on hot hemispheres by \citet{Gro20a} investigated the effect of the orientation on the auto-ignition temperatures and ignition locations of carbon disulfide and Diethyl ether.
It was observed that the ignition location changes with the orientation of the hot surface and the highest values of the hot surface ignition temperature were observed for an orientation of 180$^{\circ}$.
The precise reasons for this behavior and the ignition location remain to date unclear.

Due to the development of numerical tools and the increase of computational power, simulations represent today an attractive approach to study ignition-related processes (for example \citet{Esm17b,Gro18i}).
As regards hot surfaces, \citet{Griffiths2019} analyzed the influence of the size of the surface on the ignition temperature.
The influence of the surface temperature to ignite the ethylene-air mixture by horizontal and vertical cylinders was the focus of the computations of \citet{Melguizo,Melg16}.
Further, the influence of viscosity and buoyancy on transient natural convection over a sphere was examined by \citet{Jia2007}.
The ignition of a cold combustible mixture of propane/air has been investigated for the stagnation region of a hot projectile by \citet{Sharma2007}.

Nevertheless, little attention has been paid so far on spherical and flat surfaces and their orientation effects on the ignition and ignition locations.
In the present study, three-dimensional numerical simulations of thermal plumes originating from hot hemispheres and hot discs with each of five different orientations are discussed.
The effects of the orientation on critical hot spots are presented by examining flow structures and the temporal evolution of the temperature and velocity fields.
Also, the results are validated against experimental data.

\section{Description of the Computational Fluid Dynamics (CFD) \textcolor{black}{simulations}}

The examination of the complex buoyancy-driven gas flow and heat transfer inside the combustion chamber is performed using the \textit{BuoyantPimpleFoam} solver which is part of the OpenFOAM toolbox.
The mathematical model and techniques to solve the equations are detailed in the following.

\subsection{Governing equations and numerical methods}

The mathematical model to predict the evolution of thermal hot spots inside the combustion chamber couples the transient heat transfer to the laminar flow.
To account for stratification, the gas mixture is considered to be quasi-compressible, including density as an explicit variable in the calculation.
Thus, the physics of the flow is described by the governing equations of CFD, namely the continuity, momentum, and energy equations.
In Cartesian coordinates, these equations read

\begin{equation} 
 \label{eq:Mass}
 \frac{\partial \rho}{\partial t} + \nabla \cdot \left( \rho {\bm u} \right) = 0  \, ,
\end{equation}
\begin{equation} 
 \label{eq:Momentum}
 \frac{\partial \left( \rho {\bm u} \right)}{\partial t} + \nabla \cdot \left(\rho {\bm u} {\bm u}\right) = -\nabla p + \rho {\bm g} + \mu\left(\nabla^2 \cdot {\bm u} \right) \, .
\end{equation}
The pressure gradient and gravity force in the momentum equation (Eq.~(\ref{eq:Momentum})) are rearranged as 
\begin{equation}
 \label{eq:Momentum2}
 \nabla p + \rho {\bm g} = \nabla p_\mathrm{rgh} - \left( {\bm g} \cdot {\bm r} \right) \nabla \rho  \, ,
\end{equation}
where $p_\mathrm{rgh}$ is the dynamic pressure which is given by
\begin{equation}
 p_\mathrm{rgh} = p - \rho {\bm g} \cdot {\bm r}  \, .
\end{equation}
In the above equations, ${\bm u}$ is the fluid velocity, $\rho$ is the density, $p$ is the static pressure, ${\bm g}$ is the gravitational acceleration, $\mu $ is the dynamic viscosity, and ${\bm r}$ is the position vector.

The thermal energy transport is described through the enthalpy equation
\begin{equation} 
 \label{eq:Energy}
 \frac{\partial \left(\rho h \right)}{\partial t} + \nabla \cdot \left( \rho {\bm u} h \right) + \frac{\partial \left(\rho K \right)}{\partial t} + \nabla \cdot \left (\rho {\bm u} K \right) -\frac{\partial p}{\partial t} = \nabla \cdot \left( \alpha \nabla h \right) + \rho {\bm u} \cdot {\bm g} \, ,
\end{equation}
where $K$ and $h$ are the kinetic energy and enthalpy per unit mass, respectively.
The thermal diffusivity, $\alpha$, is estimated based on the ratio of viscosity and  Prandtl number, i.e.
\begin{equation}
 \alpha =  \frac{\mu}{Pr} \, .
\end{equation}

Finally, to close the system of equations the ideal gas law is solved, which reads
\begin{equation}
 \label{eq:Ideal gas}
 p=\rho R T  \, ,
\end{equation}
where the specific gas constant is $R=$~287~J/(kg\,K).

As regards the boundary conditions, no-slip velocity and zero pressure gradient are assumed at all surfaces, i.e.~both the hot surface and the chamber walls.
Initially, the temperature at the hot surface is set to room temperature, 293.65~K.
Then, it is gradually and uniformly increased with time according to the procedure applied in the experiments.
At the walls of the combustion chamber, a uniform fixed temperature of 293.65~K is applied.

\textcolor{black}{
The divergence terms in Eqs.~(\ref{eq:Mass}) and~(\ref{eq:Momentum}) are discretized by first-order upwind schemes.
The gradients in Eqs.~(\ref{eq:Momentum}) and~(\ref{eq:Momentum2}) and the Laplacian in Eq.~(\ref{eq:Momentum}) are approximated by second-order central differences.
In Eq.~(\ref{eq:Energy}), the divergence terms involving $K$ are discretized by second-order central and those involving $h$ by first-order upwind schemes.
Further, all temporal derivatives in the governing equations are approximated by a first-order accurate Euler scheme.
For all simulations we used a numerical time-step size of 0.1~ms which corresponds to a very small mean Courant number of~0.004.
Thus, the time resolution of the simulations is very high.
}

As a final remark, it is noted that ignition and combustion are not considered in the present simulations.

\subsection{Geometry and mesh generation}

\begin{figure}[b]
\centering
\subfigure[]{
\includegraphics[trim=10mm 0mm 10mm 0mm,clip=true,height=6.1cm]{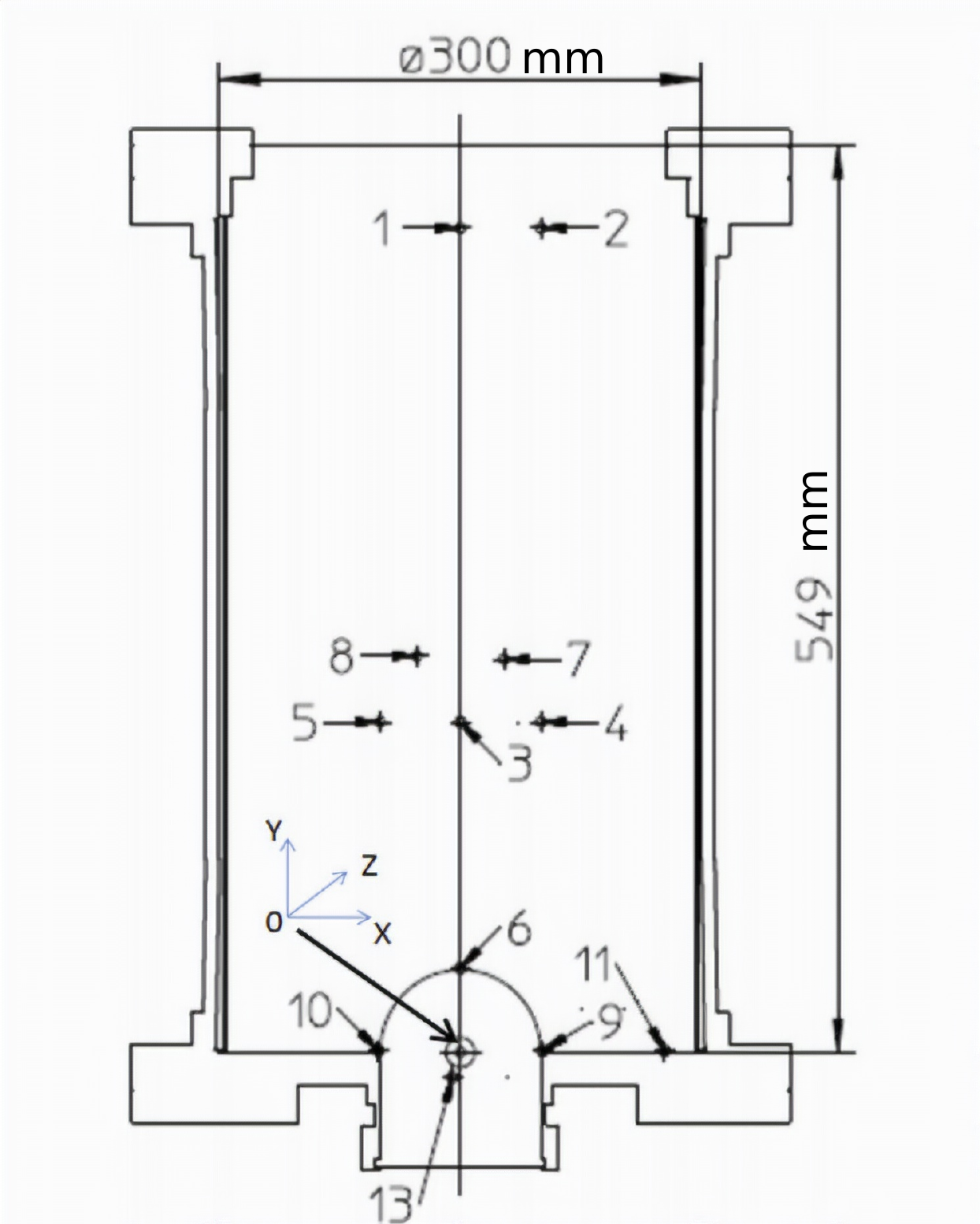}
\includegraphics[trim=10mm 0mm 10mm 0mm,clip=true,height=6cm]{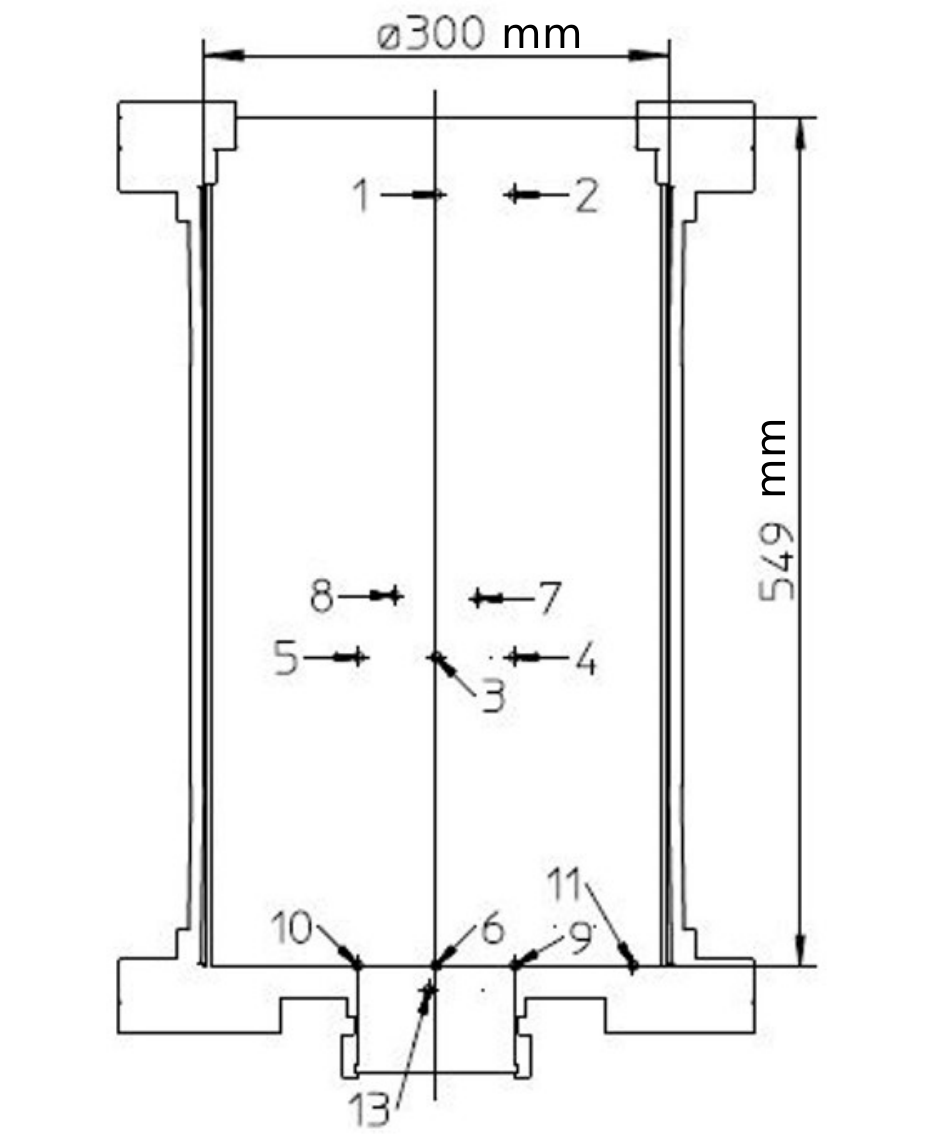}
\label{fig:HemiSphere_thermocouples}
}
\qquad
\subfigure[]{
\begin{tikzpicture}[thick]
\node [anchor=south,inner sep=0] at (0,0) {\includegraphics[trim=192mm 20mm 192mm 16mm,clip=true,height=4.6cm]{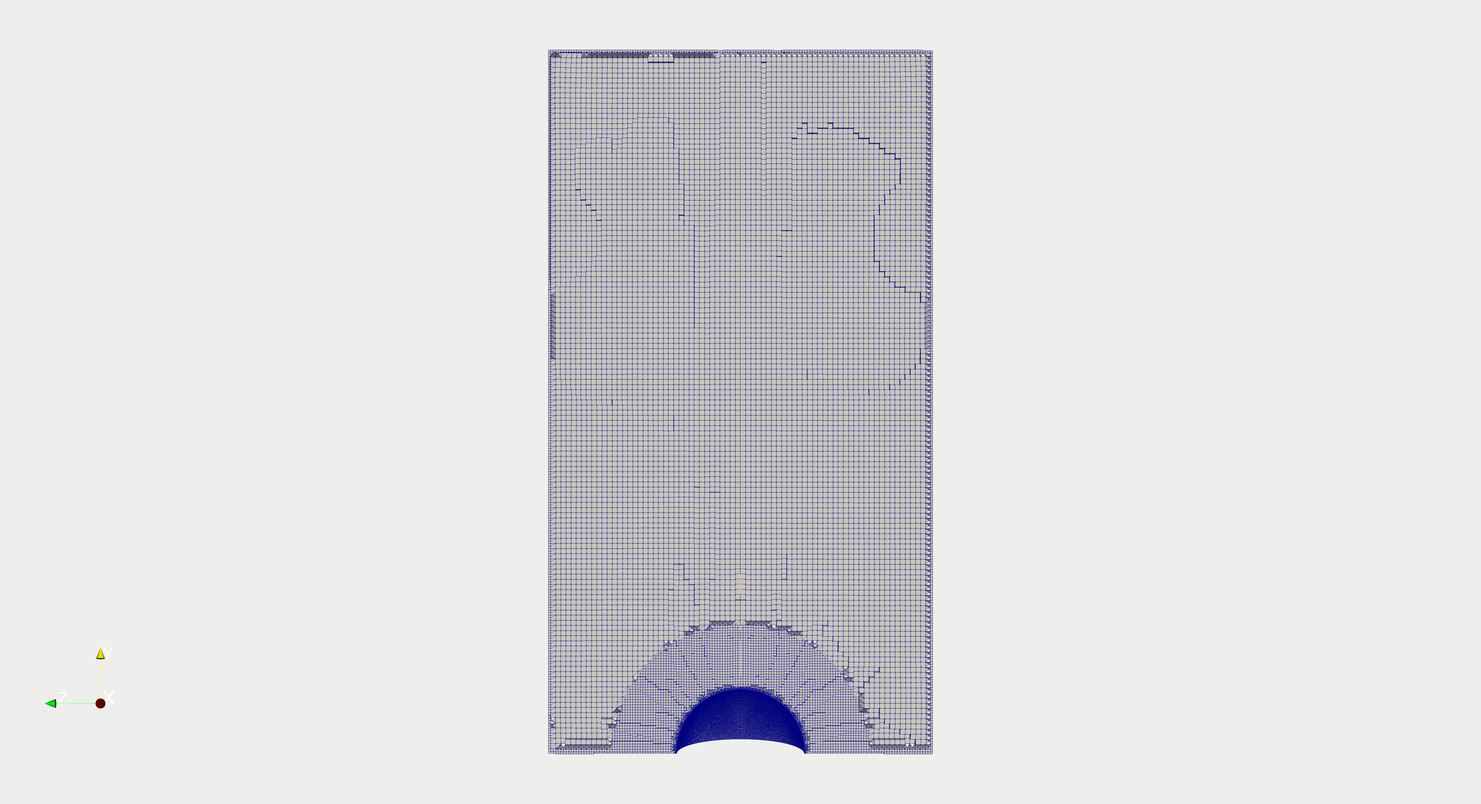}};
\draw [->] (0,0) -- (0,1.2) node[anchor=west]{\small $x$};         
\draw [->] (0,0) -- (1.1,0) node[anchor=south]{\small $r$};         
\begin{scope}[shift={(0,-.8)}]
 \draw [white] (0,0) -- (0,0);         
\end{scope}
\end{tikzpicture}
\label{fig:1}}
\caption[]{(a) Schematic representation of the experimental setup and the location of thermocouples of the hemisphere and disc geometry~(reprinted from~\citet{Priyank2019}).
\textcolor{black}{(b) Numerical grid for the simulations of the hemisphere and orientation of the coordinate system.}}
\label{fig:Schematic_rep}
\end{figure}
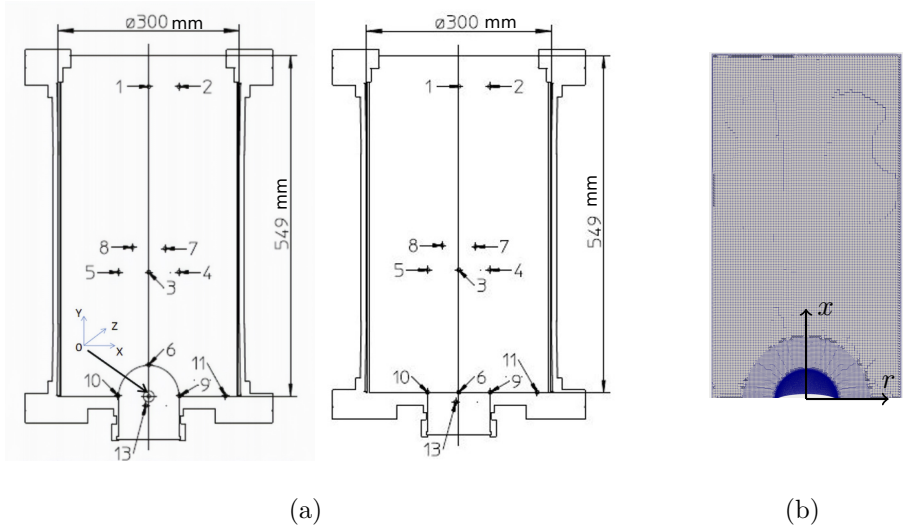

The geometry is created using the open-source CAE software Salome based on the experimental setup which is sketched in Fig.~\ref{fig:HemiSphere_thermocouples}.
The geometry consists of a cylinder of a height of 549~mm and a diameter of 300~mm and contains a hot surface at the bottom, which is either a hemisphere or a flat plate.
\textcolor{black}{Both hot surfaces have a diameter of 100~mm.
Then, the complete space filled with fluid is discretized by a computational mesh on which the governing equations are solved.
Since the vessel has no inlet and outlet during the experimental procedure, the simulation domain entirely covers the flow.}
The meshing of the geometry is done using the blockMesh and SnappyHexMesh utilities provided in OpenFOAM.
In most of the domain, hexahedral elements are generated.
The mesh near the hot surfaces is refined in order to capture the thermal and velocity boundary layers, as visualized in Fig.~\ref{fig:1} for the hemisphere geometry.
\textcolor{black}{The origin of the coordinate system lies in the centre of the hemisphere, respectively the disc.
The $x$-axis points upwards along the symmetry line of the cylinder, and the $r$-axis directs in radial direction.}

\section{Results}

\subsection{\textcolor{black}{Experimental procedure and numerical set-up}}

\textcolor{black}{
In the present study, extensive numerical simulations of transient laminar natural convection of gaseous mixture from gradually heated surfaces were carried out.
Our simulations aimed to resemble the previous experiments of \citet{Priyank2019} who used the vessels and the thermocouples as depicted in Fig.~\ref{fig:HemiSphere_thermocouples}.
In these experiment, the working fluid  was a carbon disulfide/air mixture containing a volume fraction of 2\% of CS\textsubscript{2}.
The cylindrical part of thesse vessels is manufactured of borosilicate glass and the upper and bottom plates of aluminium.
The heating body, that means the disc or hemisphere, is made of copper, due to its high thermal conductivity, and coated with a gold layer to minimize heat radiation.
Moreover, the heating body is insulated by a ring to reduce the heat transfer to the bottom plate.
}

\textcolor{black}{In the previous experiments by \citet{Priyank2019}, the temperature of the body was increased linearly in time until the mixture ignited.
It was observed that the start of ignition time in the combustion chamber changes with orientation and also with the kind of hot surface being used.
Thus, the main focus of our current simulations was on answering the questions of why the ignition location changes with the orientation of the surface and at which location ignition initiates.}
The simulations were performed for five different orientations ranging from 0$^{\circ}$ to 180$^{\circ}$ with an interval of 45$^{\circ}$.
The temperature of the hot surface was gradually increased with time at the same rate as that of experiments.
All the simulations are run until the time instance when, according to the experiments conducted by \citet{Priyank2019}, ignition takes place.
The simulations were performed using the Linux high-performance computing cluster at Physikalisch-Technische Bundesanstalt (PTB), Berlin.
Each case was distributed on 12 processors.

\subsection{Validation of the \textcolor{black}{simulations}}

The major contributors to the numerical error of our simulation results are the truncation error of the employed spatial discretization schemes and the possible inaccuracies in describing the real physical problem through our mathematical model.
As regards the former, the truncation error diminishes with the successive refinement of the mesh.
Thus, we first carried out a grid refinement study for a representative flow case.
The tested meshes consisted of 0.25~million to 2.28~million cells.
The maximum observed difference in the results between the cases of 1.17~million and 2.28~million cells is depending on the spatial location between 0.009\% and 0.34\% in terms of the temperatures at the locations of the thermocouples.
Thus, we can safely assume that the results obtained from our simulations conducted on a grid of 1.17 million cells are grid-independent.

Further, we use the experimental data of \citet{Priyank2019} to validate our mathematical model.
This data represents the temporal evolution measured by the thermocouples at the locations depicted in Fig.~\ref{fig:HemiSphere_thermocouples}.
The transient CFD model was validated for the 0$^{\circ}$ orientation of the hemisphere surface.

\begin{figure}[tb]
\centering
\includegraphics[trim=5mm 2mm 10mm 5mm,clip=true,width=12cm]{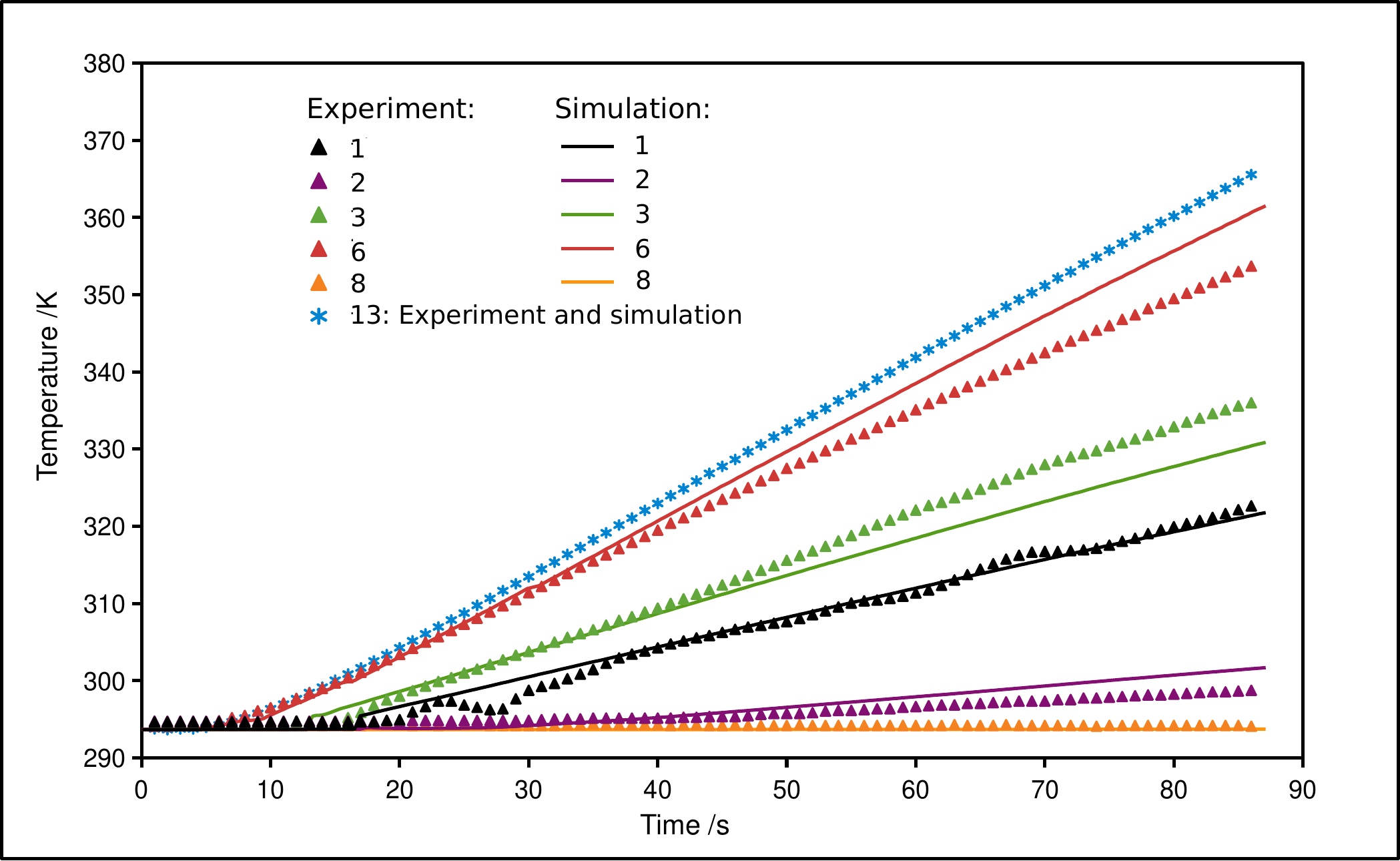}
\caption{Comparison of the simulation results for a hot hemisphere of 0$^{\circ}$ orientation with experimental data by~\citet{Priyank2019} \textcolor{black}{at the locations of the thermocouples depicted in Fig.~\ref{fig:HemiSphere_thermocouples}.
The temperature evolution measured at location~13 serves as boundary condition to the simulation, thus, their curves collapse.}}
\label{fig:Comparison}
\end{figure}

Figure~\ref{fig:Comparison} plots the temperature evolution obtained from simulations and experiments considering five chosen thermocouple locations, namely numbers 1, 2, 3, 6, and 8.
\textcolor{black}{The deviations between these curves grow in time due to the accumulation of temporal errors, which is natural for instationary simulations.
The maximum difference is 1.66\% which appears after a simulated time of 85~s at thermocouple~6.
Thermocouple number~13 is directly located on the hot surface.
The temperature evolution measured at this position was applied as boundary condition to the simulations.
Moreover, the nearly linear increase of the temperature in the experiments confirms our assumption of a laminar flow when choosing the simulation model.}
Thus, the simulations are in very close agreement with the experiments and our model was judged to be suitable for carrying out a detailed analysis.

\subsection{Analysis of the thermal flow field originating from a hot hemisphere}

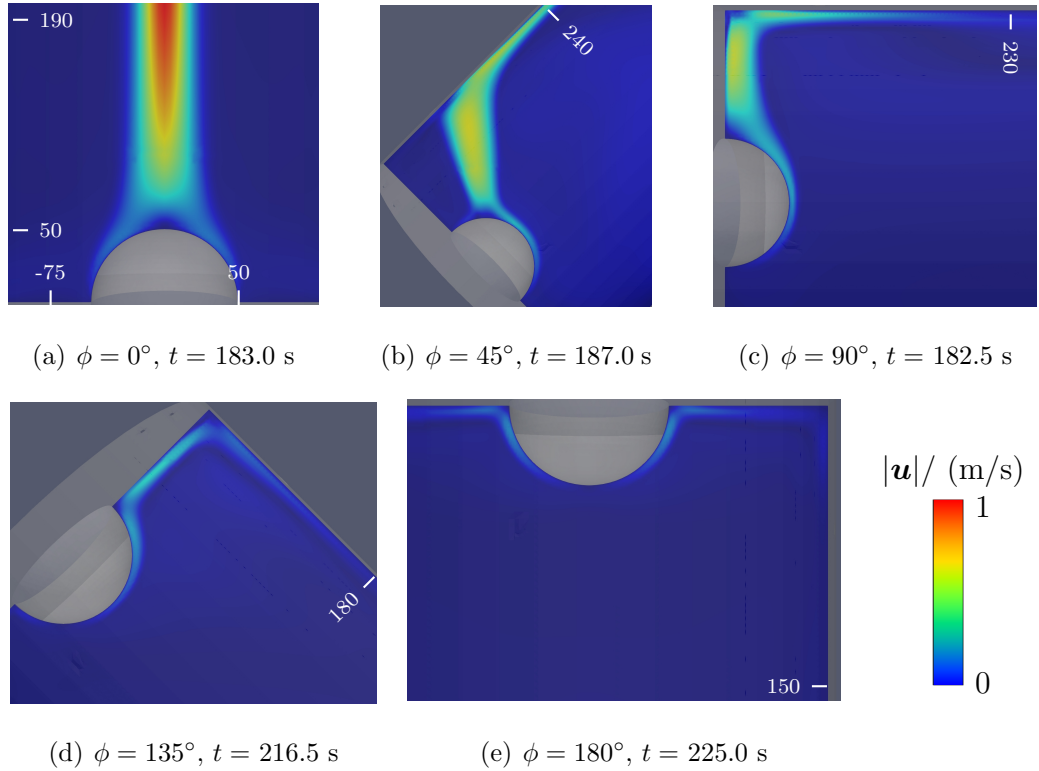
\begin{figure}[b]
\begin{center}
\subfigure[$\phi=0^{\circ}$, $t=$~183.0~s]{%
\begin{tikzpicture}[thick,white]
\node [anchor=south,inner sep=0] at (0,0) {
\includegraphics[trim=270mm 95mm 310mm 0mm,clip=true,height=4cm]{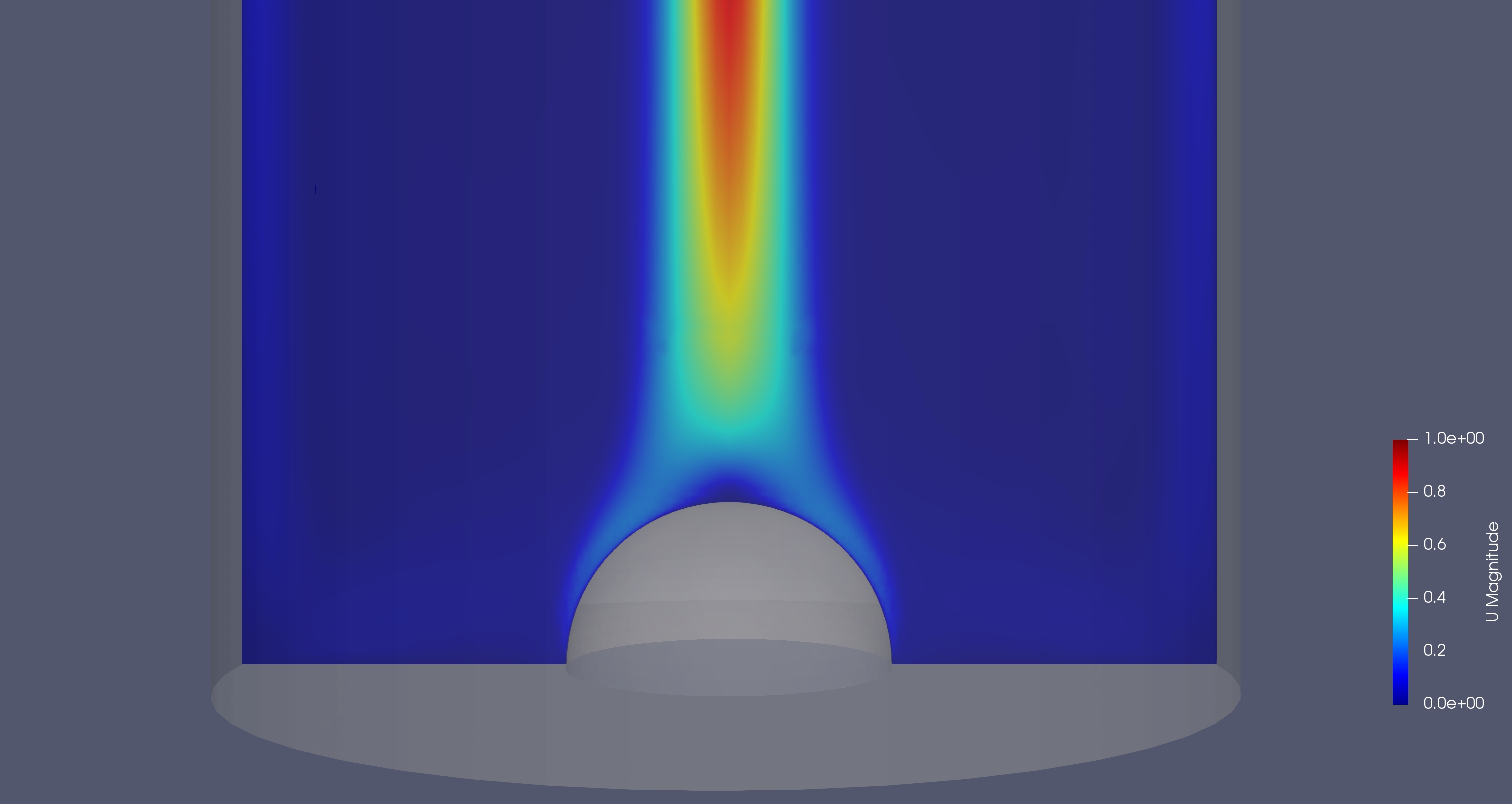}};
\draw [-] (1,0) -- (1,.2) node[anchor=south]{\scriptsize 50};
\draw [-] (-1.5,0) -- (-1.5,.2) node[anchor=south]{\scriptsize -75};
\draw [-] (-2.0,1) -- (-1.8,1) node[anchor=west]{\scriptsize 50};
\draw [-] (-2.0,3.8) -- (-1.8,3.8) node[anchor=west]{\scriptsize 190};
\end{tikzpicture}%
\label{fig:0_U}}%
\qquad
\subfigure[$\phi=45^{\circ}$, $t=$~187.0~s]{%
\begin{tikzpicture}[thick,white]
\node [anchor=south,inner sep=0] at (0,0) {
\includegraphics[trim=220mm 80mm 480mm 80mm,clip=true,height=4cm]{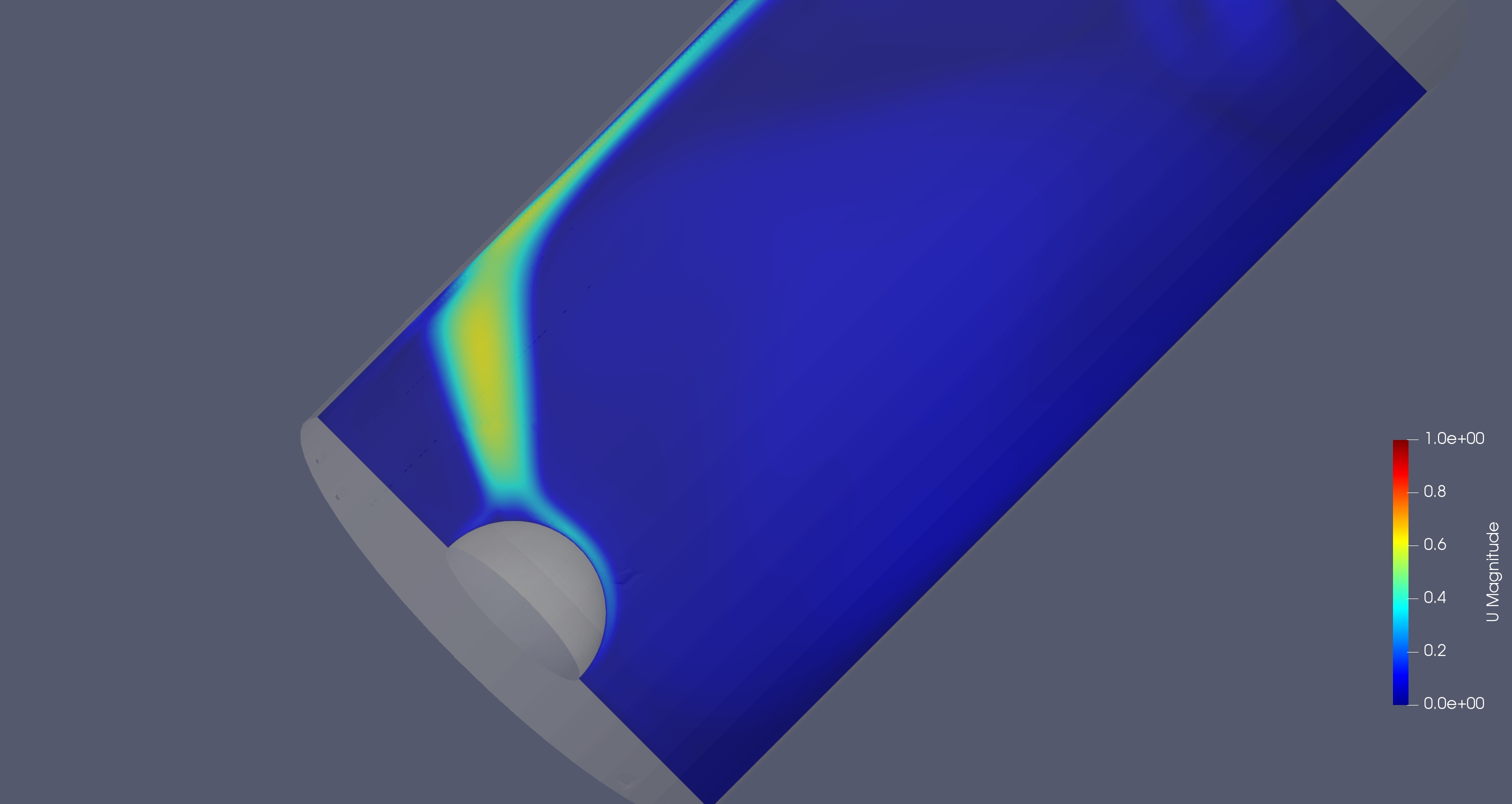}};
\begin{scope}[rotate=-45]
 \draw [-] (-2.5,3.1) -- (-2.3,3.1) node[anchor=west,rotate=-45]{\scriptsize 240};
\end{scope}
\end{tikzpicture}%
\label{fig:45_U}}%
\qquad
\subfigure[$\phi=90^{\circ}$, $t=$~182.5~s]{%
\begin{tikzpicture}[thick,white]
\node [anchor=south,inner sep=0] at (0,0) {
\includegraphics[trim=240mm 150mm 400mm 20mm,clip=true,height=4cm]{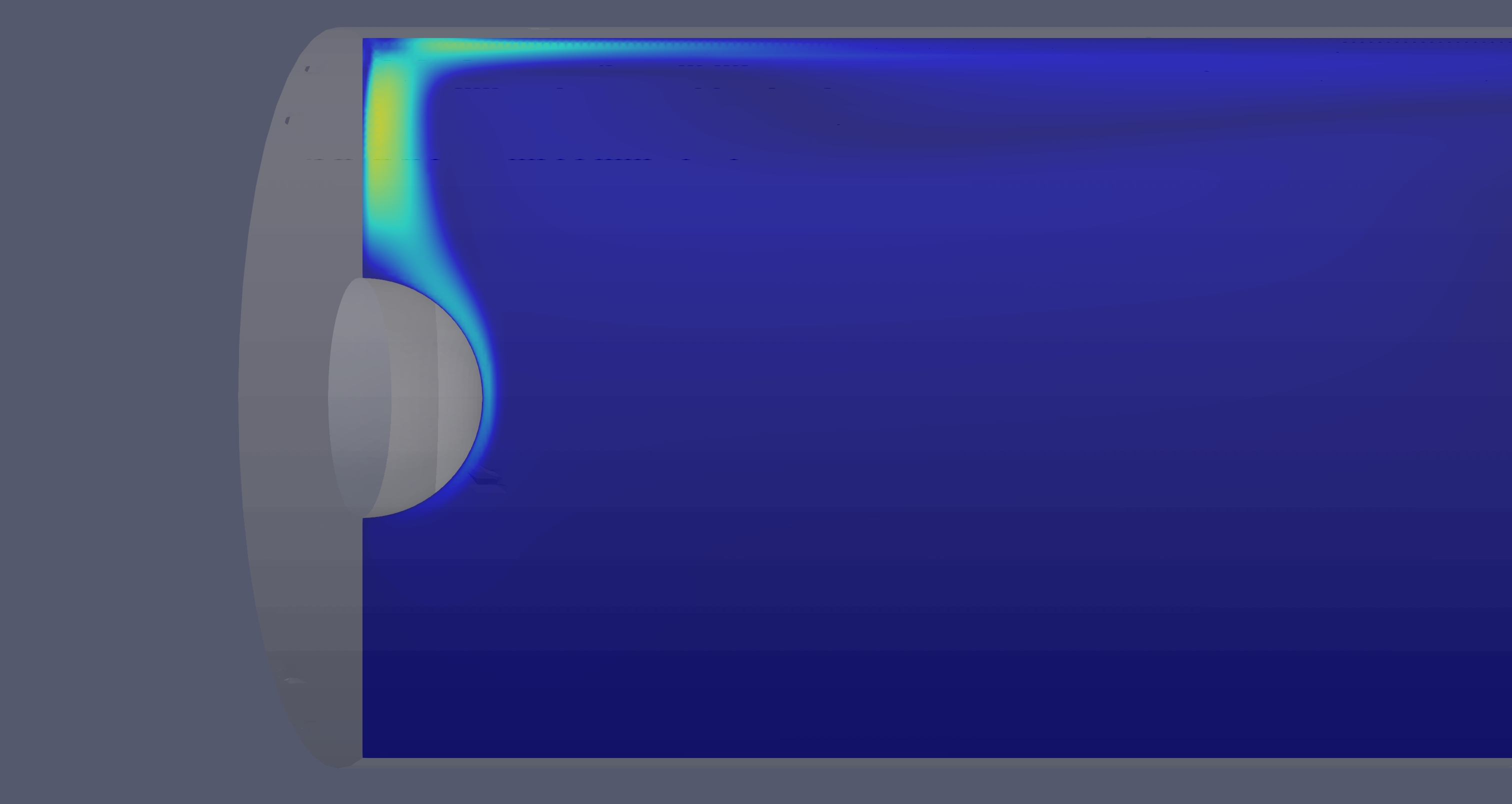}};
\begin{scope}[rotate=-90]
 \draw [-] (-3.9,1.8) -- (-3.7,1.8) node[anchor=west,rotate=-90]{\scriptsize 230};
\end{scope}
\end{tikzpicture}%
\label{fig:90_U}}%
\\
\subfigure[$\phi=135^{\circ}$, $t=$~216.5~s]{%
\begin{tikzpicture}[thick,white]
\node [anchor=south,inner sep=0] at (0,0) {
\includegraphics[trim=260mm 200mm 360mm 0mm,clip=true,height=4cm]{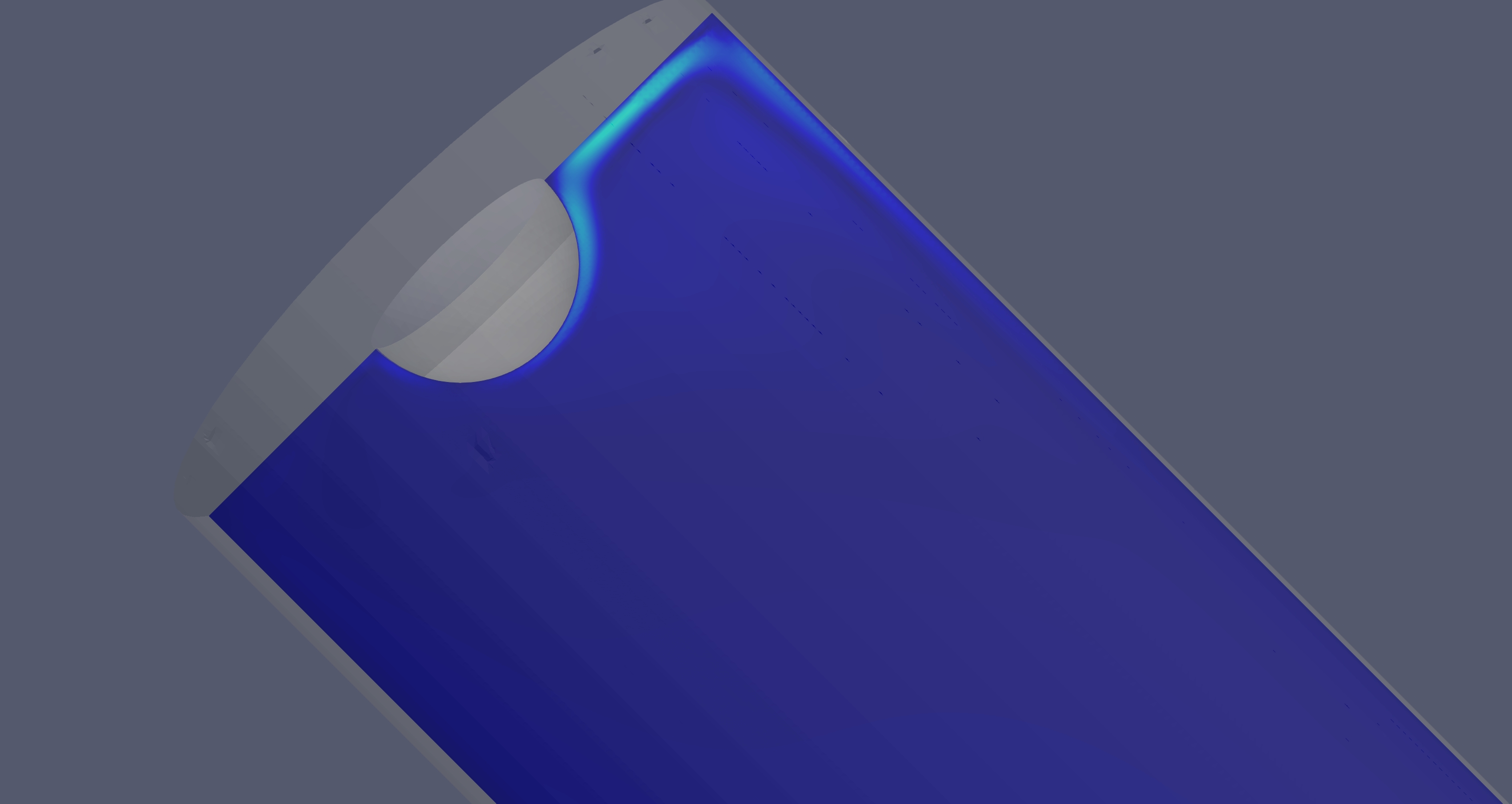}};
\begin{scope}[rotate=-135]
 \draw [-] (-2.9,0.5) -- (-2.7,0.5) node[anchor=east,rotate=-315]{\scriptsize 180};
\end{scope}
\end{tikzpicture}%
\label{fig:135_U}}
\quad
\subfigure[$\phi=180^{\circ}$, $t=$~225.0~s]{%
\begin{tikzpicture}[thick,white]
\node [anchor=south,inner sep=0] at (0,0) {
\includegraphics[trim=260mm 80mm 250mm 100mm,clip=true,height=4cm]{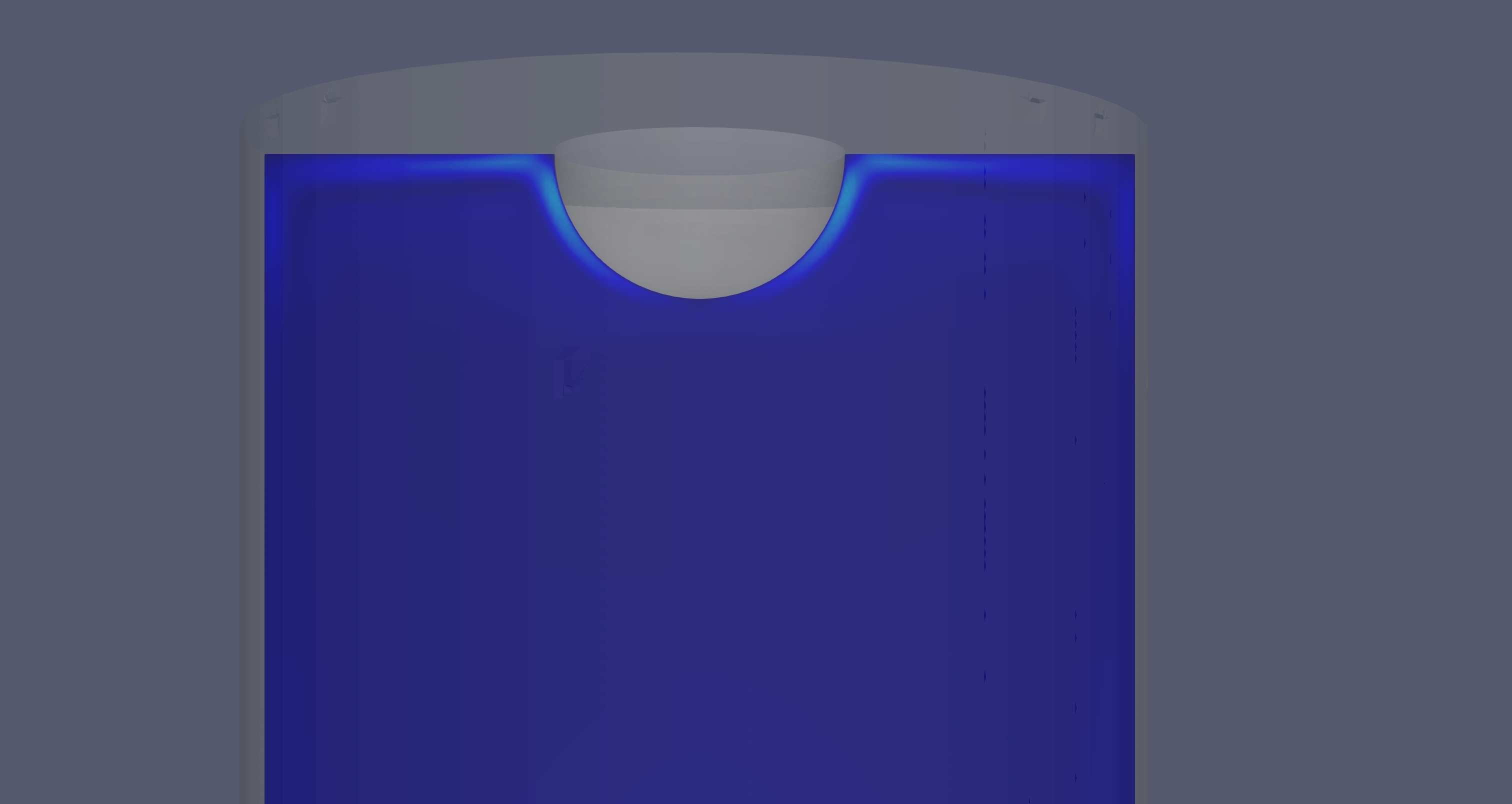}};
\begin{scope}[rotate=0]
 \draw [-] (2.5,0.2) node[anchor=east,rotate=0]{\scriptsize 150} -- (2.7,0.2);
\end{scope}
\end{tikzpicture}%
\label{fig:180_U}}
\quad
\begin{tikzpicture}
\node[anchor=south west,inner sep=0] (Bild) at (0,0)
{\includegraphics[trim=0cm 0cm 0cm 0cm,clip=true,width=5mm,height=25mm]{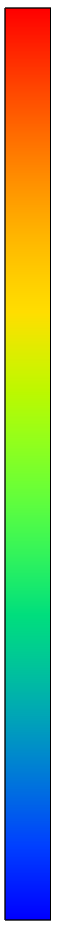}};
\begin{scope}[x=(Bild.south east),y=(Bild.north west)]
\draw [](0.6,1.0) node[anchor=south]{$|{\bm u}|/$~(m/s)};
\draw [](0.7,0.95) node[anchor=west,xshift=1mm]{1}; 
\draw [](0.7,0.02) node[anchor=west,xshift=1mm]{0};         
\end{scope}
\end{tikzpicture}
\end{center}
\caption{Magnitude of the velocity of the buoyant flow induced by the hot hemisphere.
The depicted instantaneous snapshots correspond for each case to the time instance of ignition as observed in the experiments.
\textcolor{black}{(All labels in mm.)}}
\label{fig:U}
\end{figure}

\begin{figure}[tb]
\begin{center}
\subfigure[$\phi=0^{\circ}$, $t=$~183.0~s]{\includegraphics[trim=75mm 0mm 75mm 25mm,clip=true,height=4cm]{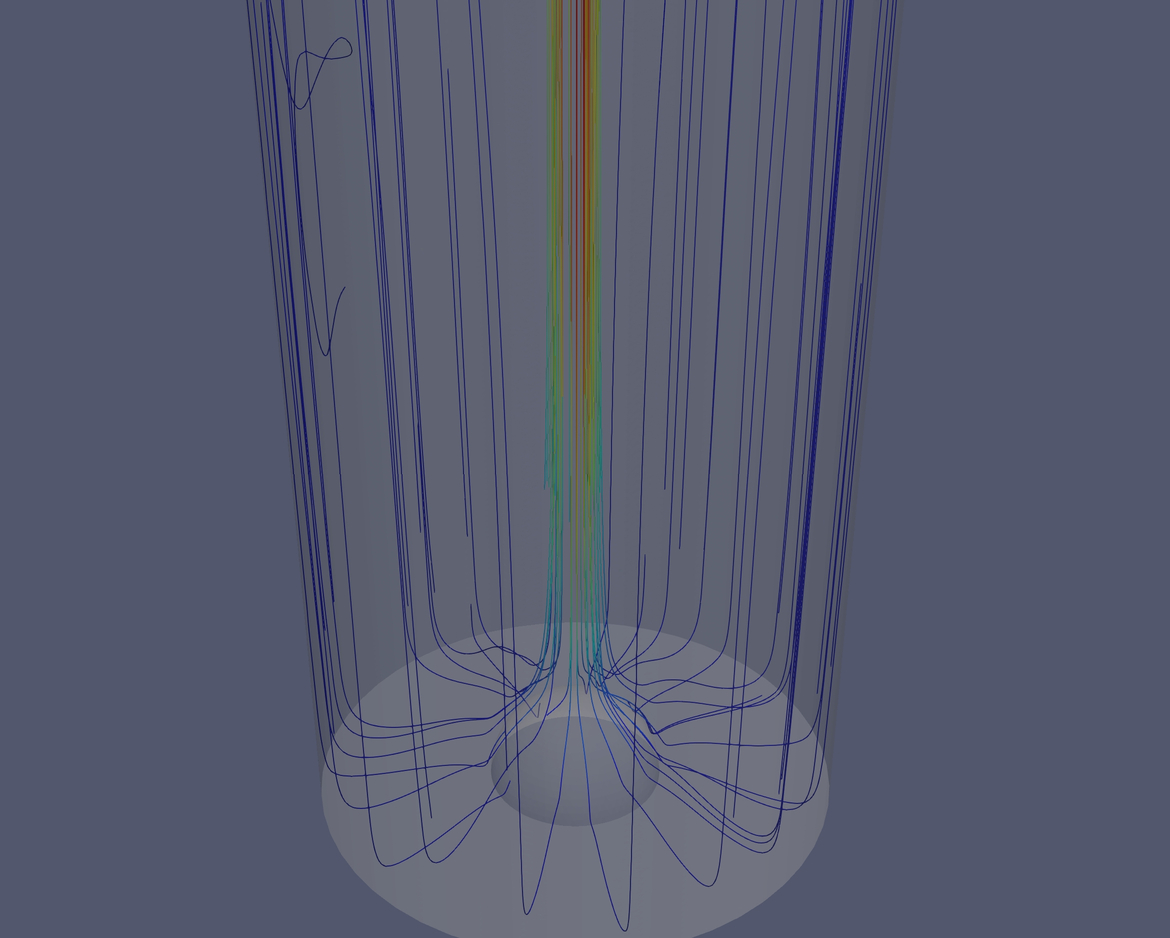}\label{fig:0_S}}
\quad
\subfigure[$\phi=45^{\circ}$, $t=$~187.0~s]{\includegraphics[trim=50mm 0mm 50mm 0mm,clip=true,height=4cm]{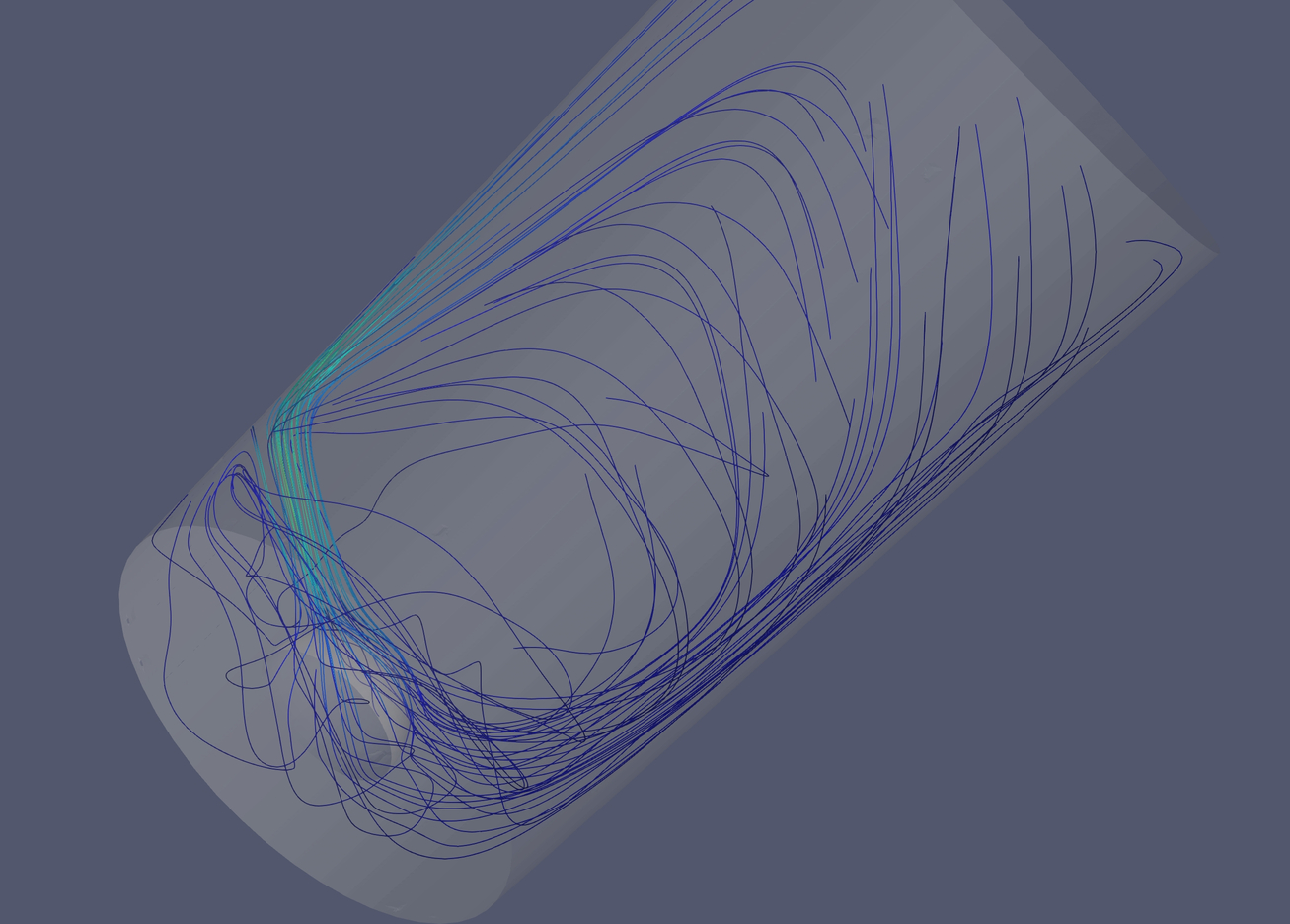}\label{fig:45_S}}
\quad
\subfigure[$\phi=90^{\circ}$, $t=$~182.5~s]{\includegraphics[trim=0mm 40mm 160mm 40mm,clip=true,height=4cm]{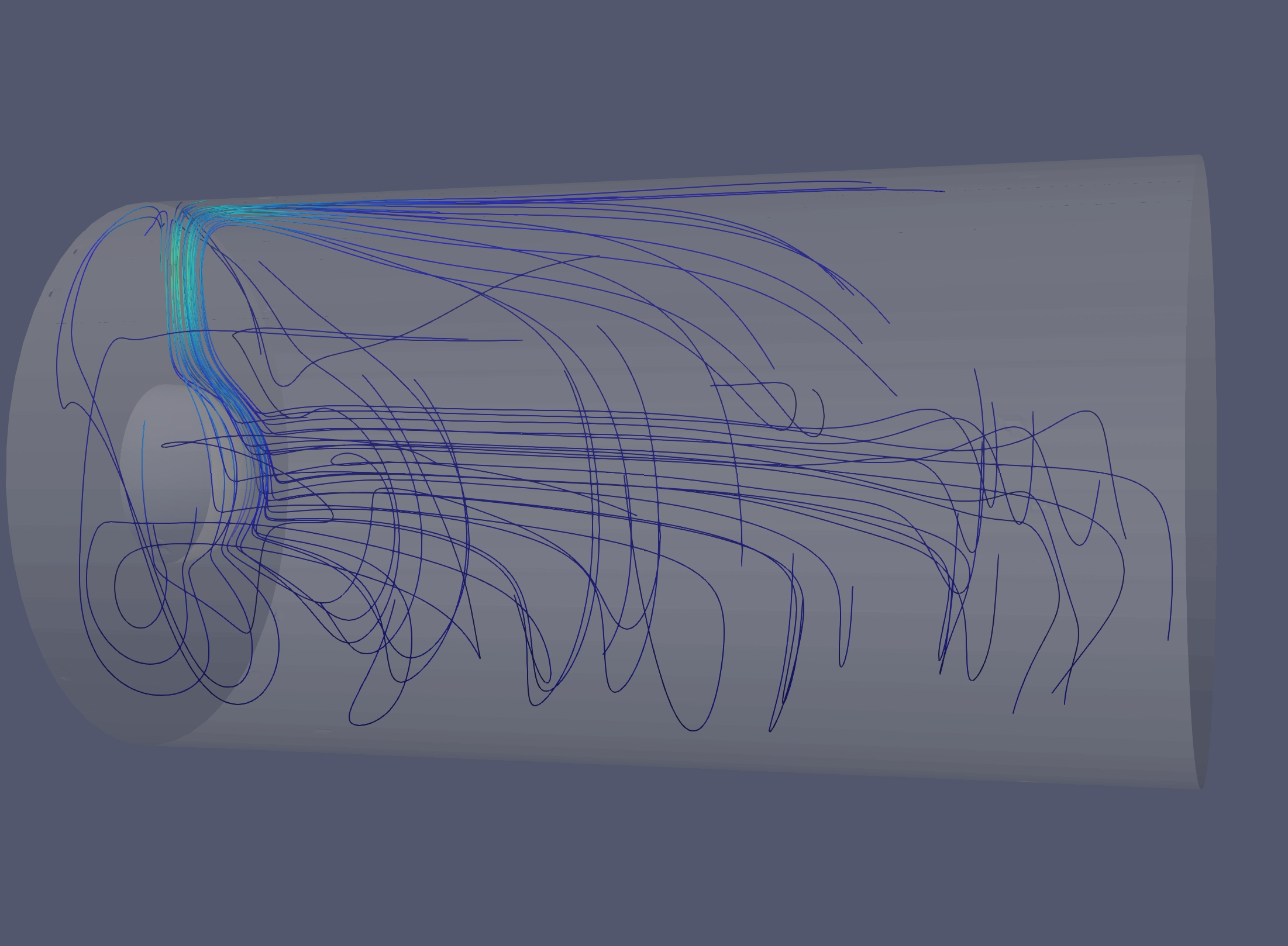}\label{fig:90_S}}
\\
\subfigure[$\phi=135^{\circ}$, $t=$~216.5~s]{\includegraphics[trim=110mm 50mm 200mm 0mm,clip=true,height=4cm]{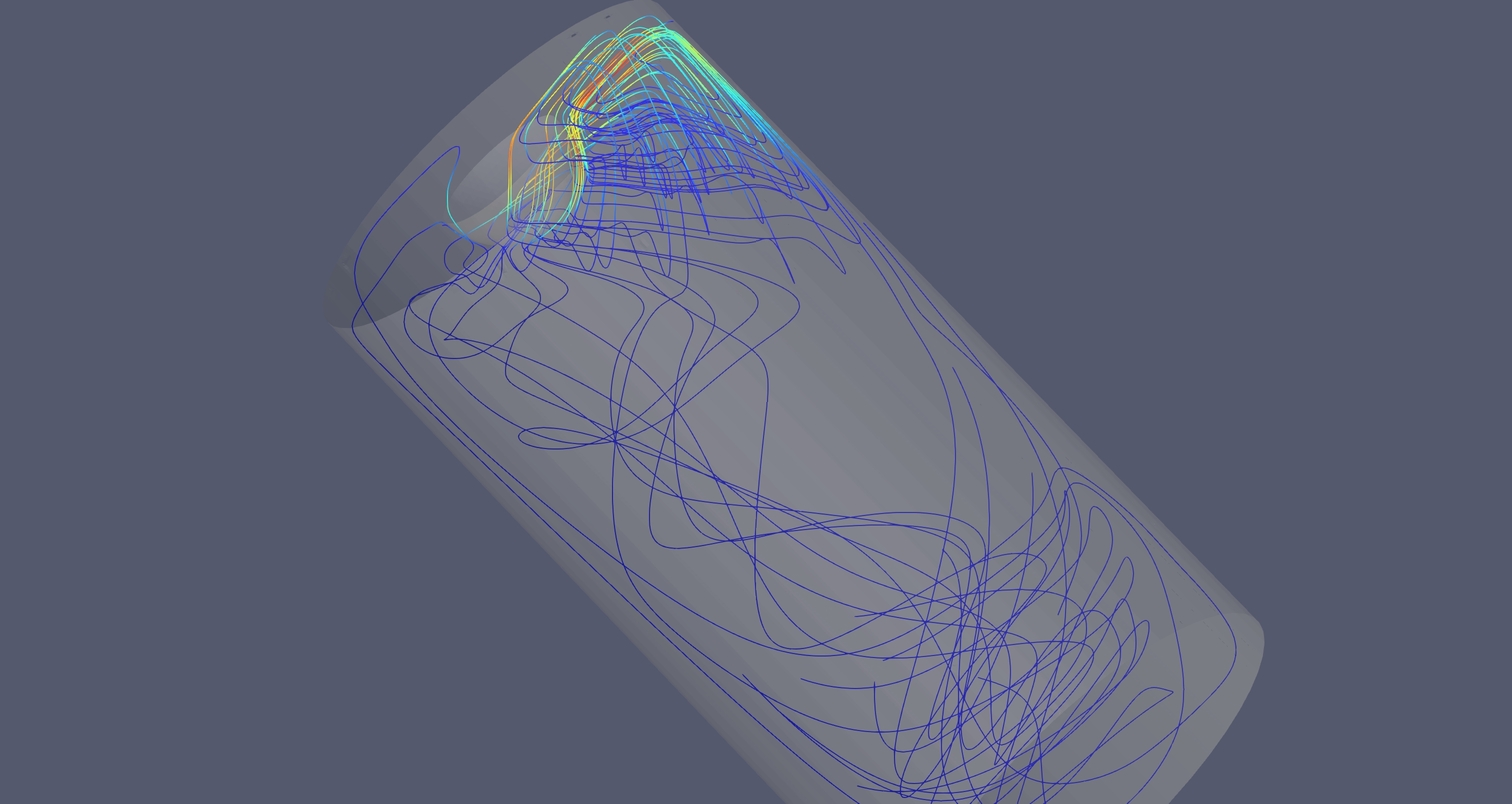}\label{fig:135_S}}
\qquad
\subfigure[$\phi=180^{\circ}$, $t=$~225.0~s]{\includegraphics[trim=150mm 250mm 230mm 0mm,clip=true,height=4cm]{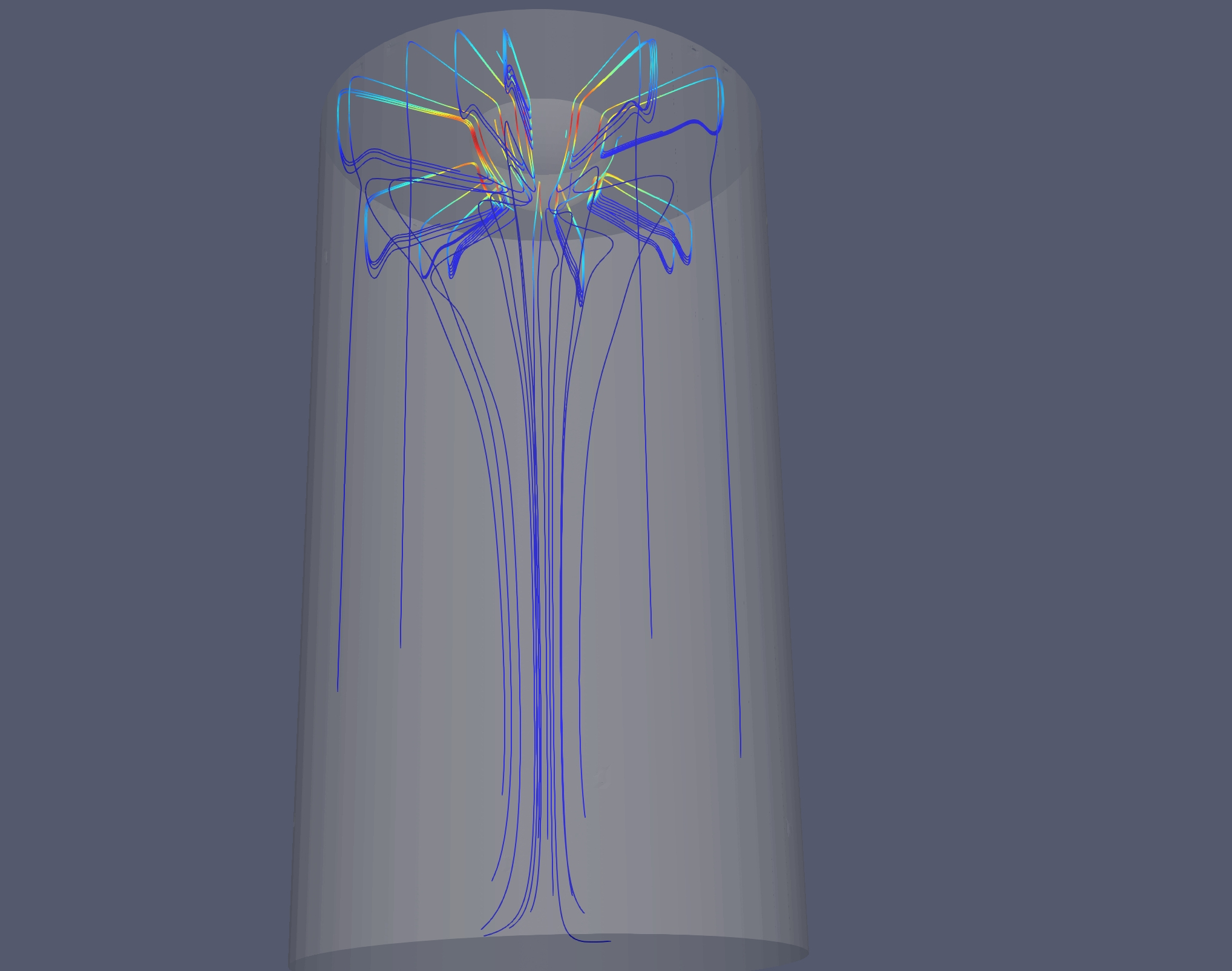}\label{fig:180_S}}
\qquad
\begin{tikzpicture}
\node[anchor=south west,inner sep=0] (Bild) at (0,0)
{\includegraphics[trim=0cm 0cm 0cm 0cm,clip=true,width=5mm,height=25mm]{fig/colorbar}};
\begin{scope}[x=(Bild.south east),y=(Bild.north west)]
\draw [](0.6,1.0) node[anchor=south]{$|{\bm u}|/$~(m/s)};
\draw [](0.7,0.95) node[anchor=west,xshift=1mm]{1}; 
\draw [](0.7,0.02) node[anchor=west,xshift=1mm]{0};         
\end{scope}
\end{tikzpicture}
\end{center}
\caption{Streamlines of the buoyant flow induced by the hot hemisphere, colored by the magnitude of the velocity.
The depicted instantaneous snapshots correspond for each case to the time instance of ignition as observed in the experiments.}
\label{fig:S}
\end{figure}

\begin{figure}[t]
\begin{center}
\subfigure[$\phi=0^{\circ}$, $t=$~183.0~s]{%
\begin{tikzpicture}[thick,white]
\node [anchor=south,inner sep=0] at (0,0) {
\includegraphics[trim=250mm 95mm 290mm 0mm,clip=true,height=4cm]{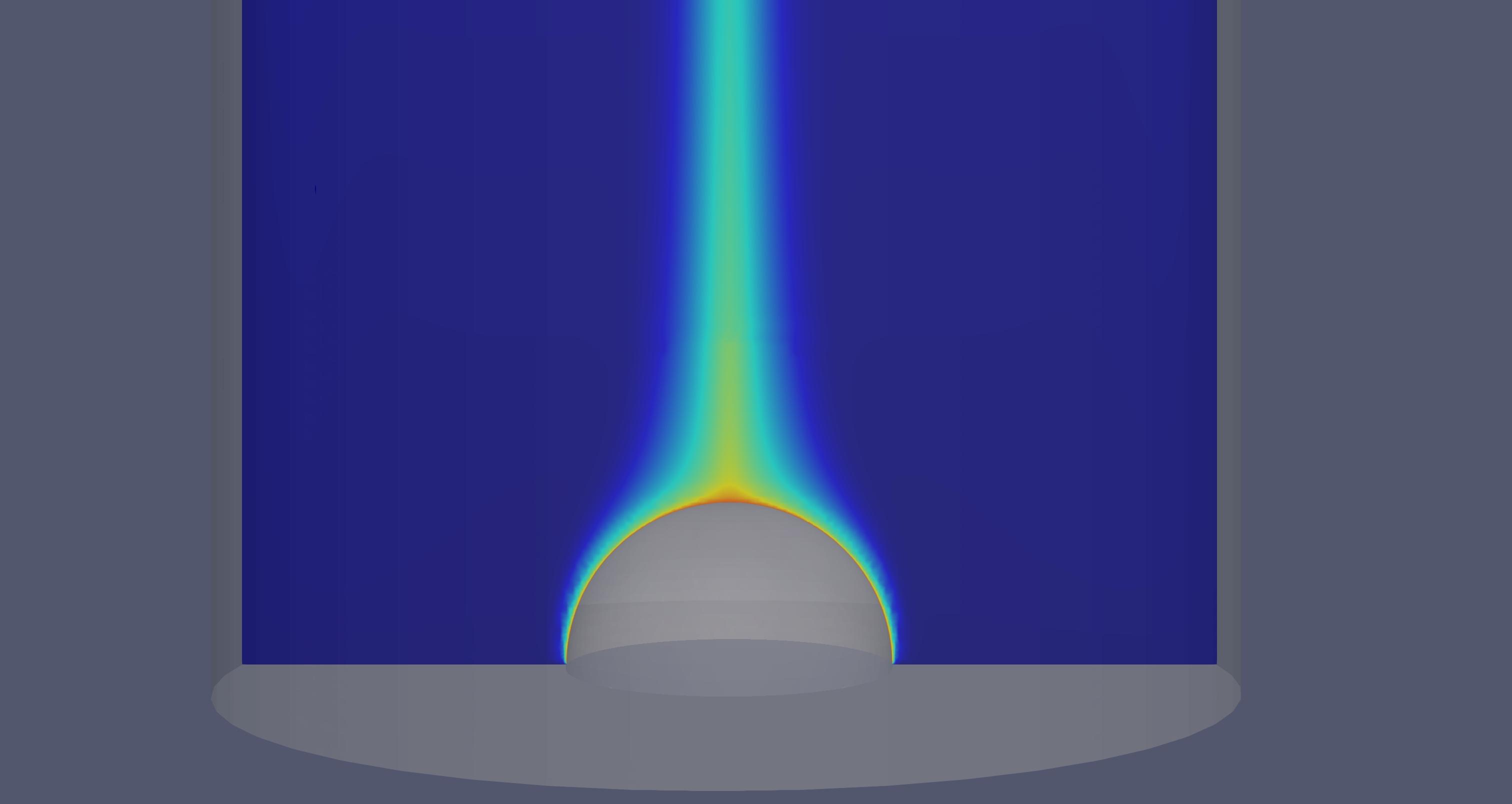}};
\draw [-] (1,0) -- (1,.2) node[anchor=south]{\scriptsize 50};
\draw [-] (-1.5,0) -- (-1.5,.2) node[anchor=south]{\scriptsize -75};
\draw [-] (-2.2,1) -- (-2.0,1) node[anchor=west]{\scriptsize 50};
\draw [-] (-2.2,3.8) -- (-2.0,3.8) node[anchor=west]{\scriptsize 190};
\end{tikzpicture}%
\label{fig:0_T}}%
\quad
\subfigure[$\phi=45^{\circ}$, $t=$~187.0~s]{%
\begin{tikzpicture}[thick,white]
\node [anchor=south,inner sep=0] at (0,0) {
\includegraphics[trim=220mm 80mm 420mm 80mm,clip=true,height=4cm]{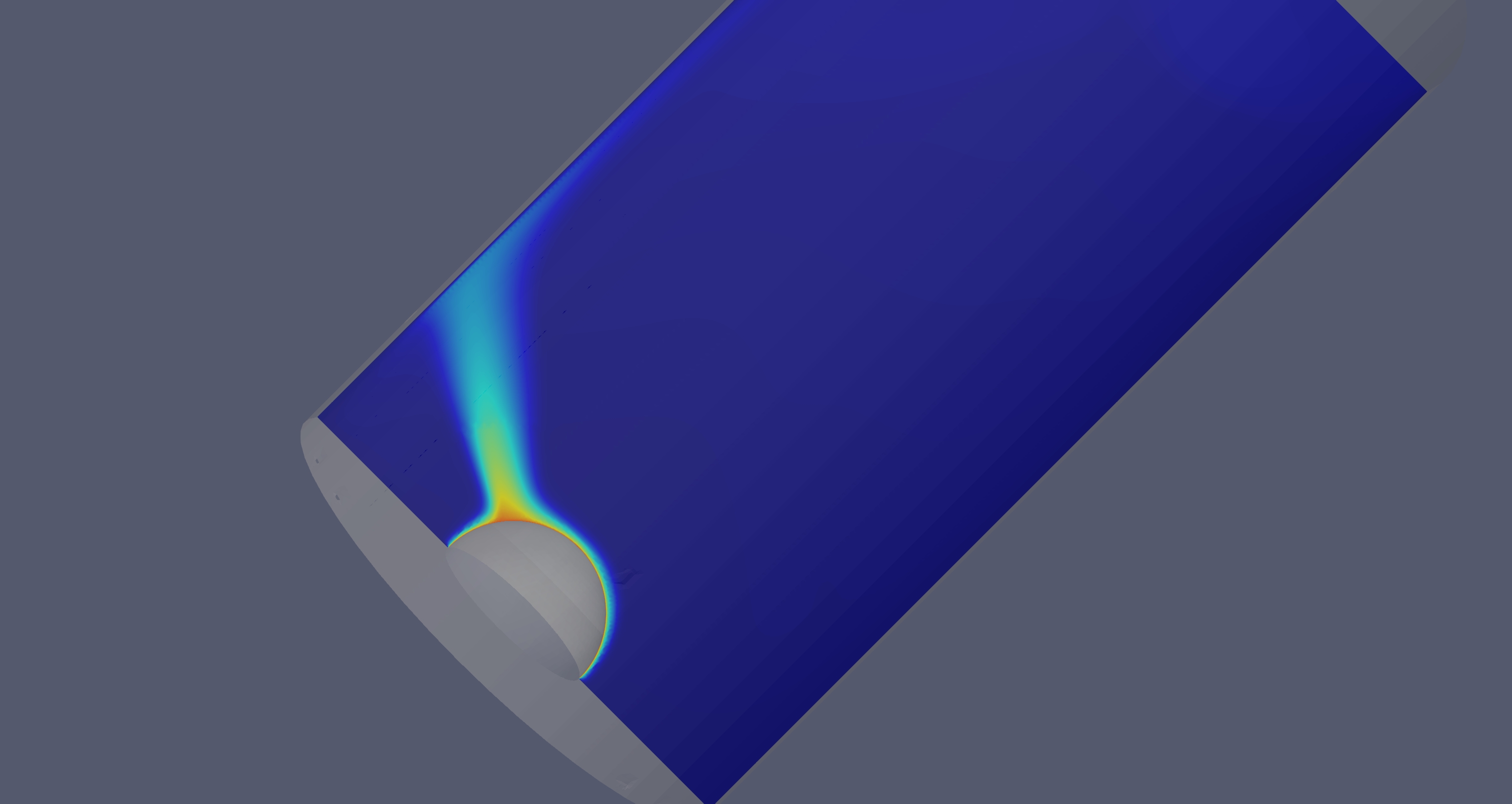}};
\begin{scope}[rotate=-45]
 \draw [-] (-2.8,2.9) -- (-2.6,2.9) node[anchor=west,rotate=-45]{\scriptsize 240};
\end{scope}
\end{tikzpicture}%
\label{fig:45_T}}%
\quad
\subfigure[$\phi=90^{\circ}$, $t=$~182.5~s]{%
\begin{tikzpicture}[thick,white]
\node [anchor=south,inner sep=0] at (0,0) {
\includegraphics[trim=240mm 150mm 410mm 20mm,clip=true,height=4cm]{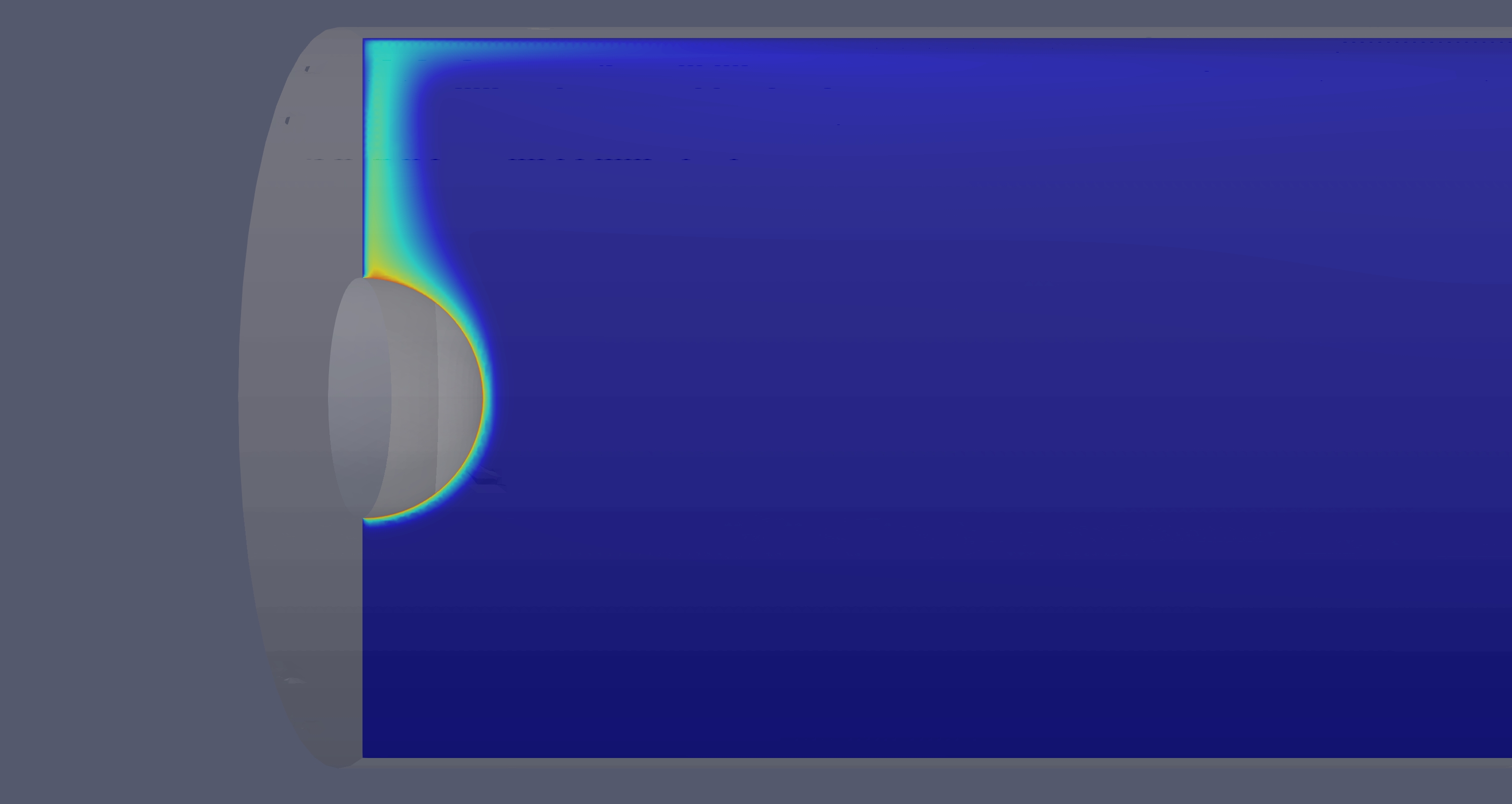}};
\begin{scope}[rotate=-90]
 \draw [-] (-3.9,1.8) -- (-3.7,1.8) node[anchor=west,rotate=-90]{\scriptsize 220};
\end{scope}
\end{tikzpicture}%
\label{fig:90_T}}%
\\
\subfigure[$\phi=135^{\circ}$, $t=$~216.5~s]{%
\begin{tikzpicture}[thick,white]
\node [anchor=south,inner sep=0] at (0,0) {
\includegraphics[trim=240mm 200mm 410mm 0mm,clip=true,height=4cm]{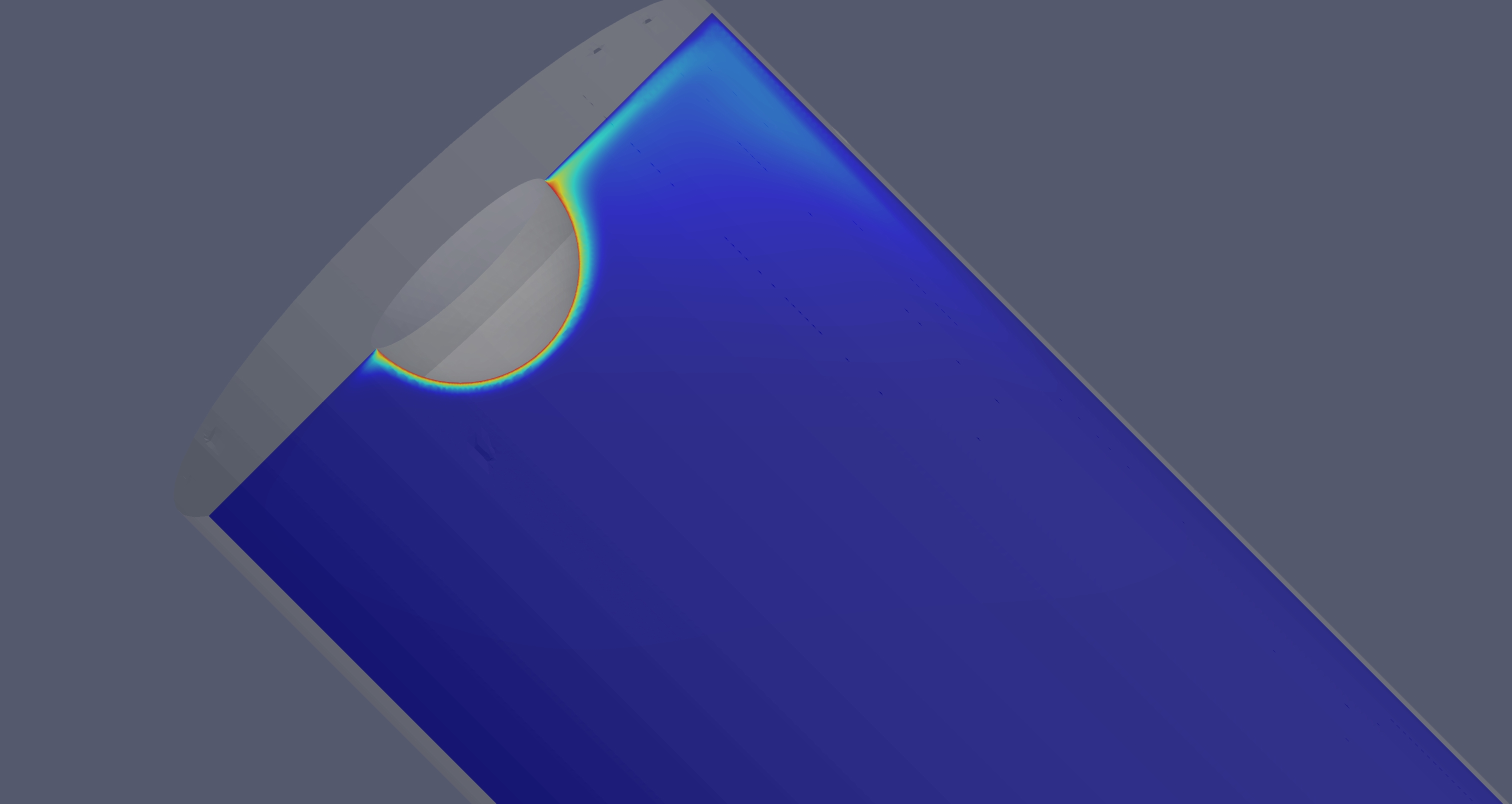}};
\begin{scope}[rotate=-135]
 \draw [-] (-3.1,0.0) -- (-2.9,0.0) node[anchor=east,rotate=-315]{\scriptsize 120};
\end{scope}
\end{tikzpicture}%
\label{fig:135_T}}%
\quad
\subfigure[$\phi=180^{\circ}$, $t=$~225.0~s]{%
\begin{tikzpicture}[thick,white]
\node [anchor=south,inner sep=0] at (0,0) {
\includegraphics[trim=180mm 60mm 250mm 100mm,clip=true,height=4cm]{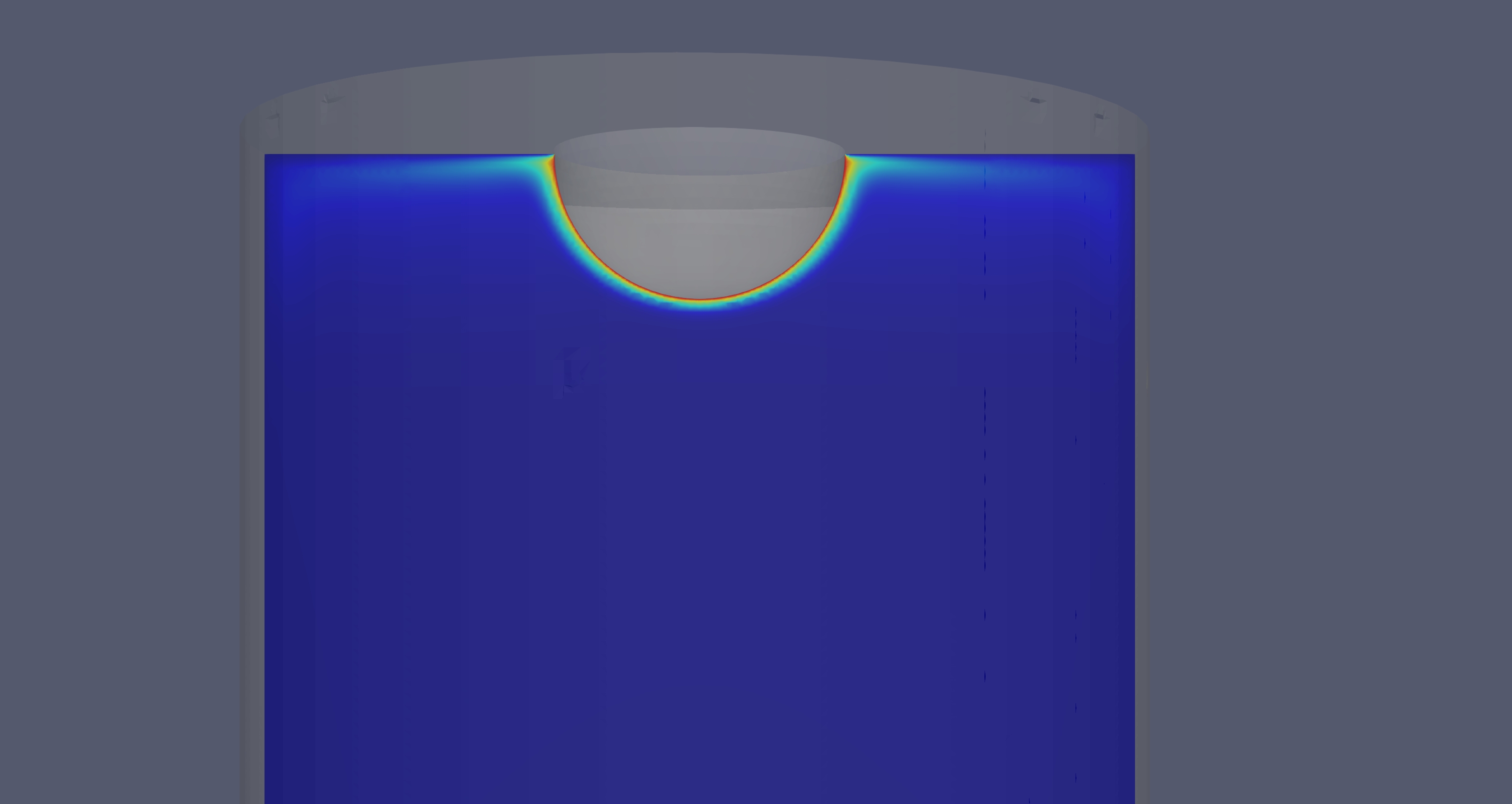}};
\begin{scope}[rotate=0]
 \draw [-] (2.8,0.2) node[anchor=east,rotate=0]{\scriptsize 180} -- (3.0,0.2);
\end{scope}
\end{tikzpicture}%
\label{fig:180_T}}%
\quad
\begin{tikzpicture}
\node[anchor=south west,inner sep=0] (Bild) at (0,0)
{\includegraphics[trim=0cm 0cm 0cm 0cm,clip=true,width=5mm,height=25mm]{fig/colorbar}};
\begin{scope}[x=(Bild.south east),y=(Bild.north west)]
\draw [](0.6,1.0) node[anchor=south]{$T /$~K};
\draw [](0.7,0.95) node[anchor=west,xshift=1mm]{490}; 
\draw [](0.7,0.02) node[anchor=west,xshift=1mm]{290};         
\end{scope}
\end{tikzpicture}
\end{center}
\caption{Temperature field of the thermal plumes induced by the hot hemisphere.
The depicted instantaneous snapshots correspond for each case to the time instance of ignition as observed in the experiments.
\textcolor{black}{(All labels in mm.)}}
\label{fig:T}
\end{figure}

In the following, we discuss the flow and thermal field induced by a hot hemisphere of the orientations $\phi=0^{\circ}$, 45$^{\circ}$, 90$^{\circ}$, 135$^{\circ}$, and 180$^{\circ}$.
The magnitude of the flow velocity for each case is depicted in Fig.~\ref{fig:U}.
We are mainly interested in the precise conditions which lead to the ignition of the explosive atmosphere.
Thus, each snapshot corresponds to the time instance where ignition was observed in the experiments.
For further visualization of the flow field, the streamlines of the flow at the same time instances are given in Fig.~\ref{fig:S} and the resulting thermal plumes in Fig.~\ref{fig:T}.

As it can be observed in these figures, the buoyancy induced by the heat transfer from the hemisphere results in thermal and velocity boundary layers.
\textcolor{black}{Even though the streamlimes indicate vortices, they are of a large scale and their formation is determenistic, which supports our assumption of a laminar flow.}
In the experiments, it was found that the ignition location for the case of $\phi=0^{\circ}$ is always directly above the hemisphere.
In the early stage of the simulation the buoyancy induced motion is restricted along the hot surface.
Afterward, a steady velocity boundary layer is formed that propagates towards the top of the hemisphere.
Then the separation of the boundary layer takes place at the top of the hemisphere creating a stagnation zone, as can be seen in Fig.~\ref{fig:0_U}.
Also, the thermal boundary merges where the velocity separation takes place resulting in a relatively thick thermal flow over the hemisphere.
Here, the heat accumulates and a high-temperature region (hot spot) is observed as shown in see Fig.~\ref{fig:0_T}.
Further, the streamlines in Fig.~\ref{fig:0_S} visualize the hot gas continuing to rise to the top of the chamber as it was already observed by \citet{turner_1973}.
The heated fluid moves up in a plume and forms a stagnation zone at the top of the vessel.
From there the cooled fluid moves further upward within the boundary layer and then falls downward in the outer region.
Thus, in the case of $\phi=0^{\circ}$, the highest temperatures are reached remote from the hot surface inside the stagnation zone.
In this region, the conditions are favorable for a possible ignition of the explosive atmosphere.

\textcolor{black}{According to the velocity contours in Fig.~\ref{fig:U}, the highest velocity occurs for an orientation of $\phi= 0^{\circ}$.
In this case, a jet pointing upwards obtains a velocity of up to 1~m/s.
The corresponding jet Reynolds number, based on the diameter of the hemisphere and the viscosity of the gas, equals 6800.
Thus, the Reynolds numbers of our simulations are below the critical jet Reynolds number of about 10\,000~\citep{Mi13}.}

For the remaining orientations, namely $\phi= 45^{\circ}$, 90$^{\circ}$, and 135$^{\circ}$, the thermal plumes (Figs.~\ref{fig:45_T}--\ref{fig:180_T}) do not grow along the symmetry line but follow the direction of the flow (Figs.~\ref{fig:45_U}--\ref{fig:180_U}).
The stagnation zone is shifted towards the left on the hemisphere surface  with respect to the orientation.
This shifting occurs since the boundary layer separates at the hemisphere's surface always in the direction of buoyancy.
This, in turn, leads to the shifting of the hot spot from the top of the hemisphere to the left of the combustion chamber.
Comparison with the experiments ascertains that the ignition locations for the above-mentioned orientations of the hemisphere coincide with the location of hot spots from simulations.

Remarkably, the flow fields of the cases of $\phi=0^{\circ}$ and $\phi=180^{\circ}$ (Figs.~\ref{fig:0_S} and~\ref{fig:180_S}) are nearly completely inverse to each other.
It can be seen that for $\phi=180^{\circ}$ the velocity boundary layer forms at the top of the hemisphere.
Then, it moves upwards towards the base of the hemisphere and separates in the corner where the hemisphere connects with the wall of the combustion chamber.
This leads to the formation of a narrow ring-shaped stagnation zone around the base of the hemisphere.
In this region also the thermal boundary layer is a thick ring around the base that exhibits the highest temperature in the whole domain.
In the experiments, it was observed that the ignition takes place in a ring similar to that is observed in simulations.

\begin{figure}[tb]
\begin{center}
\subfigure[$\phi=0^{\circ}$, $t=$~183.0~s]{\includegraphics[trim=270mm 90mm 310mm 20mm,clip=true,height=4cm]{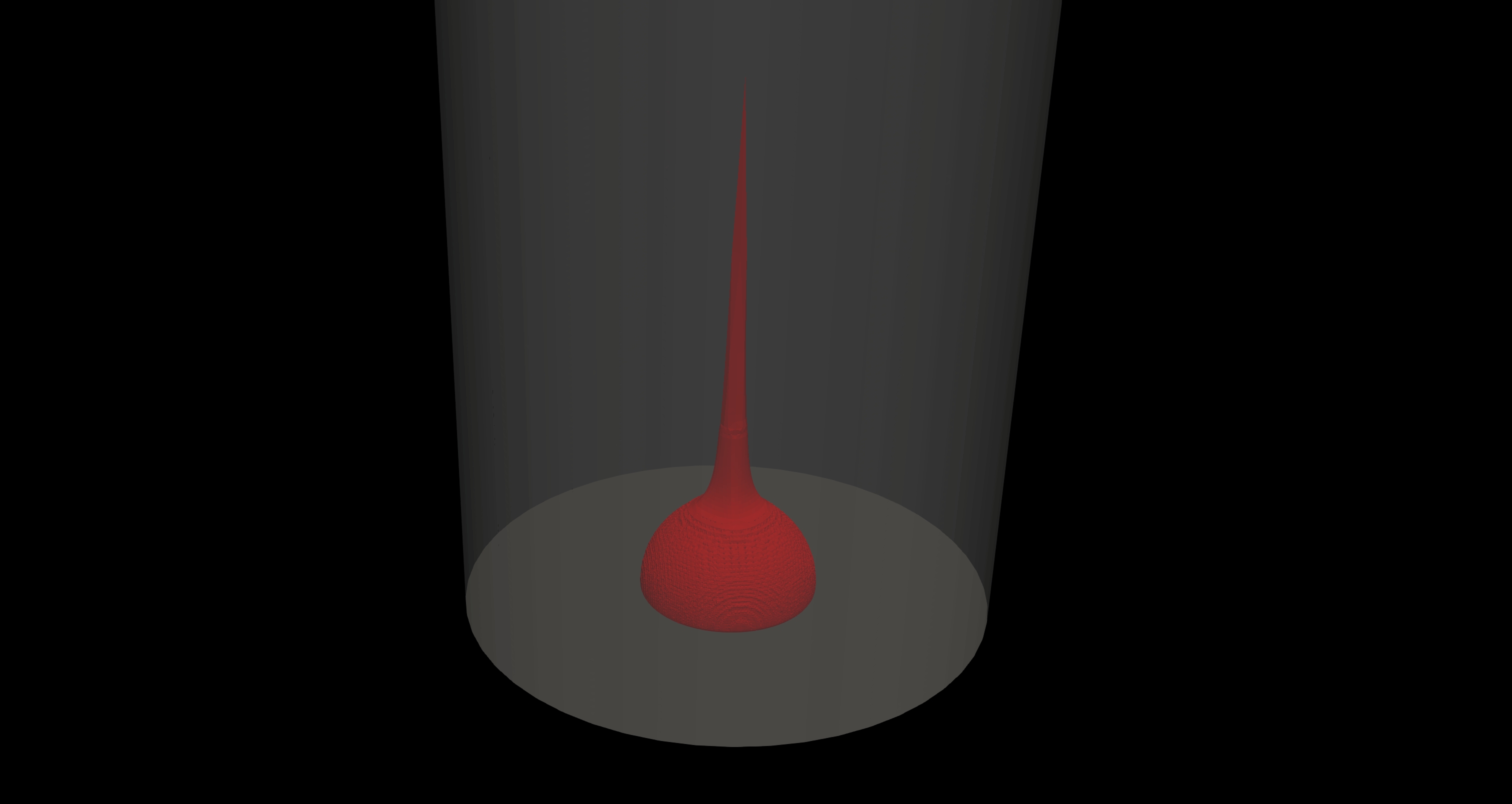}\label{fig:0_V}}
\quad
\subfigure[$\phi=45^{\circ}$, $t=$~187.0~s]{\includegraphics[trim=220mm 80mm 410mm 80mm,clip=true,height=4cm]{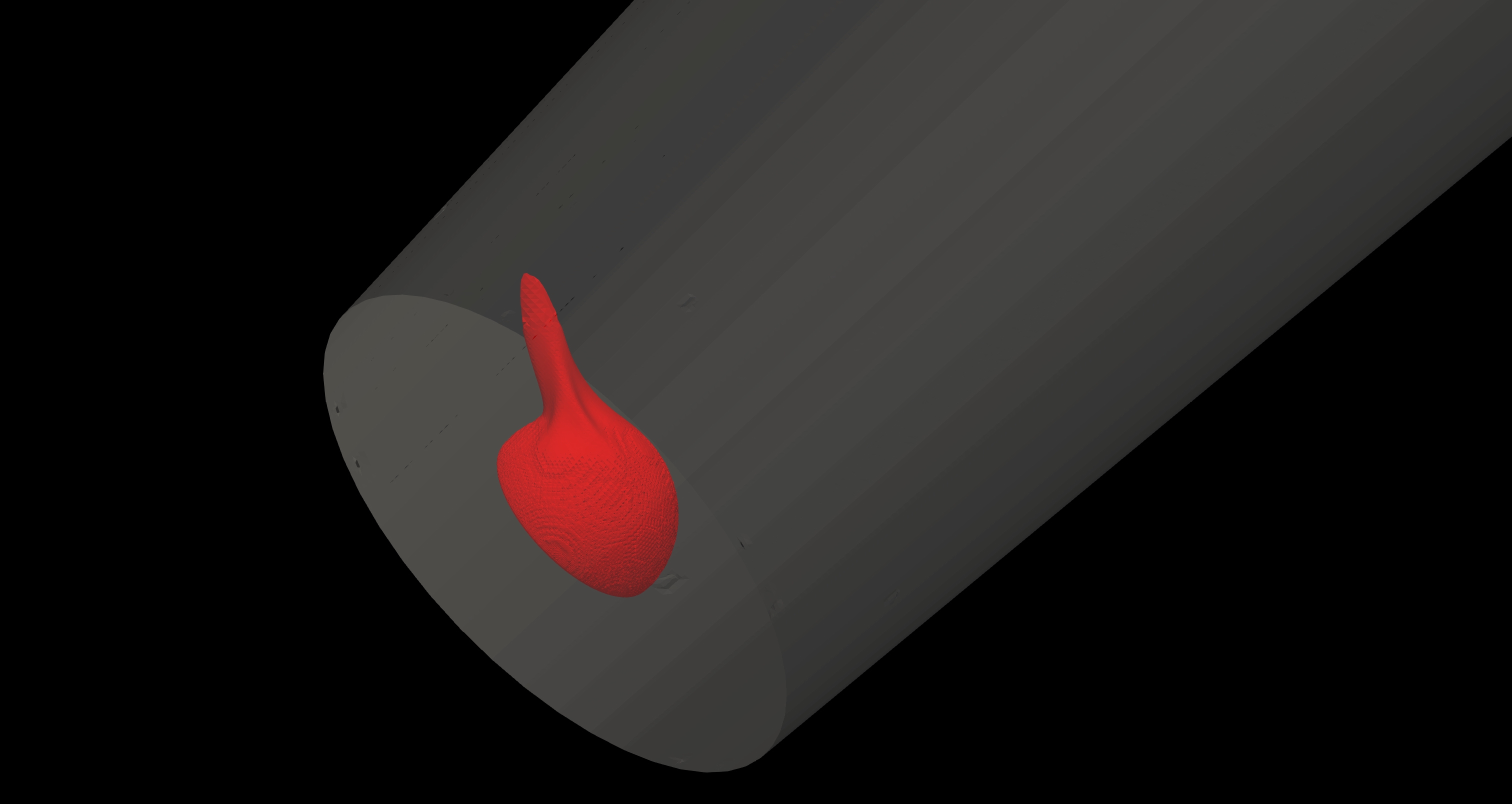}\label{fig:45_V}}
\quad
\subfigure[$\phi=90^{\circ}$, $t=$~182.5~s]{\includegraphics[trim=180mm 150mm 510mm 20mm,clip=true,height=4cm]{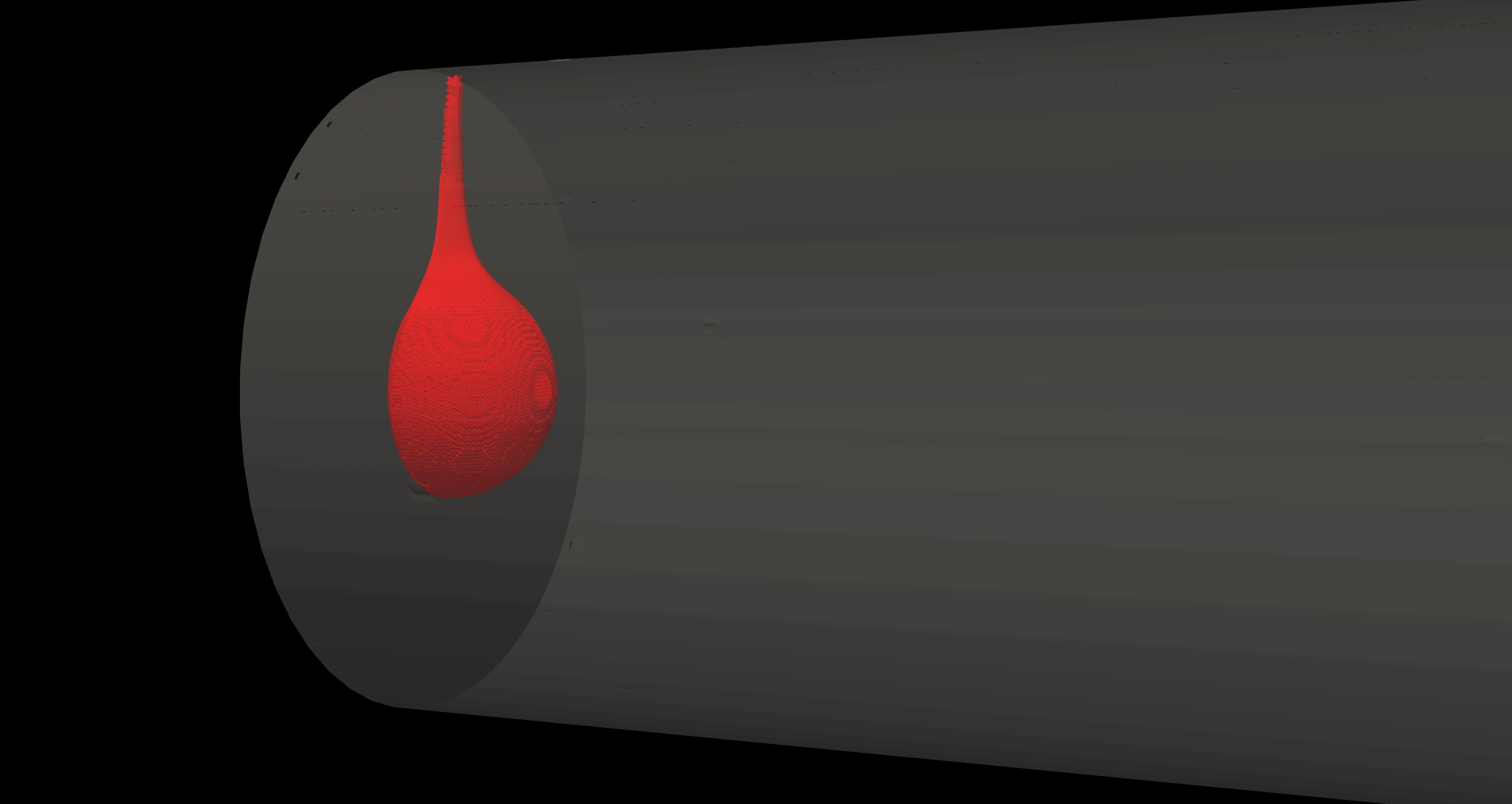}\label{fig:90_V}}
\\
\subfigure[$\phi=135^{\circ}$, $t=$~216.5~s]{\includegraphics[trim=150mm 150mm 350mm 40mm,clip=true,height=4cm]{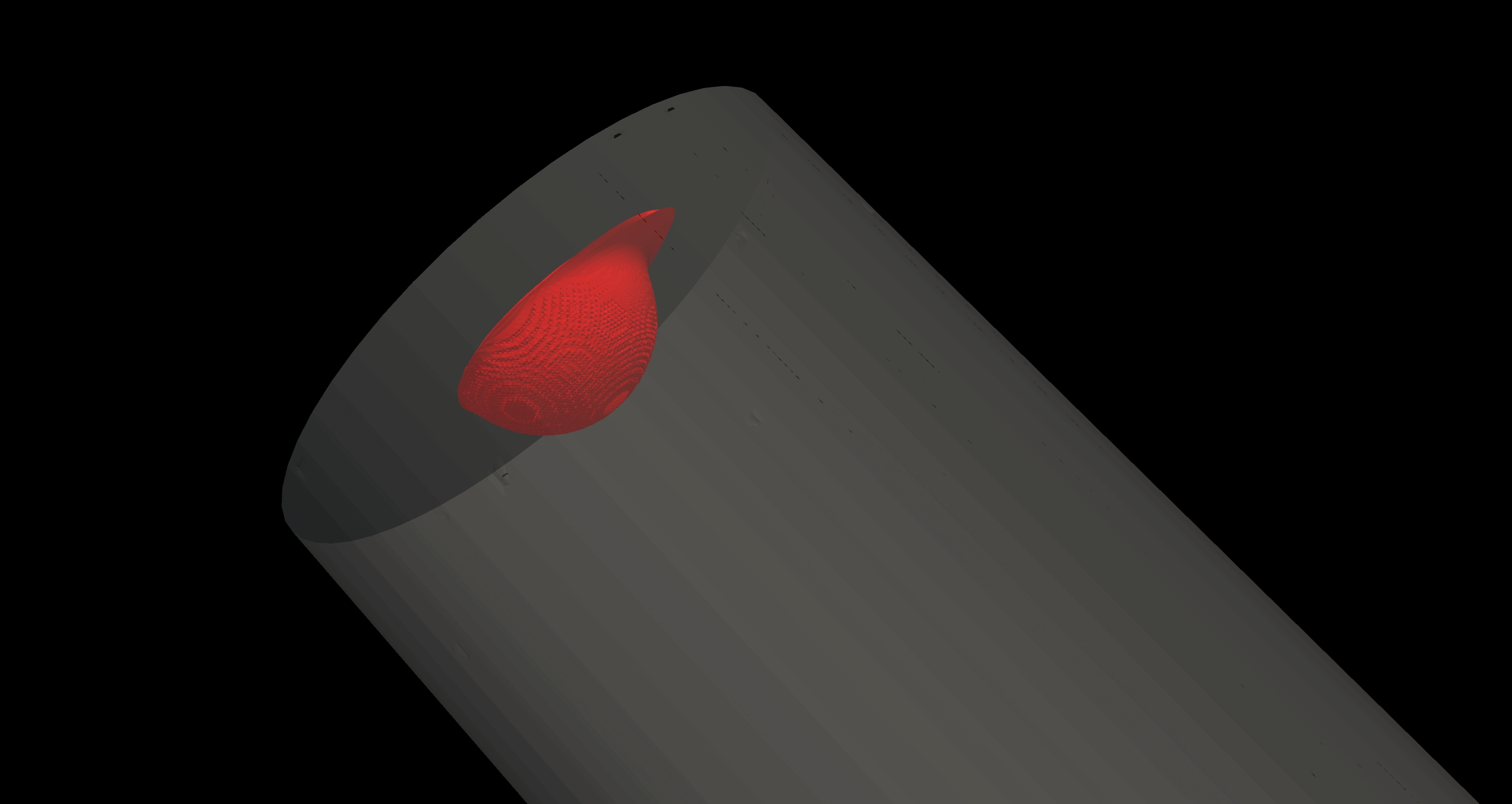}\label{fig:135_V}}
\quad
\subfigure[$\phi=180^{\circ}$, $t=$~225.0~s]{\includegraphics[trim=180mm 80mm 250mm 100mm,clip=true,height=4cm]{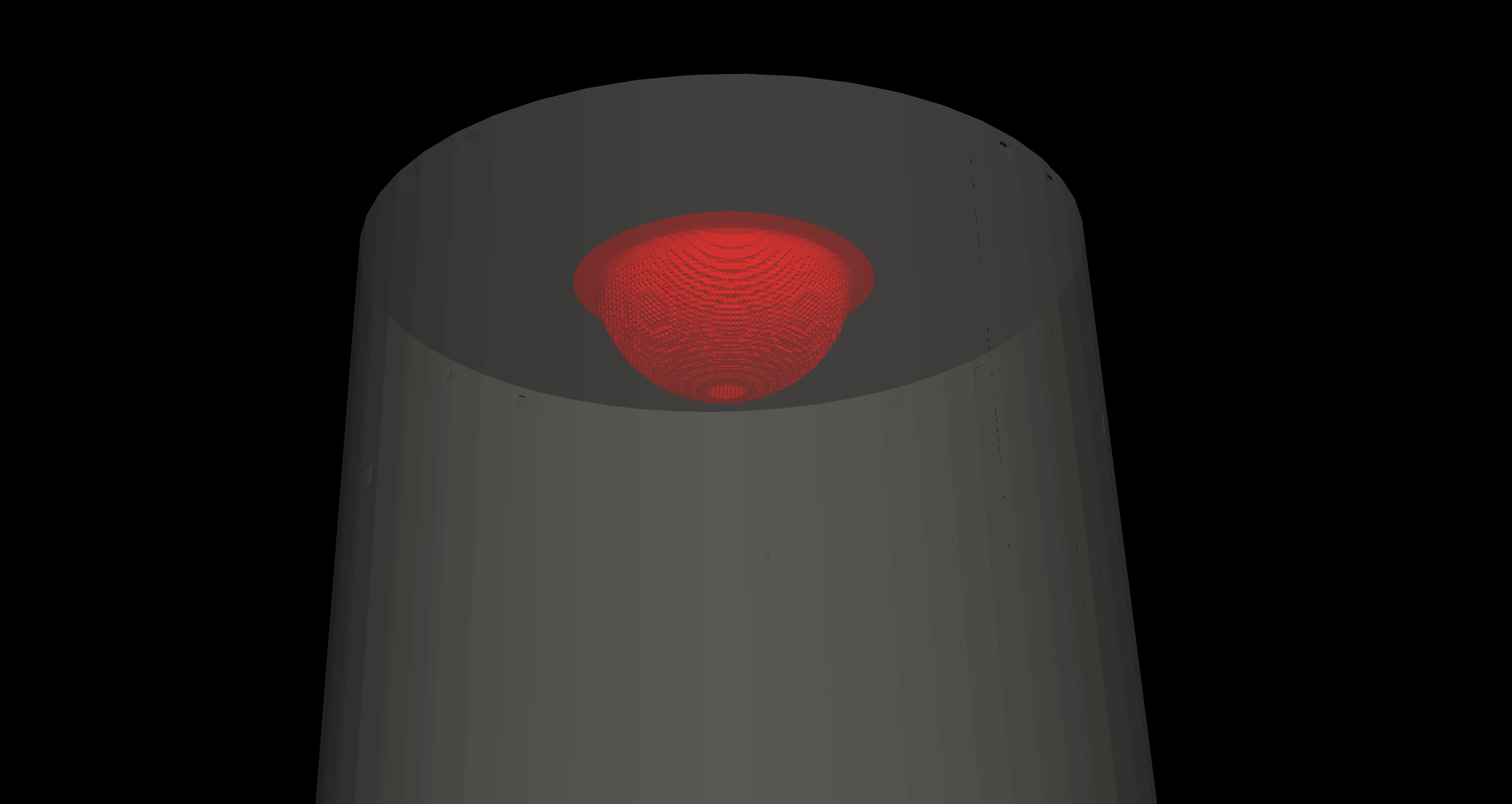}\label{fig:180_V}}
\end{center}
\caption{Fluid volume which is above the AIT of CS\textsubscript{2} of 362.15~K.
The depicted instantaneous snapshots correspond for each case to the time instance of ignition as observed in the experiments.}
\label{fig:V}
\end{figure}

Next, the volume occupied by fluid of high temperature is analyzed.
More specifically, in Fig.~\ref{fig:V} the volume of the fluid which temperature, at the time instance of ignition, is higher than the AIT is visualized.
The AIT of CS\textsubscript{2} is 362.15~K.

It is apparent from Fig.~\ref{fig:0_V} that the volume occupied by the hot gas mixture for $\phi=0^{\circ}$ is larger than for the other orientations (Figs.~\ref{fig:45_V}--(\ref{fig:180_V}).
The volume of hot gas that is accumulated in the chamber is assumed to be a major parameter of influence on the ignition times.
Interestingly, it was observed in the experiments that the ignition duration for $\phi=0^{\circ}$ is less compared to the other orientations.
In contrast, the case of $\phi=180^{\circ}$ that exhibits the second largest volumes of hot gas is found to have a longer ignition duration than the remaining orientations.
Also, for the remaining orientations, there was no obvious correlation found between the hot gas volume and the ignition time.

\begin{figure}[t]
\begin{center}
\subfigure[$\phi=0^{\circ}$, $t=$~166.0~s]{%
\begin{tikzpicture}[thick,white]
\node [anchor=south,inner sep=0] at (0,0) {
\includegraphics[trim=310mm 90mm 350mm 0mm,clip=true,height=4cm]{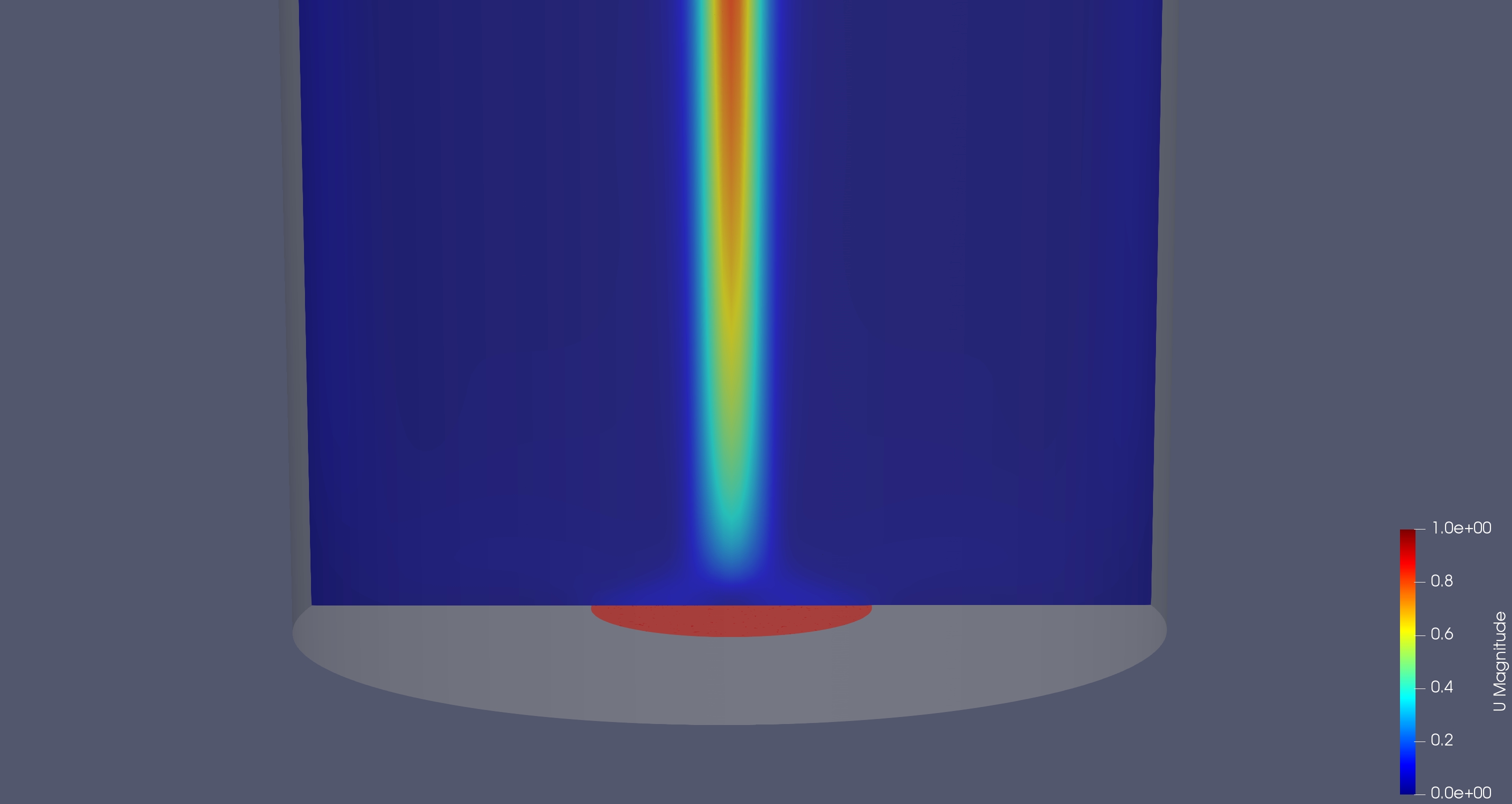}};
\draw [-] (.9,.4) -- (.9,.6) node[anchor=south]{\scriptsize 50};
\draw [-] (-1.35,.4) -- (-1.35,.6) node[anchor=south]{\scriptsize -75};
\draw [-] (-1.7,3.5) -- (-1.5,3.5) node[anchor=west]{\scriptsize 200};
\end{tikzpicture}%
\label{fig:D0_U}}%
\qquad
\subfigure[$\phi=45^{\circ}$, $t=$~212.0~s]{%
\begin{tikzpicture}[thick,white]
\node [anchor=south,inner sep=0] at (0,0) {
\includegraphics[trim=180mm 80mm 500mm 80mm,clip=true,height=4cm]{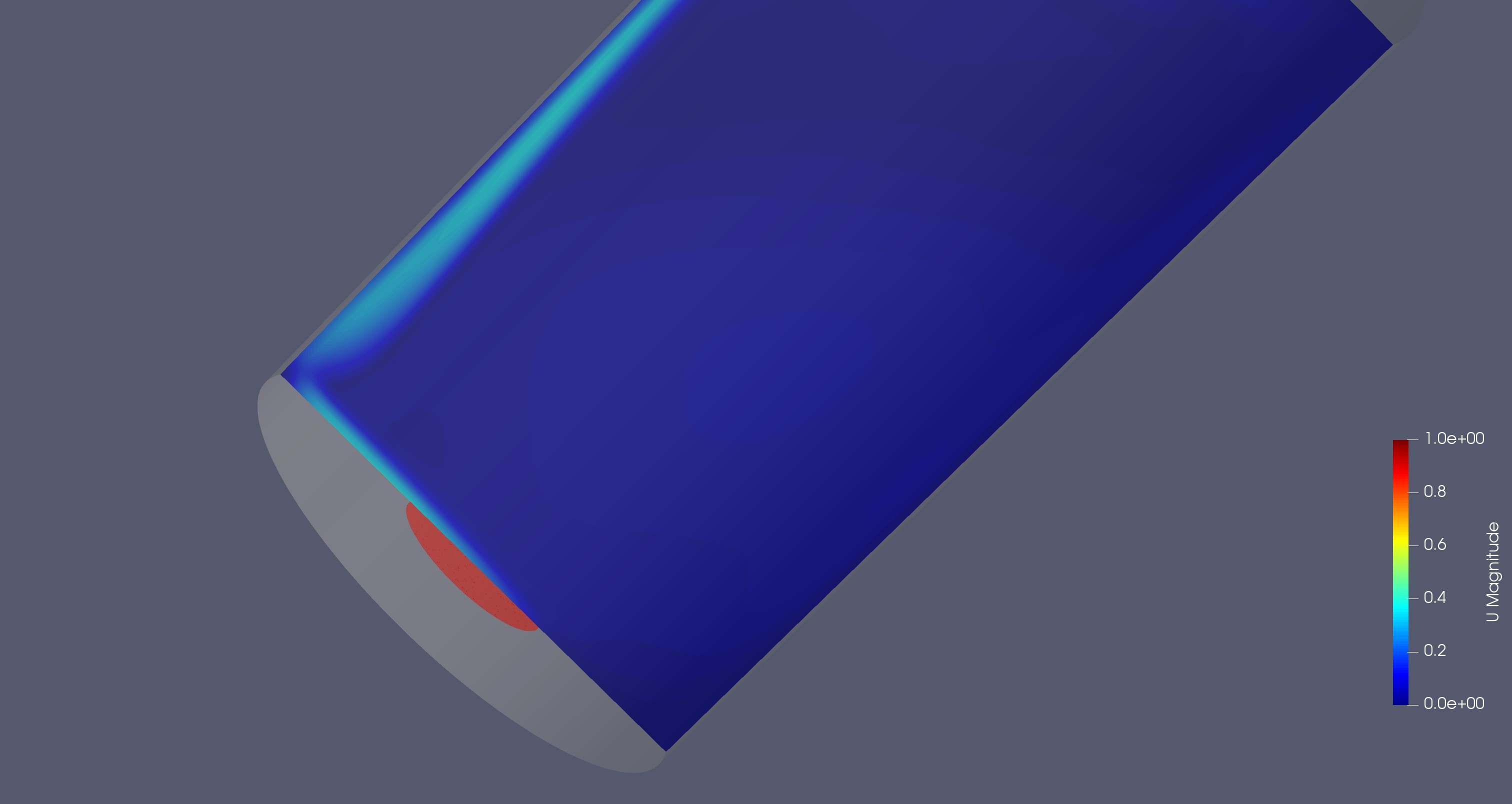}};
\begin{scope}[rotate=-45]
 \draw [-] (-2.8,2.9) -- (-2.6,2.9) node[anchor=west,rotate=-45]{\scriptsize 210};
\end{scope}
\end{tikzpicture}%
\label{fig:D45_U}}%
\qquad
\subfigure[$\phi=90^{\circ}$, $t=$~223.0~s]{%
\begin{tikzpicture}[thick,white]
\node [anchor=south,inner sep=0] at (0,0) {
\includegraphics[trim=240mm 170mm 400mm 00mm,clip=true,height=4cm]{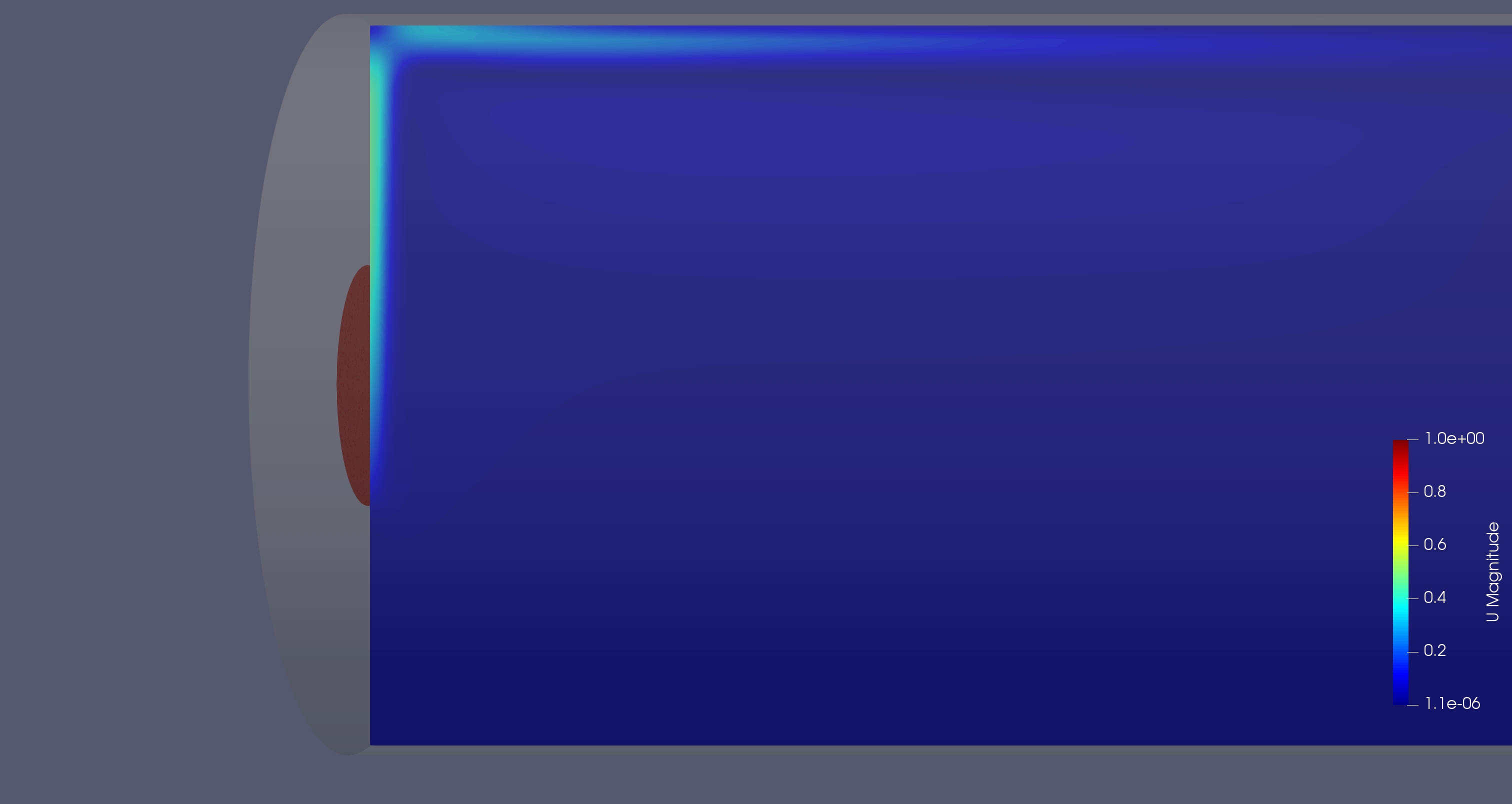}};
\begin{scope}[rotate=-90]
 \draw [-] (-3.8,1.8) -- (-3.6,1.8) node[anchor=west,rotate=-90]{\scriptsize 210};
\end{scope}
\end{tikzpicture}%
\label{fig:D90_U}}%
\\
\subfigure[$\phi=135^{\circ}$, $t=$~210.0~s]{%
\begin{tikzpicture}[thick,white]
\node [anchor=south,inner sep=0] at (0,0) {
\includegraphics[trim=240mm 200mm 420mm 0mm,clip=true,height=4cm]{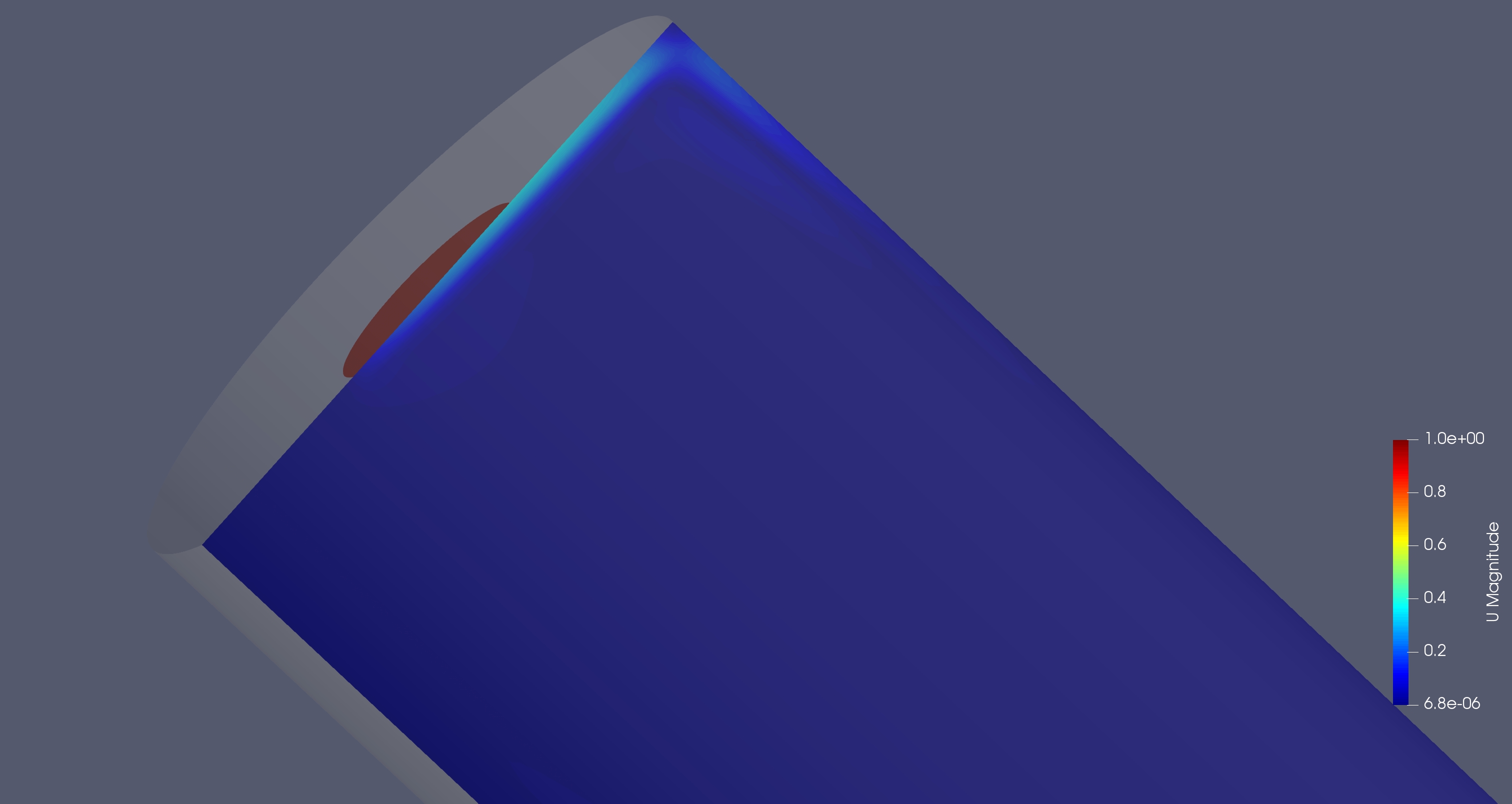}};
\begin{scope}[rotate=-135]
 \draw [-] (-3.0,0.0) -- (-2.8,0.0) node[anchor=east,rotate=-315]{\scriptsize 140};
\end{scope}
\end{tikzpicture}%
\label{fig:D135_U}}%
\qquad
\subfigure[$\phi=180^{\circ}$, $t=$~164.0~s]{%
\begin{tikzpicture}[thick,white]
\node [anchor=south,inner sep=0] at (0,0) {
\includegraphics[trim=220mm 100mm 290mm 50mm,clip=true,height=4cm]{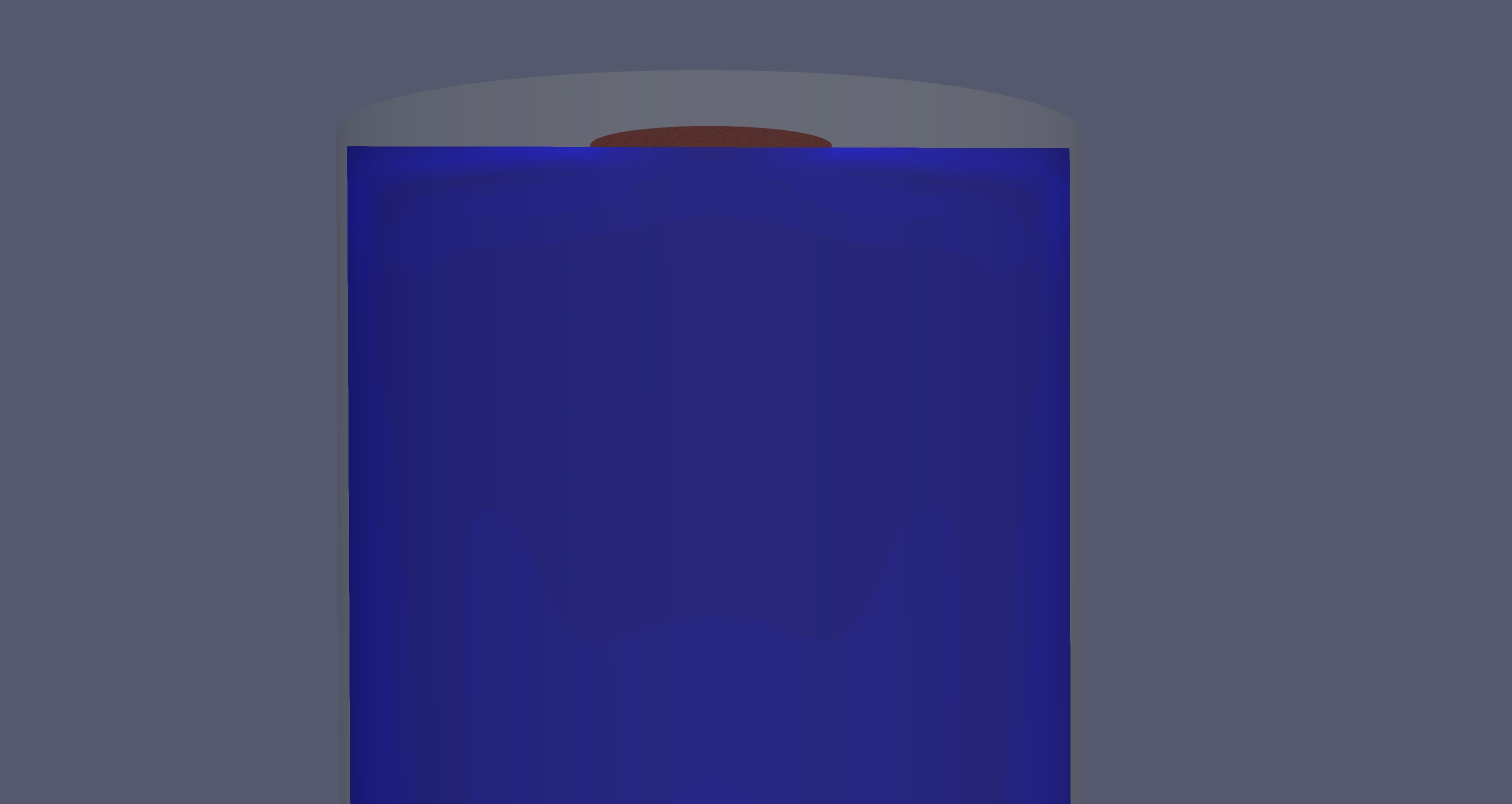}};
\begin{scope}[rotate=0]
 \draw [-] (2.2,0.5) node[anchor=east,rotate=0]{\scriptsize 200} -- (2.4,0.5);
\end{scope}
\end{tikzpicture}%
\label{fig:D180_U}}%
\quad
\begin{tikzpicture}
\node[anchor=south west,inner sep=0] (Bild) at (0,0)
{\includegraphics[trim=0cm 0cm 0cm 0cm,clip=true,width=5mm,height=25mm]{fig/colorbar}};
\begin{scope}[x=(Bild.south east),y=(Bild.north west)]
\draw [](0.6,1.0) node[anchor=south]{$|{\bm u}|/$~(m/s)};
\draw [](0.7,0.95) node[anchor=west,xshift=1mm]{1}; 
\draw [](0.7,0.02) node[anchor=west,xshift=1mm]{0};         
\end{scope}
\end{tikzpicture}
\end{center}
\caption{Magnitude of the velocity of the buoyant flow induced by the hot disc.
The depicted instantaneous snapshots correspond for each case to the time instance of ignition as observed in the experiments.
\textcolor{black}{(All labels in mm.)}}
\label{fig:DU}
\end{figure}

\begin{figure}[tb]
\begin{center}
\subfigure[$\phi=0^{\circ}$, $t=$~166.0~s]{\includegraphics[trim=270mm 0mm 350mm 50mm,clip=true,height=4cm]{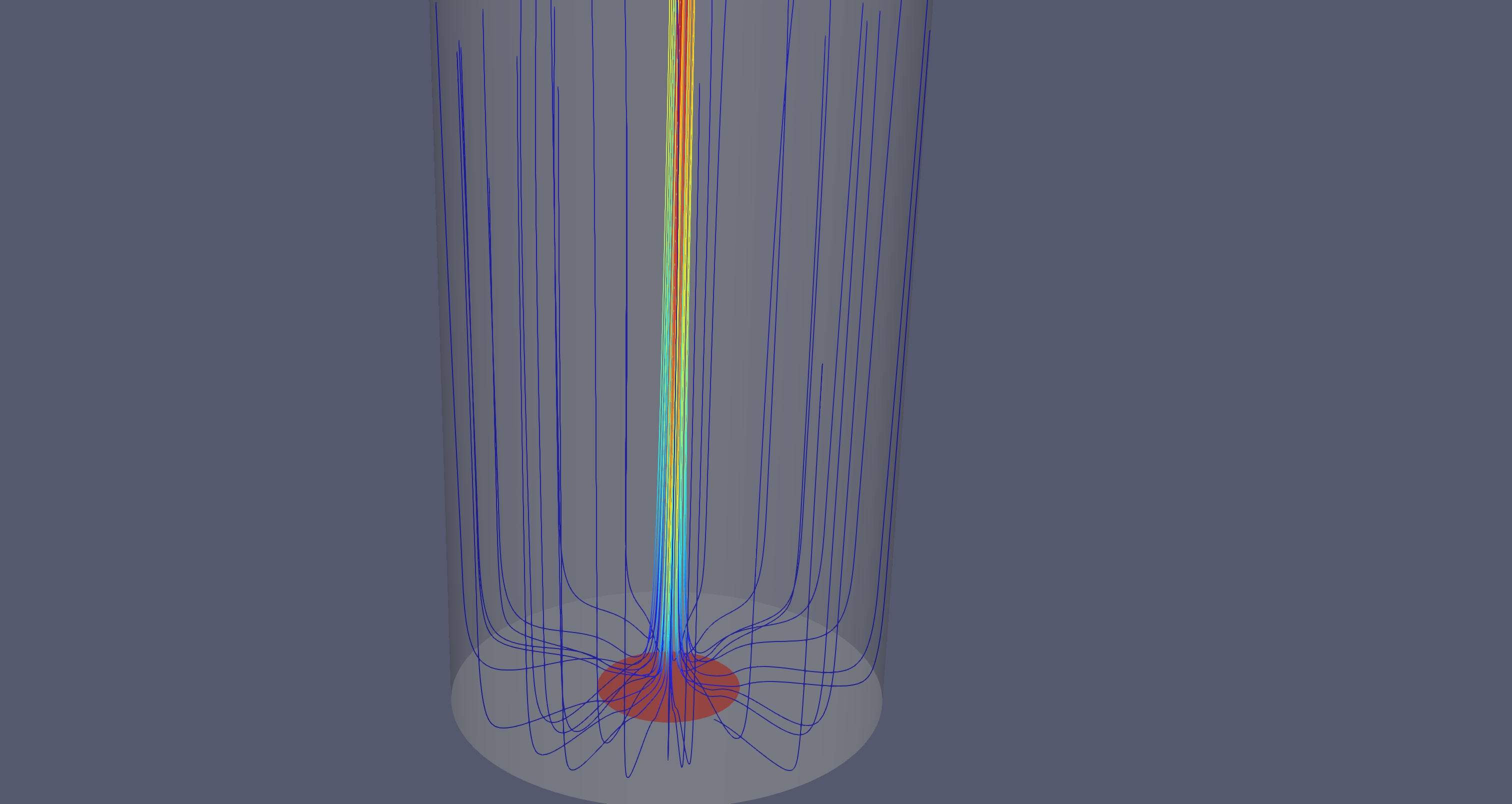}\label{fig:D0_S}}
\quad
\subfigure[$\phi=45^{\circ}$, $t=$~212.0~s]{\includegraphics[trim=50mm 0mm 110mm 0mm,clip=true,height=4cm]{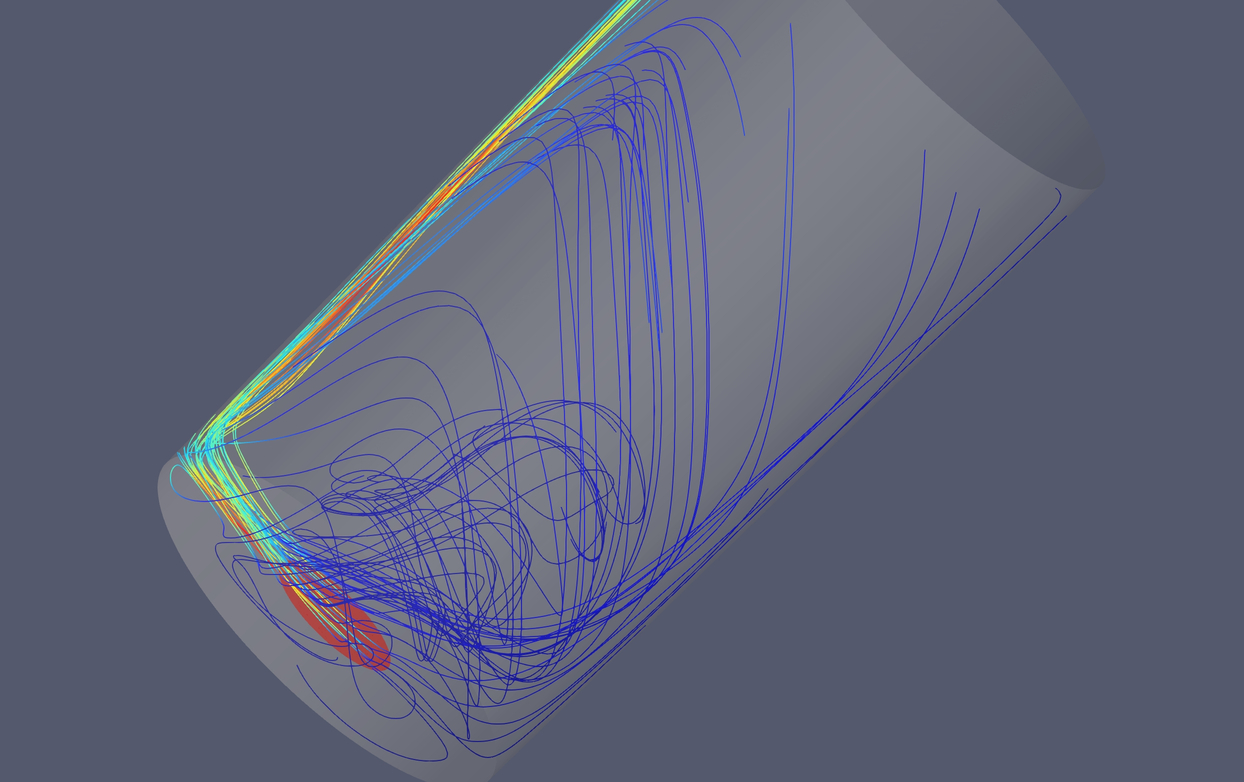}\label{fig:D45_S}}
\quad
\subfigure[$\phi=90^{\circ}$, $t=$~223.0~s]{\includegraphics[trim=0mm 20mm 100mm 00mm,clip=true,height=4cm]{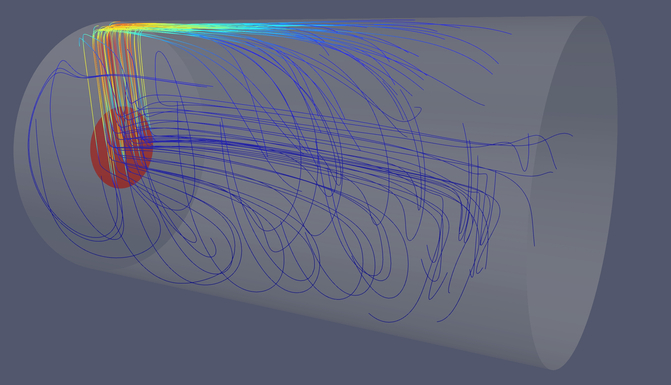}\label{fig:D90_S}}
\\
\subfigure[$\phi=135^{\circ}$, $t=$~210.0~s]{\includegraphics[trim=65mm 50mm 100mm 0mm,clip=true,height=4cm]{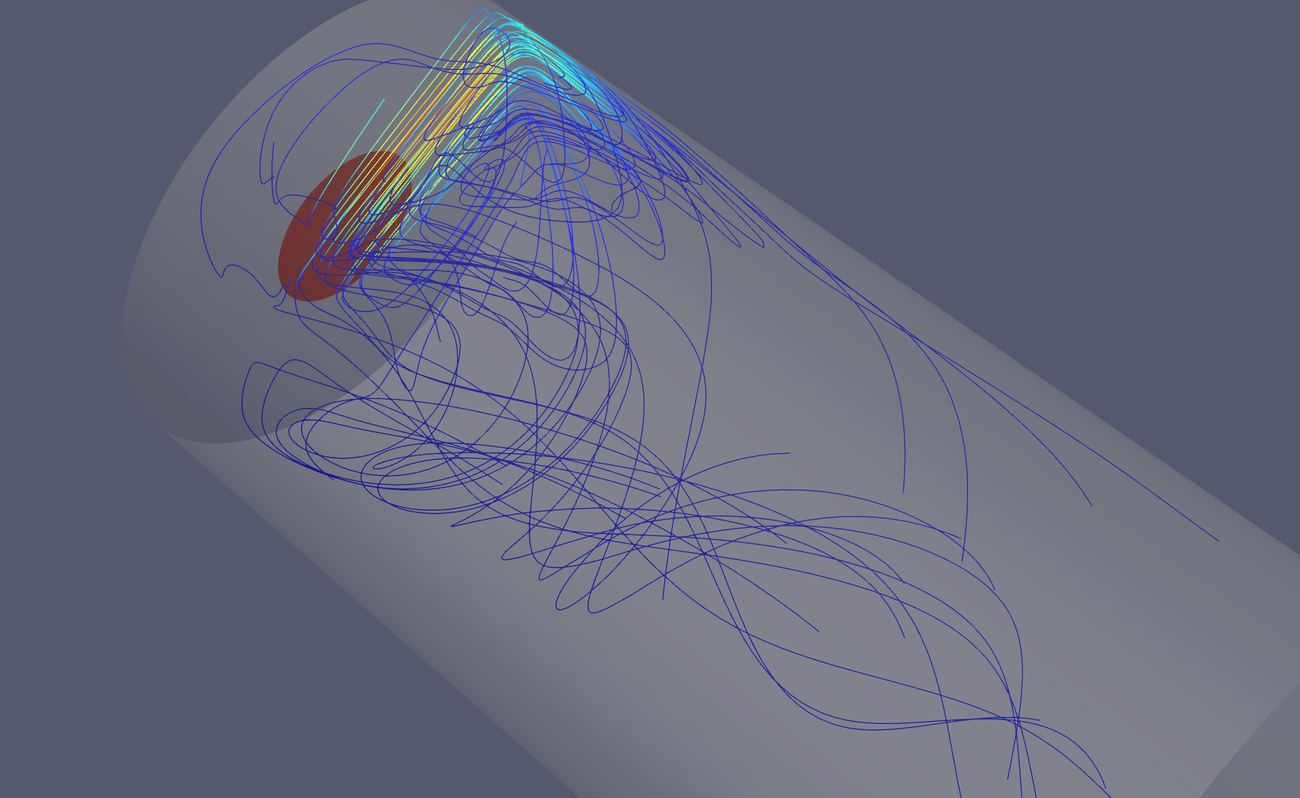}\label{fig:D135_S}}
\qquad
\subfigure[$\phi=180^{\circ}$, $t=$~164.0~s]{\includegraphics[trim=125mm 125mm 165mm 0mm,clip=true,height=4cm]{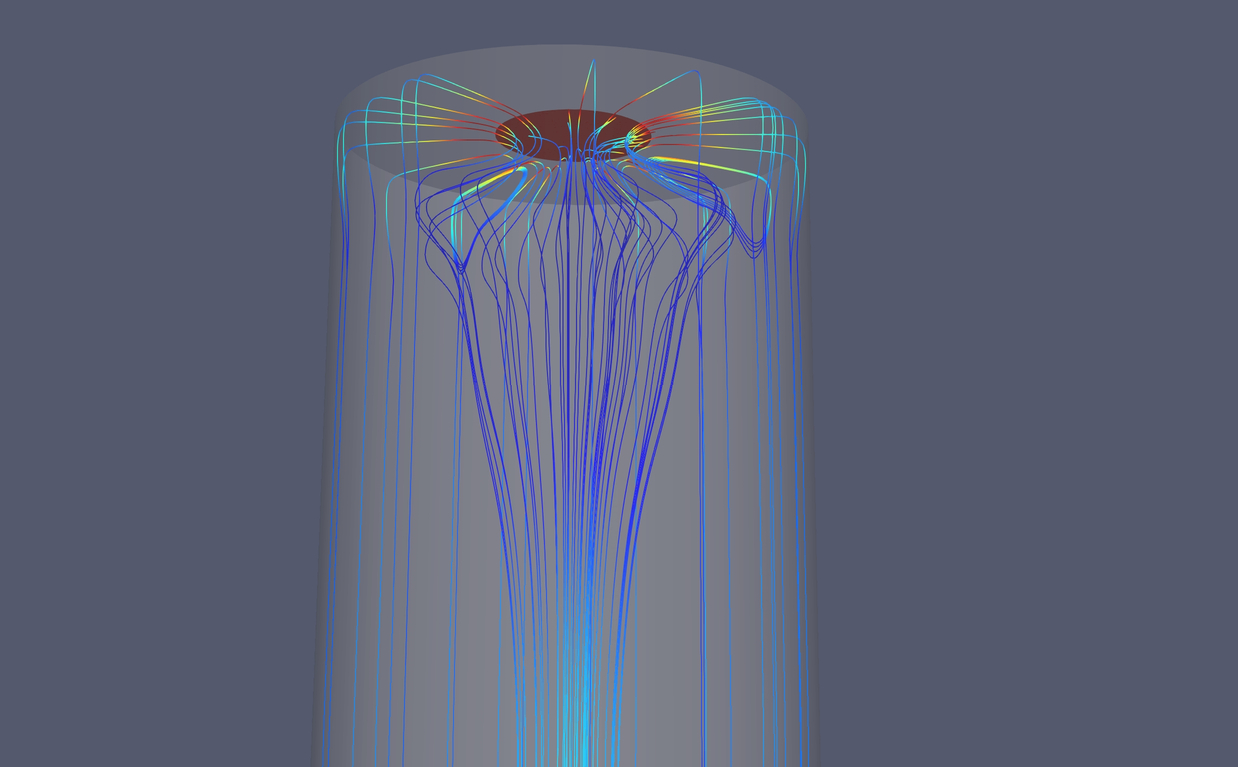}\label{fig:D180_S}}
\qquad
\begin{tikzpicture}
\node[anchor=south west,inner sep=0] (Bild) at (0,0)
{\includegraphics[trim=0cm 0cm 0cm 0cm,clip=true,width=5mm,height=25mm]{fig/colorbar}};
\begin{scope}[x=(Bild.south east),y=(Bild.north west)]
\draw [](0.6,1.0) node[anchor=south]{$|{\bm u}|/$~(m/s)};
\draw [](0.7,0.95) node[anchor=west,xshift=1mm]{1}; 
\draw [](0.7,0.02) node[anchor=west,xshift=1mm]{0};         
\end{scope}
\end{tikzpicture}
\end{center}
\caption{Streamlines of the buoyant flow induced by the hot disc, colored by the magnitude of the velocity.
The depicted instantaneous snapshots correspond for each case to the time instance of ignition as observed in the experiments.}
\label{fig:DS}
\end{figure}

\subsection{Analysis of the thermal flow field originating from a hot disc}

\begin{figure}[t]
\begin{center}
\subfigure[$\phi=0^{\circ}$, $t=$~166.0~s]{%
\begin{tikzpicture}[thick,white]
\node [anchor=south,inner sep=0] at (0,0) {
\includegraphics[trim=200mm 60mm 240mm 0mm,clip=true,height=4cm]{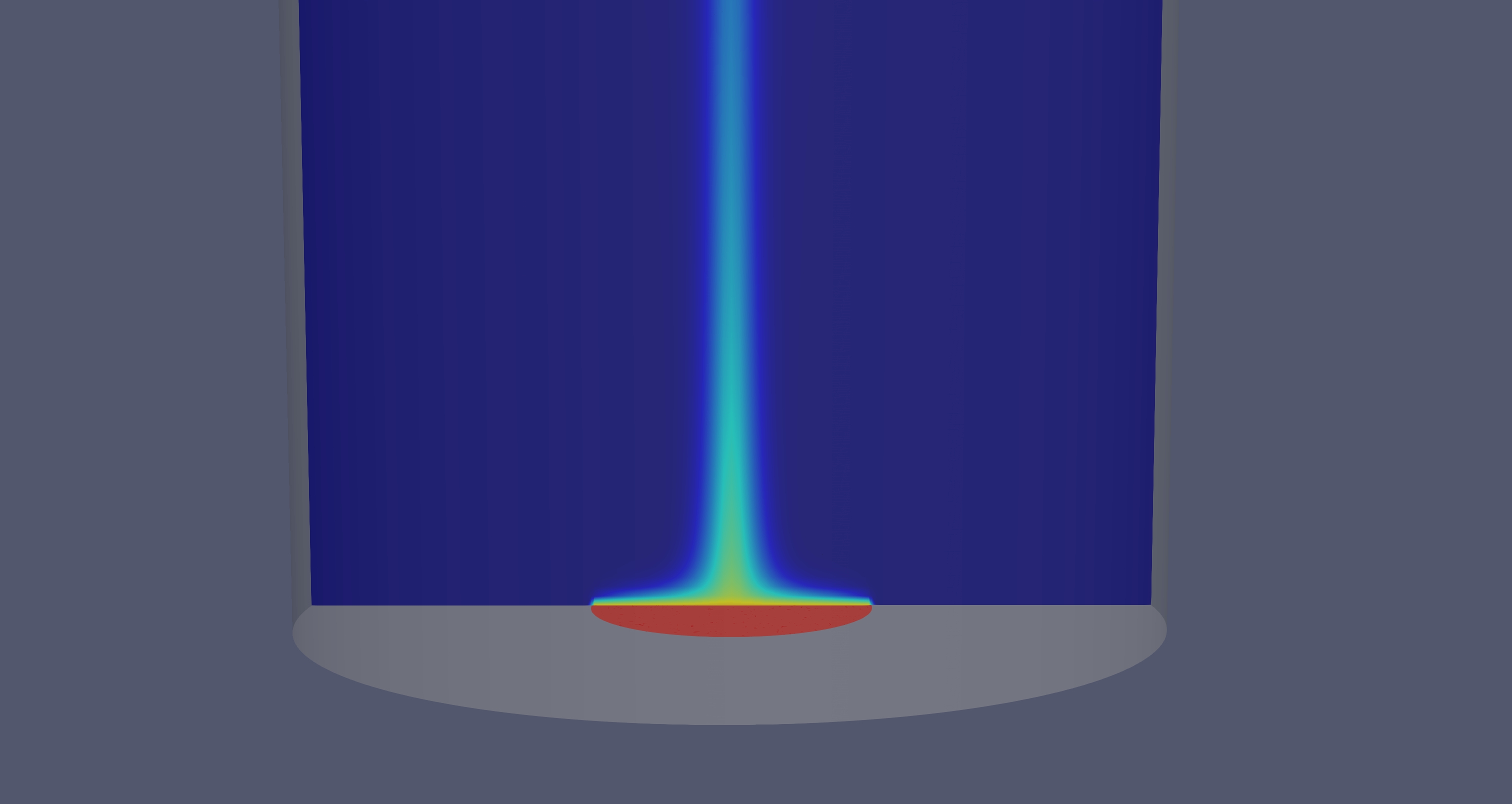}};
\draw [-] (.8,.6) -- (.8,.8) node[anchor=south]{\scriptsize 50};
\draw [-] (-1.6,.6) -- (-1.6,.8) node[anchor=south]{\scriptsize -100};
\draw [-] (-2.4,3.2) -- (-2.2,3.2) node[anchor=west]{\scriptsize 200};
\end{tikzpicture}%
\label{fig:D0_T}}%
\quad
\subfigure[$\phi=45^{\circ}$, $t=$~212.0~s]{%
\begin{tikzpicture}[thick,white]
\node [anchor=south,inner sep=0] at (0,0) {
\includegraphics[trim=180mm 80mm 480mm 80mm,clip=true,height=4cm]{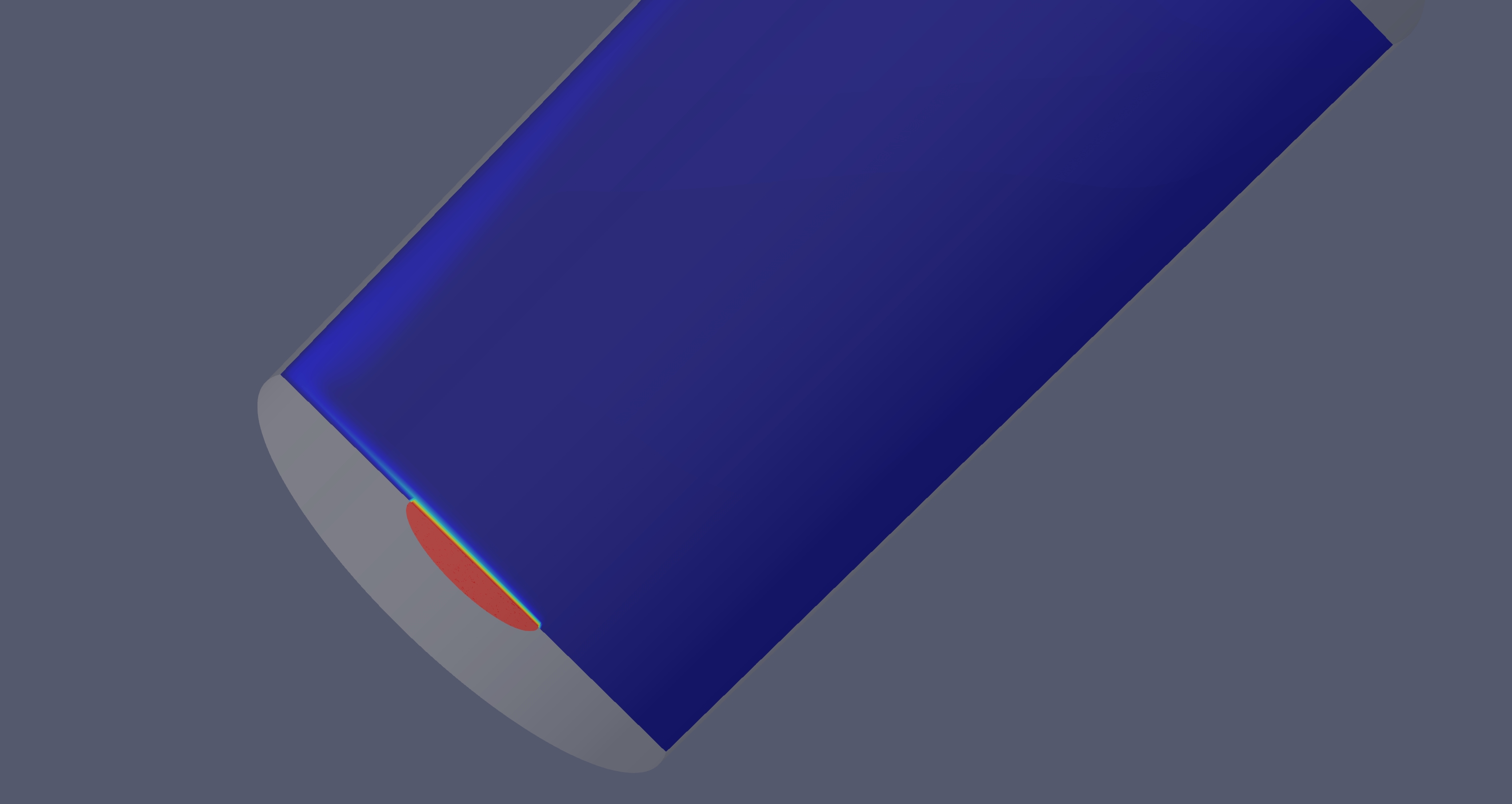}};
\begin{scope}[rotate=-45]
 \draw [-] (-2.8,2.7) -- (-2.6,2.7) node[anchor=west,rotate=-45]{\scriptsize 210};
\end{scope}
\end{tikzpicture}%
\label{fig:D45_T}}%
\quad
\subfigure[$\phi=90^{\circ}$, $t=$~223.0~s]{%
\begin{tikzpicture}[thick,white]
\node [anchor=south,inner sep=0] at (0,0) {
\includegraphics[trim=150mm 100mm 490mm 20mm,clip=true,height=4cm]{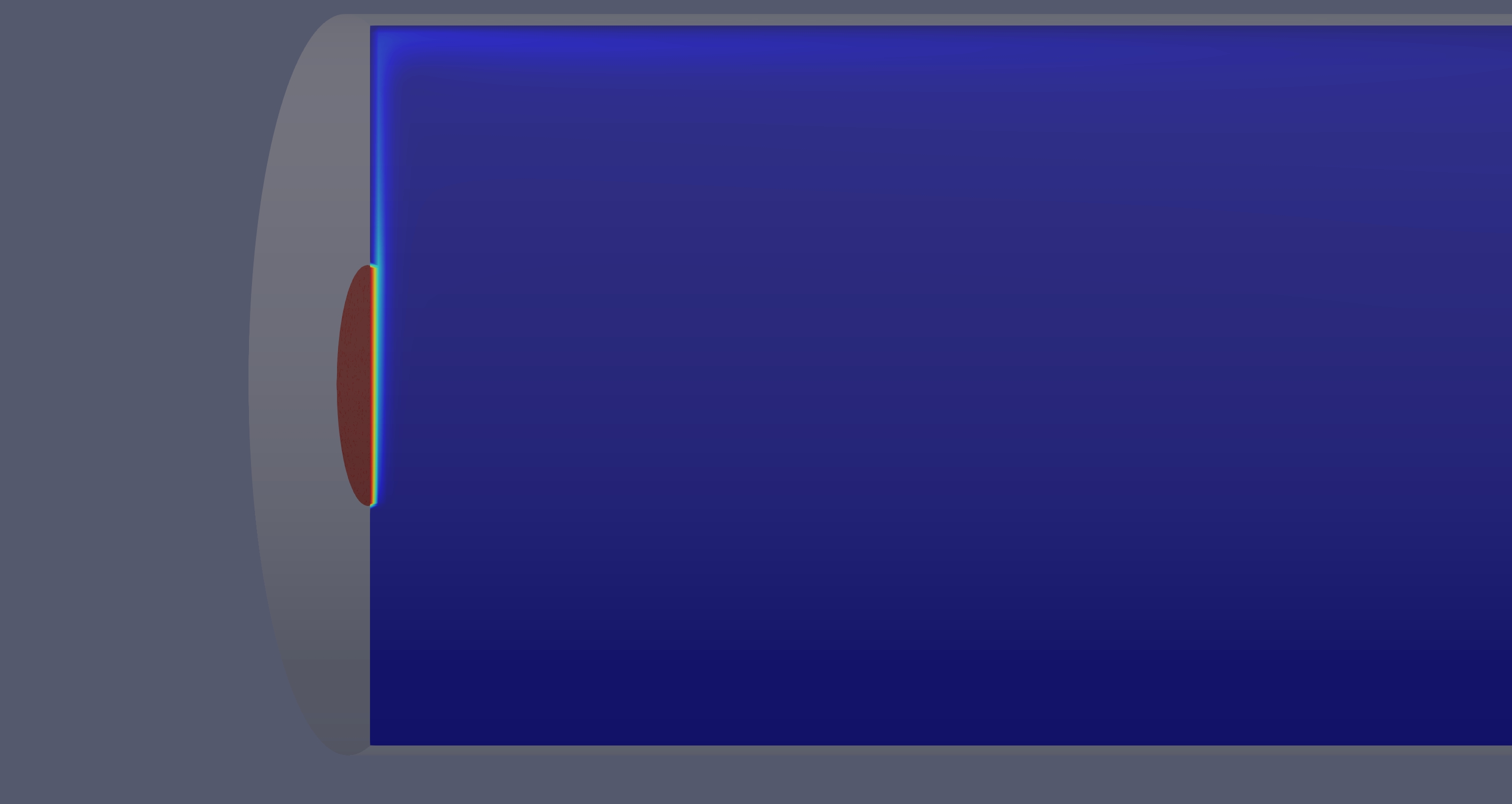}};
\begin{scope}[rotate=-90]
 \draw [-] (-4.0,1.4) -- (-3.8,1.4) node[anchor=west,rotate=-90]{\scriptsize 150};
\end{scope}
\end{tikzpicture}%
\label{fig:D90_T}}%
\\
\subfigure[$\phi=135^{\circ}$, $t=$~210.0~s]{%
\begin{tikzpicture}[thick,white]
\node [anchor=south,inner sep=0] at (0,0) {
\includegraphics[trim=200mm 200mm 350mm 0mm,clip=true,height=4cm]{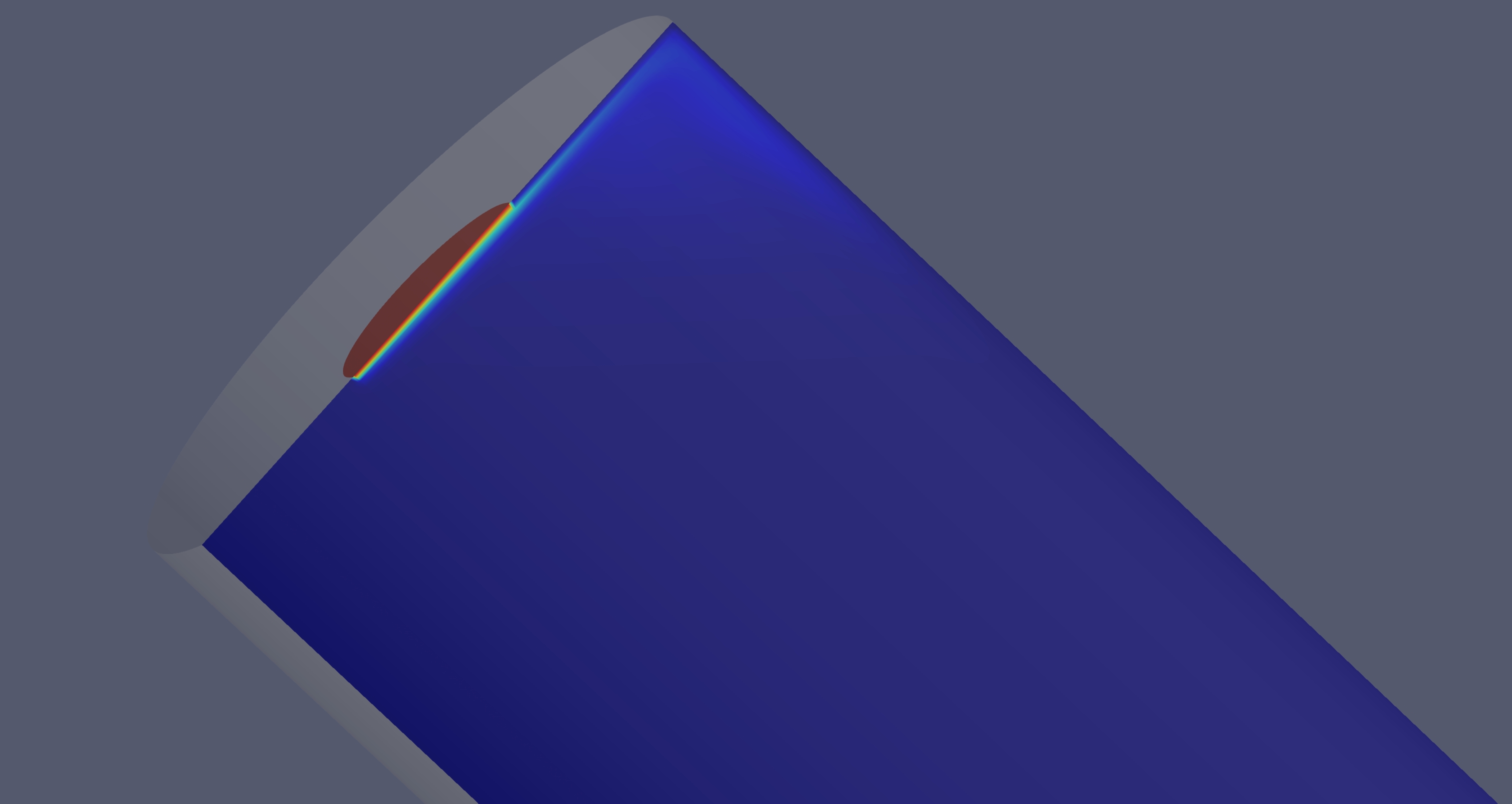}};
\begin{scope}[rotate=-135]
 \draw [-] (-2.9,0.8) -- (-2.7,0.8) node[anchor=east,rotate=-315]{\scriptsize 190};
\end{scope}
\end{tikzpicture}%
\label{fig:D135_T}}%
\quad
\subfigure[$\phi=180^{\circ}$, $t=$~164.0~s]{%
\begin{tikzpicture}[thick,white]
\node [anchor=south,inner sep=0] at (0,0) {
\includegraphics[trim=200mm 60mm 250mm 0mm,clip=true,height=4cm]{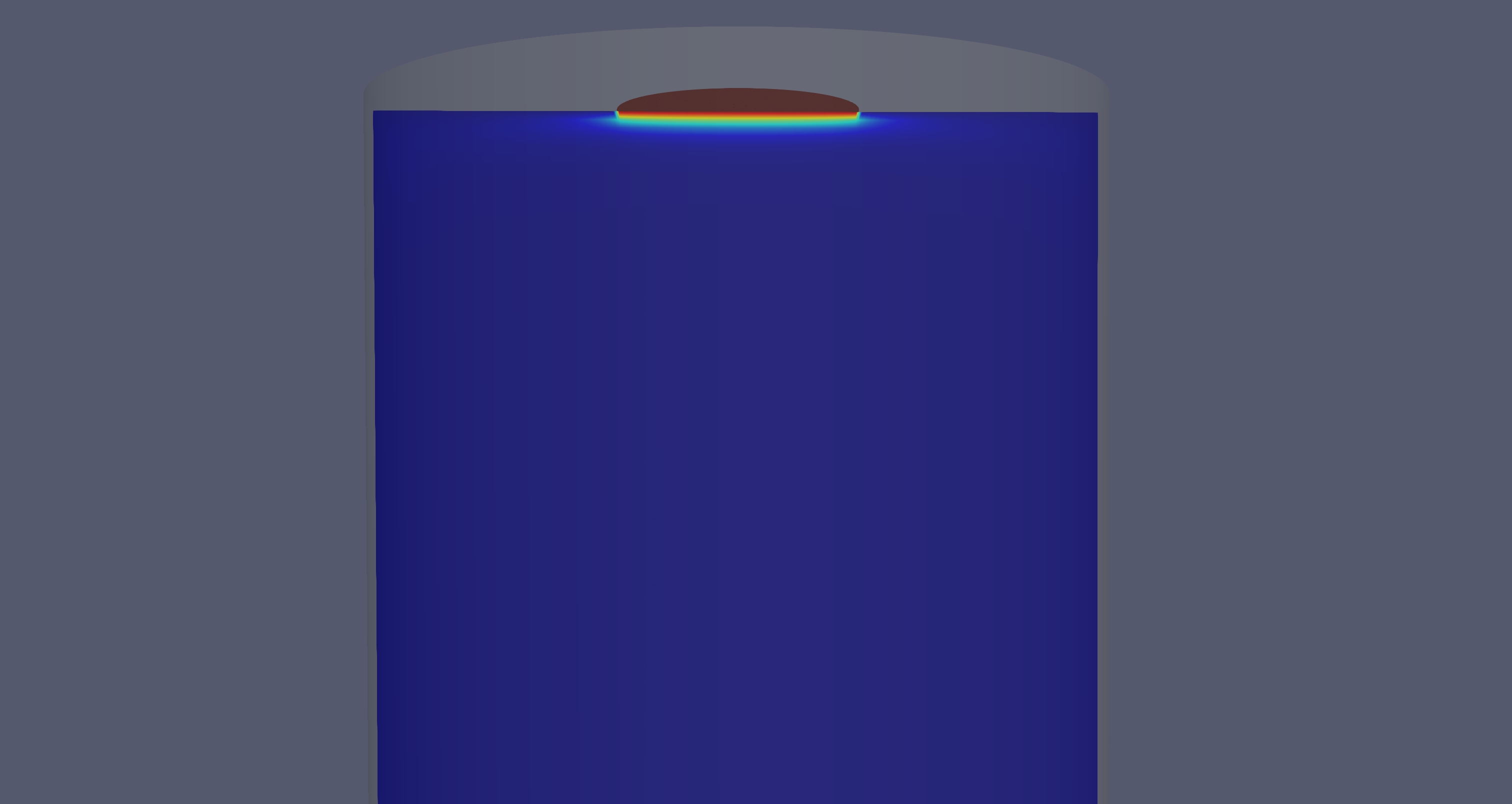}};
\begin{scope}[rotate=0]
 \draw [-] (1.9,0.5) node[anchor=east,rotate=0]{\scriptsize 210} -- (2.1,0.5);
\end{scope}
\end{tikzpicture}%
\label{fig:D180_T}}
\quad
\begin{tikzpicture}
\node[anchor=south west,inner sep=0] (Bild) at (0,0)
{\includegraphics[trim=0cm 0cm 0cm 0cm,clip=true,width=5mm,height=25mm]{fig/colorbar}};
\begin{scope}[x=(Bild.south east),y=(Bild.north west)]
\draw [](0.6,1.0) node[anchor=south]{$T /$~K};
\draw [](0.7,0.95) node[anchor=west,xshift=1mm]{480}; 
\draw [](0.7,0.02) node[anchor=west,xshift=1mm]{290};         
\end{scope}
\end{tikzpicture}
\end{center}
\caption{Temperature field of the thermal plumes induced by the hot disc.
The depicted instantaneous snapshots correspond for each case to the time instance of ignition as observed in the experiments.
\textcolor{black}{(All labels in mm.)}}
\label{fig:DT}
\end{figure}

In this section, the flow and thermal fields induced by the hot disc are discussed and compared to those of the hemisphere.
Analogous to the previous section, Fig.~\ref{fig:DU} presents the magnitude of the flow velocity for different vessel orientations, and Figs.~\ref{fig:DS} and~\ref{fig:DT} their corresponding streamlines and temperature fields.
Again, the depicted time instances correlate to those when the ignition was observed in the experiments of \citet{Priyank2019,Gro20a}.
According to these figures, for the orientation of $\phi=0^{\circ}$ the flow separation occurs at the center of the disc.
This is the same location where the thermal boundary layer merges, leading to a hot spot directly above the disc (Fig.~\ref{fig:D0_T}).
This hot spot formation is similar to that observed for the hemispherical geometry of the same orientation (cf.~Fig~\ref{fig:0_T}).

Also for all the remaining orientations, with the exception of $\phi=45^{\circ}$ which is discussed later, the flow separation takes place in qualitative agreement with the hemisphere case.
More specifically, the thermal plume propagates either in the opposite direction of the gravitational vector ($\phi=90^{\circ}$) or as restrained by the boundary of the vessel ($\phi=135^{\circ}$ and~$180^{\circ}$).
This leads to the shift of the hot spot to the left corner of the disc.
Comparison with the experiments reveals that these numerically predicted hot spot locations indeed coincide with the ignition locations.

Surprisingly, the thermal plume formed through the hot disc orientation of $\phi=45^{\circ}$ is different from all other cases, see Figs.~\ref{fig:D45_U}, \ref{fig:D45_S}, and~\ref{fig:D45_T}.
In the hemisphere case (cf.~Figs.~\ref{fig:45_U}, \ref{fig:45_S}, and~\ref{fig:45_T}), the flow separates from the disk and propagates upwards.
Contrary, the flow originating from the flat disk separates at the left corner of the disc and convects along the bottom wall of the vessel towards its corner, see Fig.~\ref{fig:D45_U}.
This behavior is unexpected since for all other simulated cases the flow separation always occurs in the direction of buoyancy.

\begin{figure}[tb]
\begin{center}

\subfigure[$\phi=45^{\circ}$, $t=$~10~s]{%
\begin{tikzpicture}[thick,white]
\node [anchor=south,inner sep=0] at (0,0) {
\includegraphics[trim=60mm 0mm 140mm 40mm,clip=true,height=4cm]{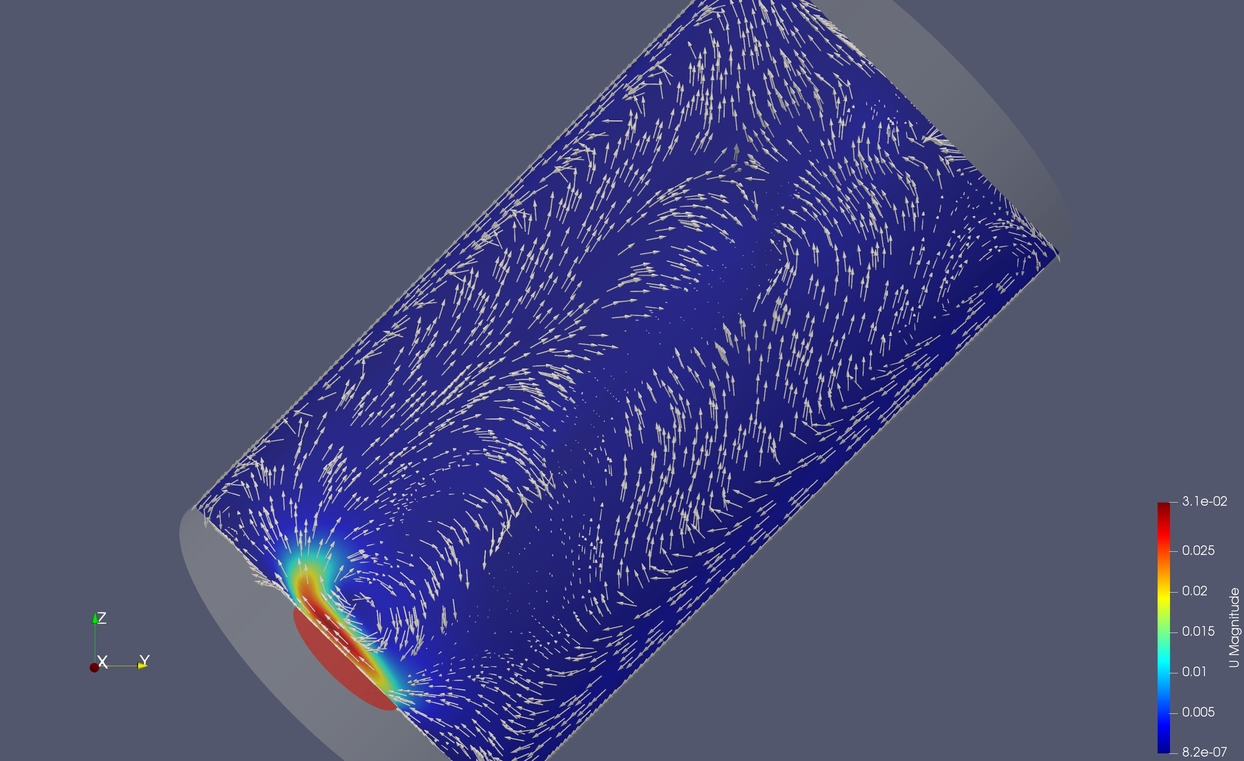}};
\begin{scope}[rotate=-45]
 \draw [-] (-2.5,3.2) -- (-2.3,3.2) node[anchor=west,rotate=-45]{\scriptsize 400};
\end{scope}
\end{tikzpicture}%
\label{fig:D45_U_10s}}%
\quad
\subfigure[$\phi=45^{\circ}$, $t=$~15~s]{\includegraphics[trim=60mm 0mm 140mm 40mm,clip=true,height=4cm]{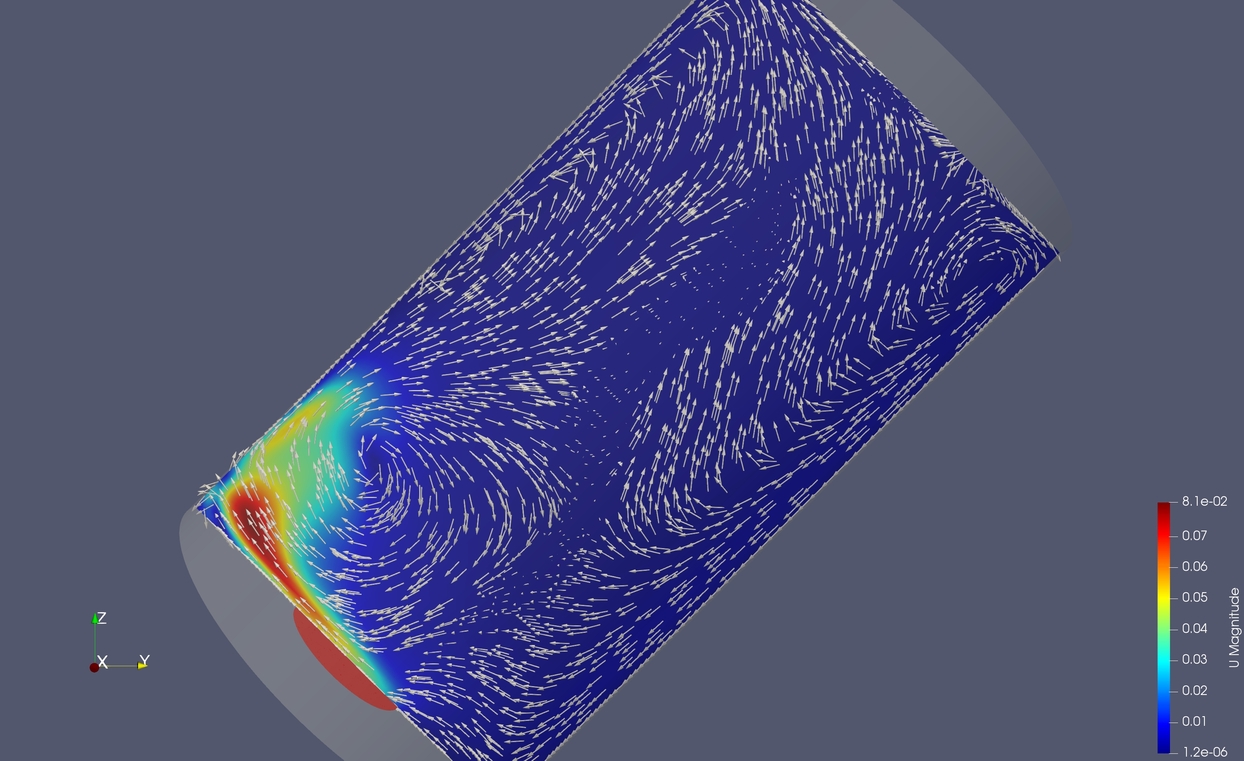}\label{fig:D45_U_15s}}
\quad
\subfigure[$\phi=45^{\circ}$, $t=$~21~s]{\includegraphics[trim=60mm 0mm 140mm 40mm,clip=true,height=4cm]{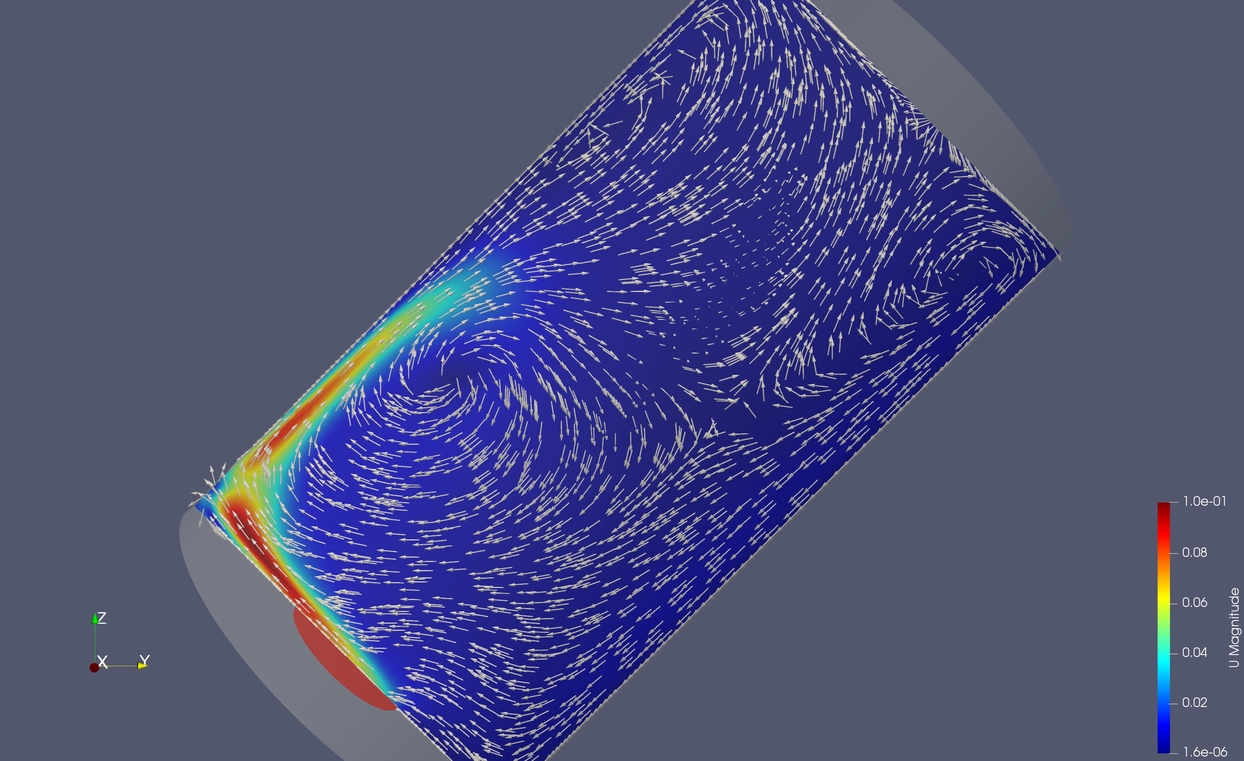}\label{fig:D45_U_21s}}
\\
\subfigure[$\phi=45^{\circ}$, $t=$~27~s]{\includegraphics[trim=60mm 0mm 140mm 40mm,clip=true,height=4cm]{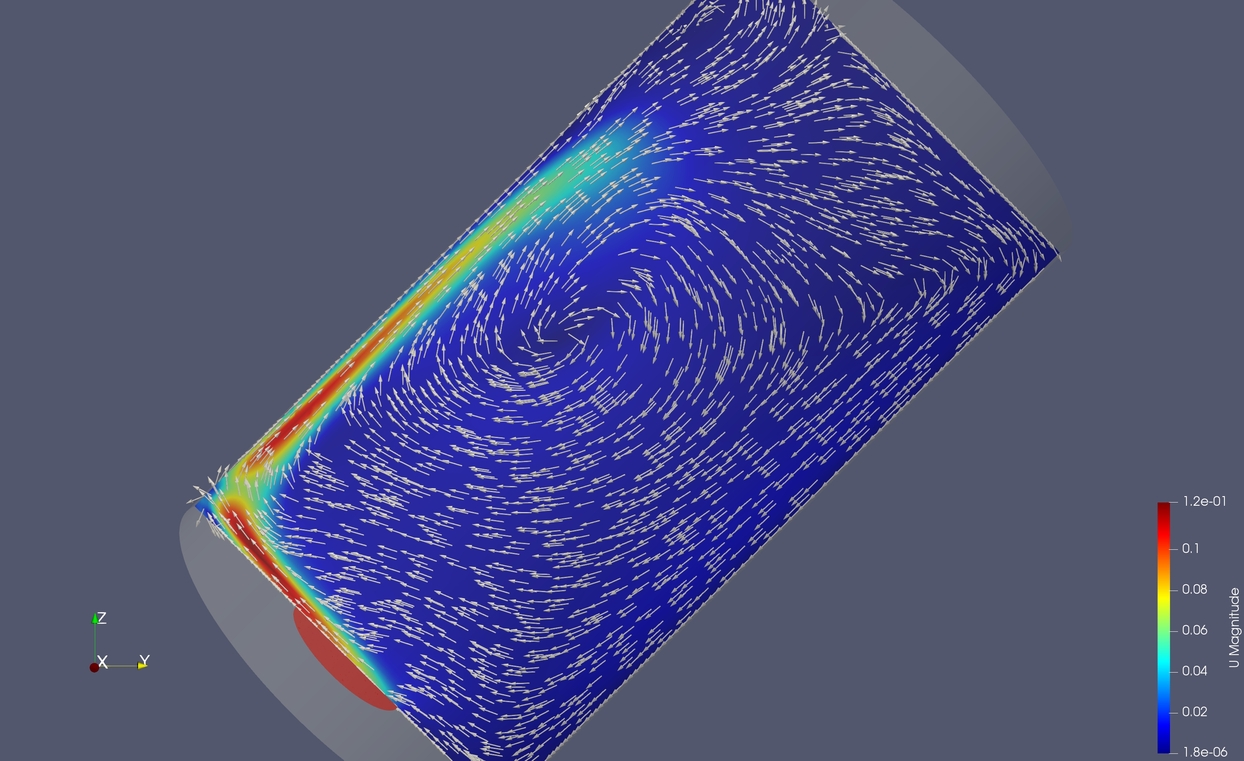}\label{fig:D45_U_27s}}
\quad
\subfigure[$\phi=45^{\circ}$, $t=$~33~s]{\includegraphics[trim=60mm 0mm 140mm 40mm,clip=true,height=4cm]{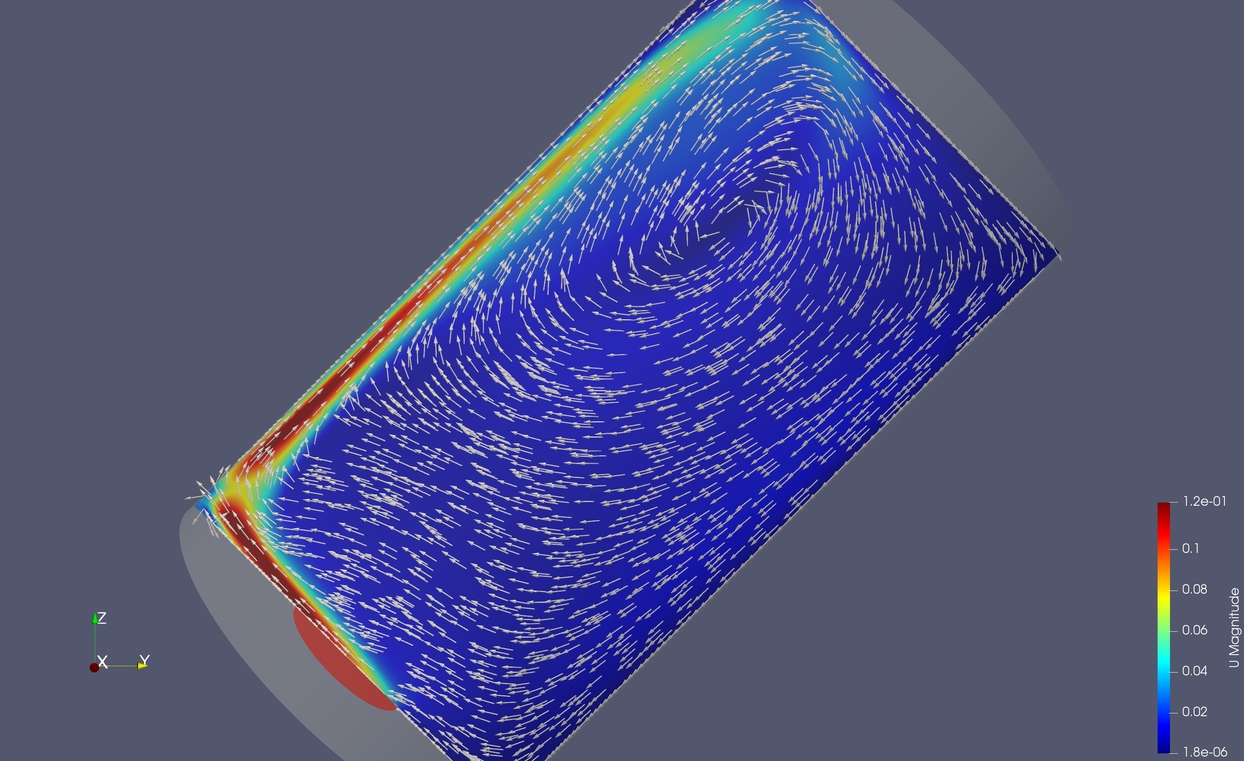}\label{fig:D45_U_33s}}
\quad
\begin{tikzpicture}
\node[anchor=south west,inner sep=0] (Bild) at (0,0)
{\includegraphics[trim=0cm 0cm 0cm 0cm,clip=true,width=5mm,height=25mm]{fig/colorbar}};
\begin{scope}[x=(Bild.south east),y=(Bild.north west)]
\draw [](0.6,1.0) node[anchor=south]{$|{\bm u}|/$~(m/s)};
\draw [](0.7,0.95) node[anchor=west,xshift=1mm]{0.3}; 
\draw [](0.7,0.02) node[anchor=west,xshift=1mm]{0};         
\end{scope}
\end{tikzpicture}
\end{center}
\caption{Magnitude of the velocity and the uniform velocity vectors of the buoyant flow induced by the hot disc at different time instances.
}
\label{fig:DU_45_t}
\end{figure}

\begin{figure}[tb]
\begin{center}
\subfigure[$\phi=0^{\circ}$, $t=$~166.0~s]{\includegraphics[trim=260mm 90mm 300mm 20mm,clip=true,height=4cm]{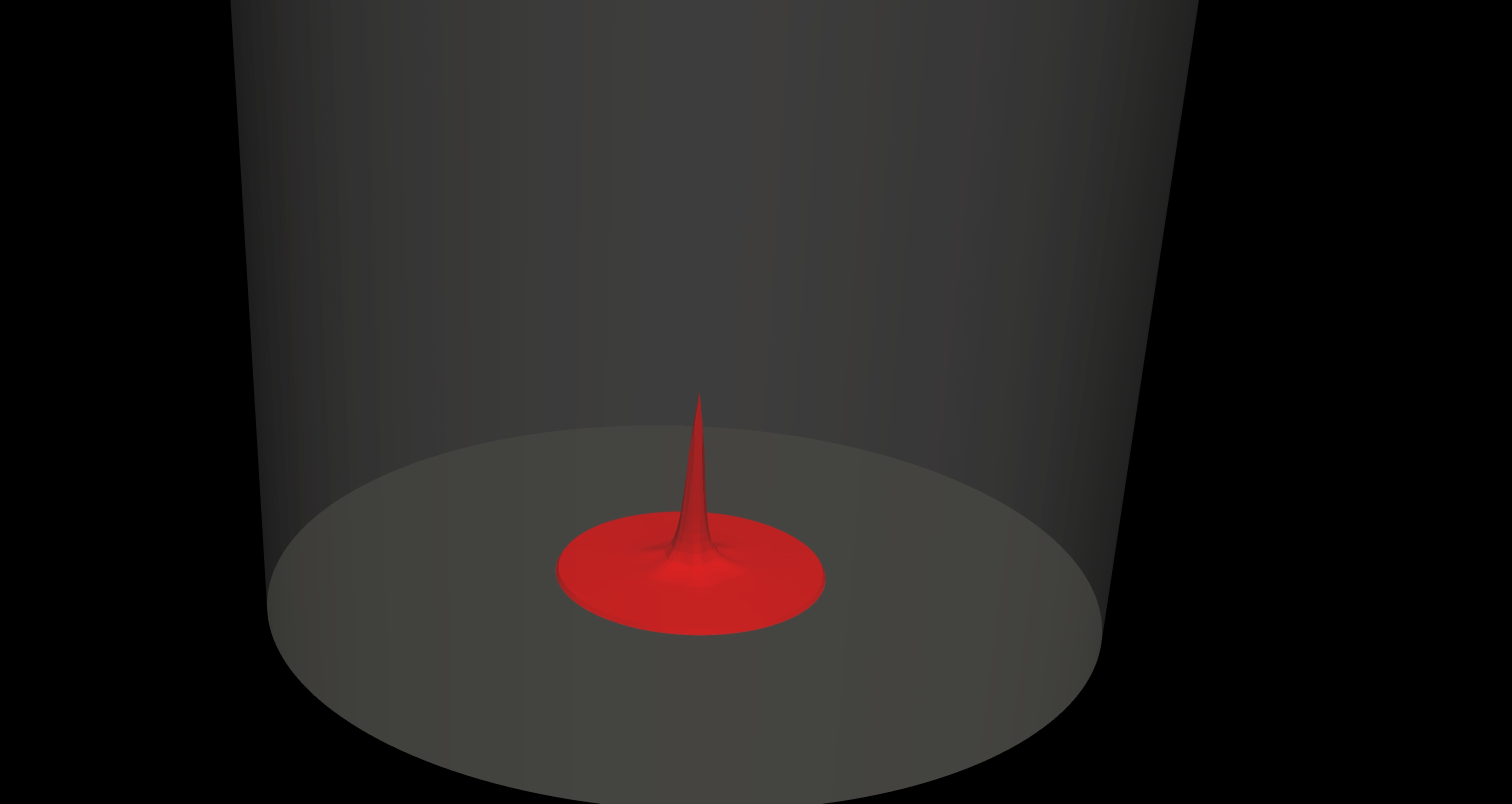}\label{fig:D0_V}}
\quad
\subfigure[$\phi=45^{\circ}$, $t=$~212.0~s]{\includegraphics[trim=220mm 80mm 400mm 80mm,clip=true,height=4cm]{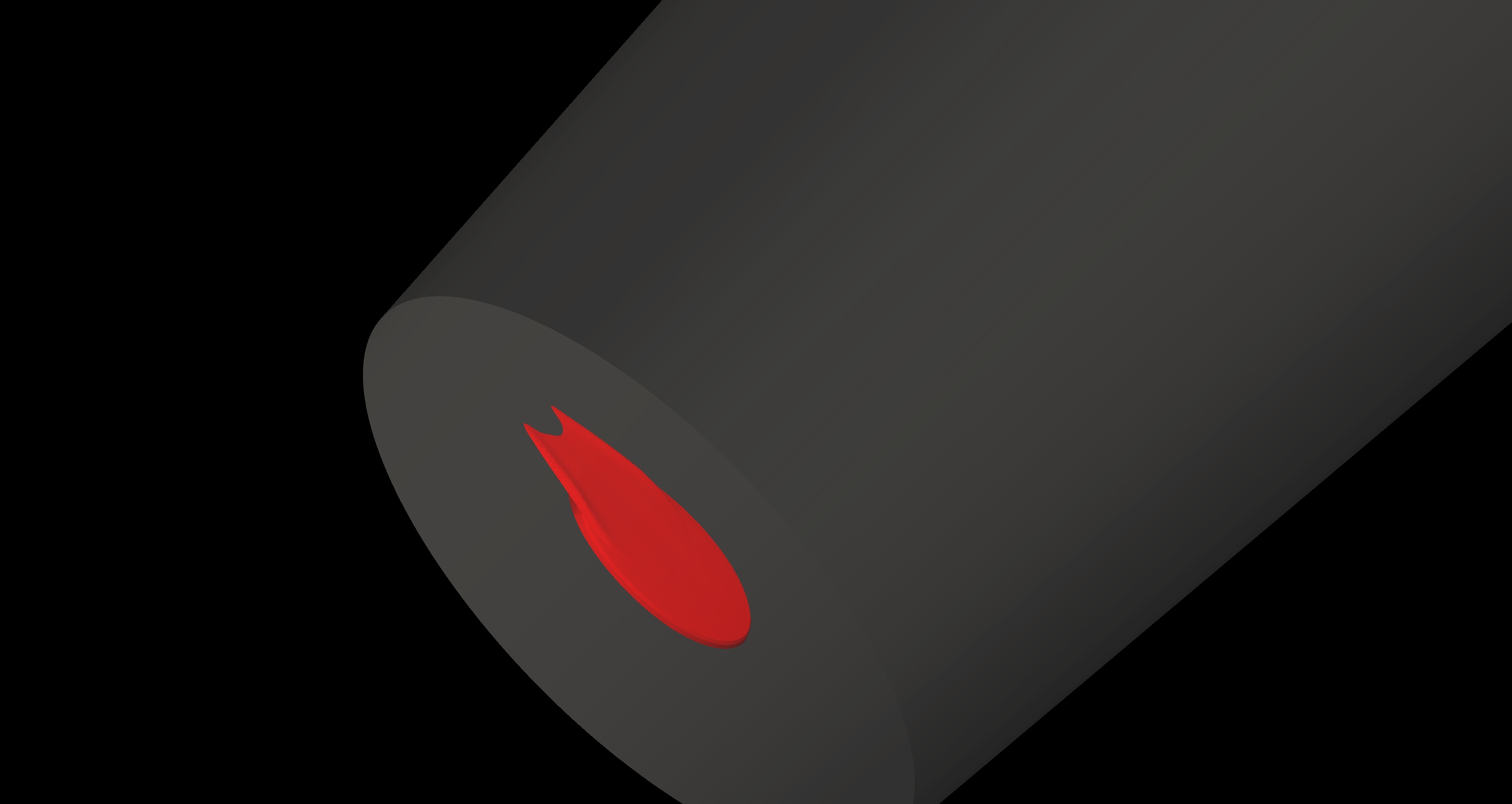}\label{fig:D45_V}}
\quad
\subfigure[$\phi=90^{\circ}$, $t=$~223.0~s]{\includegraphics[trim=180mm 150mm 500mm 20mm,clip=true,height=4cm]{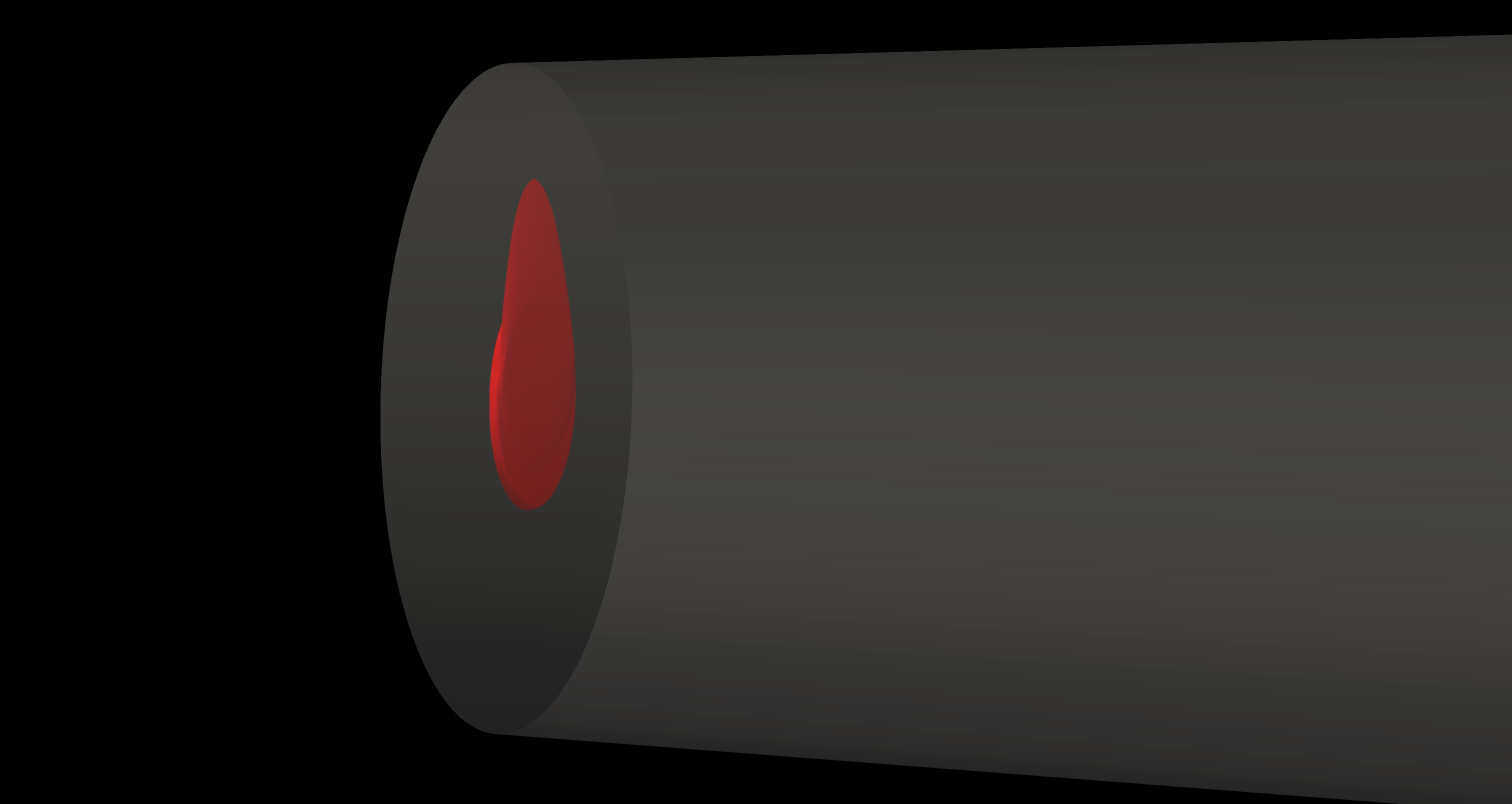}\label{fig:D90_V}}
\\
\subfigure[$\phi=135^{\circ}$, $t=$~210.0~s]{\includegraphics[trim=150mm 150mm 350mm 40mm,clip=true,height=4cm]{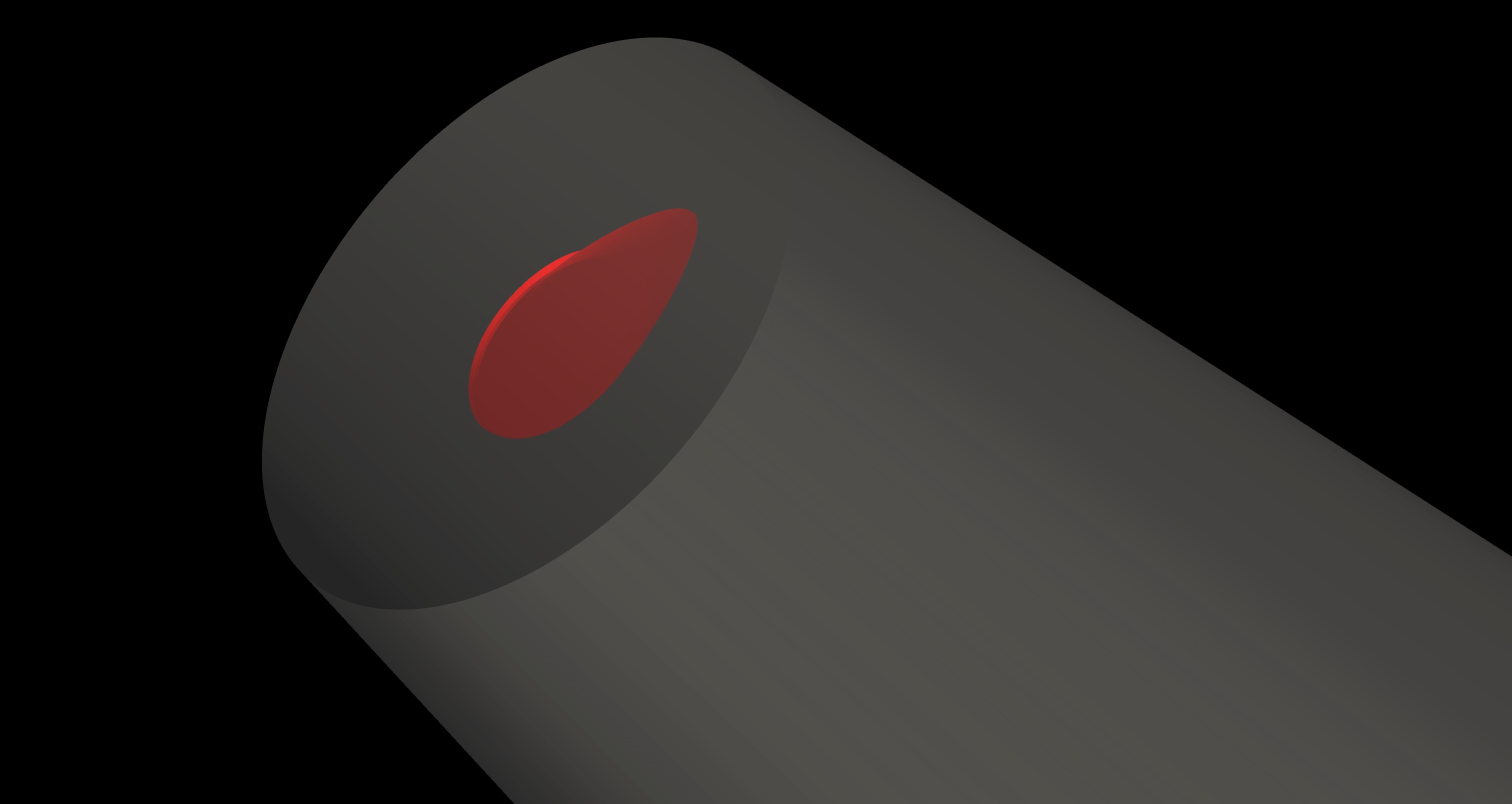}\label{fig:D135_V}}
\qquad
\subfigure[$\phi=180^{\circ}$, $t=$~164.0~s]{\includegraphics[trim=180mm 80mm 250mm 100mm,clip=true,height=4cm]{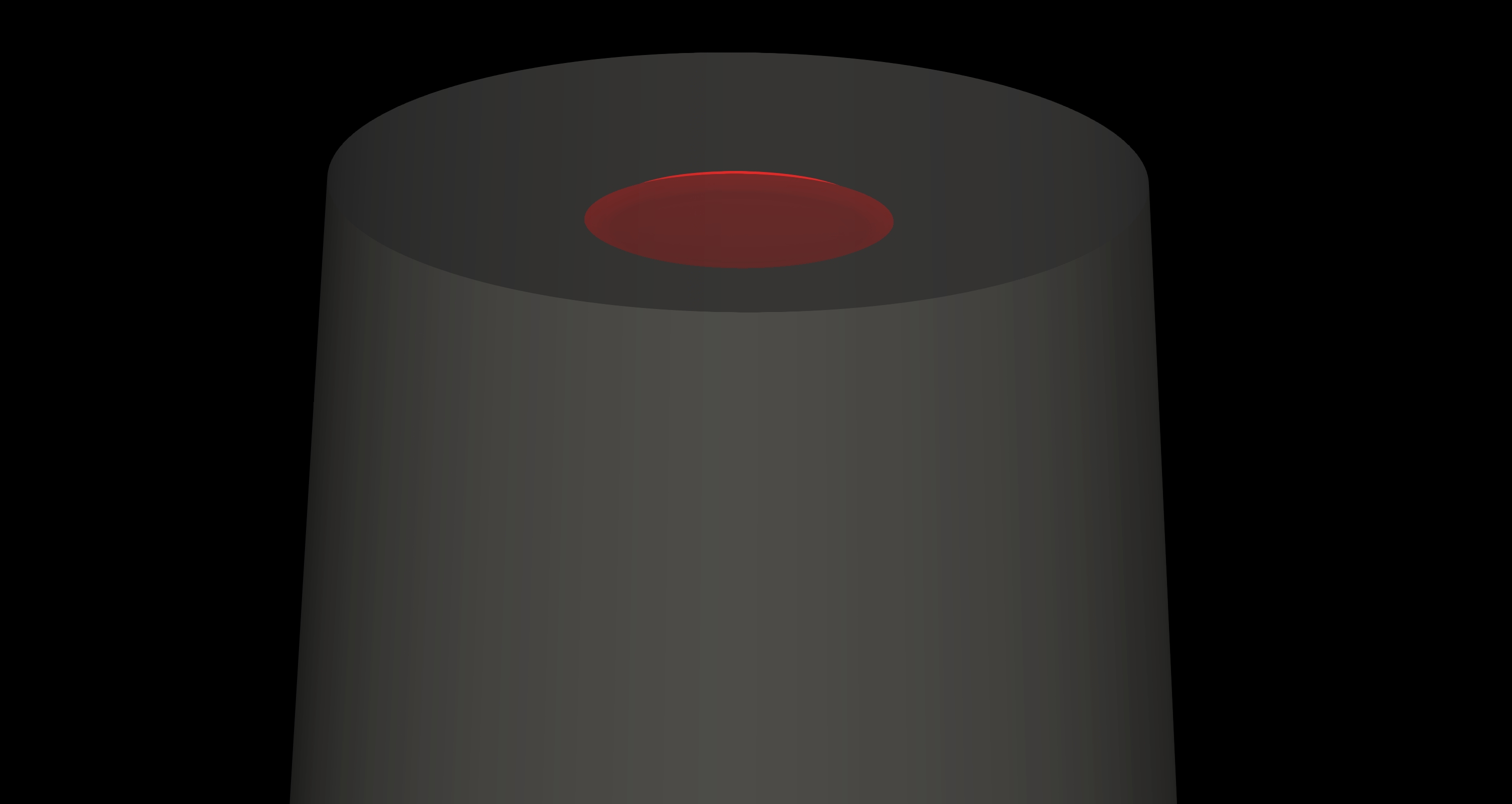}\label{fig:D180_V}}
\end{center}
\caption{Fluid volume which is above the AIT of CS\textsubscript{2} of 362.15~K.
The depicted instantaneous snapshots correspond for each case to the time instance of ignition as observed in the experiments.}
\label{fig:DV}
\end{figure}

For a better understanding of this phenomenon, we plot a sequence of snapshots of the temporal development of the velocity magnitude and vectors in Fig.~\ref{fig:DU_45_t}.
It can be seen that at the first depicted time instance ($t=10$~s, Fig.~\ref{fig:D45_U_10s}) the flow direction is in the direction of buoyancy as expected.
However, already at $t=15$~s (Fig.~\ref{fig:D45_U_15s}) the velocity vectors indicate the formation of a vortical flow structure just above the disc.
The following snapshots (Figs.~\ref{fig:D45_U_21s}~--~\ref{fig:D45_U_33s}) reveal that the bottom part of this vortex pushes the thermal plume to the left and forces is to flow along the bottom of the chamber.
Moreover, one can observe the upward motion of this vortex to the top of the chamber.
Nevertheless, the vortex provides sufficient kinetic energy to keep the plume directed towards the corner of the chamber even after a steady-state is reached.

\begin{figure}[tb]
\centering
\includegraphics[trim=2mm 2mm 2mm 2mm,clip=true,width=10cm]{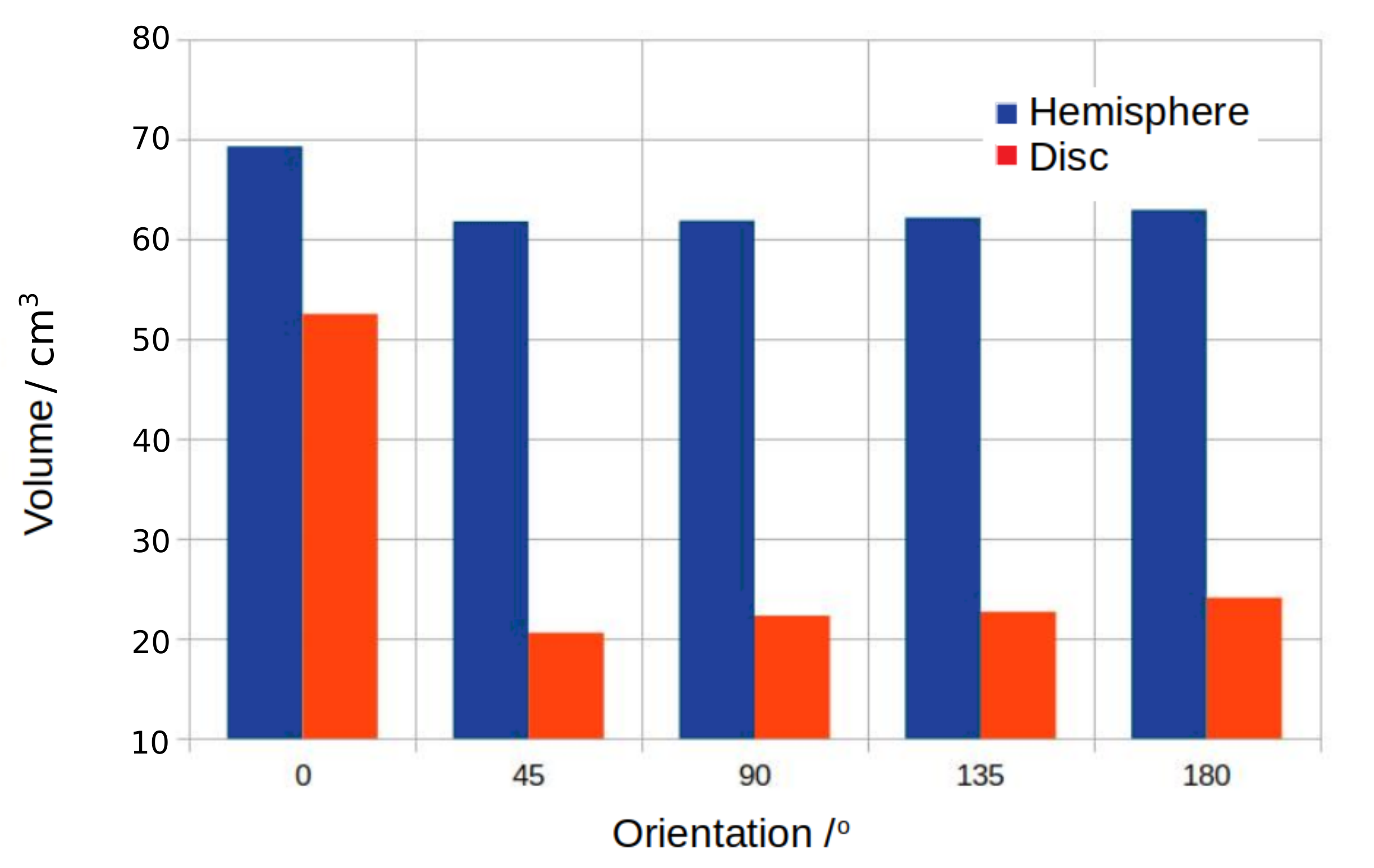}
\caption{Comparison of fluid volume above the AIT of CS\textsubscript{2} of 362.15 K for the hemisphere and the disc at different orientation at the time instance of ignition observed in the experiments}
\label{fig:Volume_Comparison}
\end{figure}

As discussed above, the volume of hot gas is an important parameter for the evaluation of ignition.
The fluid volume occupied above the AIT for all the orientations is shown in Fig.~\ref{fig:DV}.
Further, we compared the size of the volume of all simulated cases and compared them in Fig.~\ref{fig:Volume_Comparison}.

In general, the volumes relating to the disc are smaller than those of the hemisphere for all orientations.
As can be seen, the volume occupied at $\phi=0^{\circ}$ is the largest compared to all other orientations for both disc and hemisphere, followed by the orientation $\phi=180^{\circ}$.
Interestingly, even though the difference in the volume occupied for $\phi=0^{\circ}$ and $\phi=180^{\circ}$ is large, the ignition commences for both orientations nearly at the same time.
Thus, with the exception of the case $\phi=180^{\circ}$, it is noticed from Figs.~\ref{fig:Volume_Comparison} and~\ref{fig:DV} that the ignition time decreases as the volume occupied above the AIT increases, especially in the case of the flat disc.

\section{Conclusion}

In this paper, the thermal flow fields originating from a hot hemispherical and hot disc surfaces of different orientations were investigated numerically through three-dimensional simulations.
\textcolor{black}{
The computational results agree well with data from previous experiments obtained via thermocouples.}
A total of five different orientations were considered ranging from 0$^{\circ}$ to 180$^{\circ}$.
The numerical data allowed a detailed explanation of the formation of the thermal hot spots.
It was observed that the critical hot spot location changes with the orientation of the hemisphere.
More precisely, the location shifts towards the left of the hemisphere with increasing orientation angle $\phi$ in clock-wise direction.
For the case of $\phi=180^{\circ}$, the critical hot spot forms a ring around the base of the hemisphere.
In general, the location of the hot spots in simulations was in good agreement with the ignition location found in the experiments.
A remarkable flow field was observed for the hot disc at an orientation of $\phi=45^{\circ}$.
In this case, a vortex formed which forced the direction of flow towards the corner of the chamber.
\textcolor{black}{
This new knowledge regarding the flow field and heat transfer helps to further understand the ignition of combustible gas mixtures by hot surfaces.
In this sense, the naturally next question arising from the presented results is under which conditions the hot-spots initiate chemical chain reactions.
Thus, in the long-term this work and following research will contribute to plant safety.}

\section*{Acknowledgements}
The authors gratefully acknowledge the financial contribution from the Berufsgenossenschaft Rohstoffe und chemische Industrie (BG RCI).
The last author acknowledges the the funding from the European Research Council~(ERC) under the European Union’s Horizon 2020 research and innovation programme~(grant agreement No.~947606 PowFEct).


\bibliography{References,\string~/essentials/publications.bib}

\end{document}